\documentstyle[PASJadd,psfig]{PASJ95}
\draft
\begin{document}
\title{Metal Abundances in the Hot Interstellar Medium in 
Early-Type Galaxies Observed with ASCA}
\author{{Kyoko} {\sc Matsushita}, {Takaya} {\sc Ohashi}\\
{\it Department of Physics, Tokyo Metropolitan University,
1-1 Minami-Ohsawa Hachioji, Tokyo 192-03}\\
{\it  matusita@phys.metro-u.ac.jp}\\
and\\
 { Kazuo} {\sc Makishima}\\
{\it Department of Physics, University of Tokyo,
7-3-1 Hongo, Bunkyo-ku, Tokyo 113}\\
{\it Research Center for the Early Universe (RESCEU),
University of Tokyo, 7-3-1 Hongo, Bunkyo-ku, Tokyo 113}}

\abst{ We have analyzed ASCA data of 27 early-type galaxies, and
 studied the properties of their X-ray emitting ISM (Inter Stellar
 Medium) in detail.  We found that overlapping lines and free-bound
 continuum cause strong coupling in the derived abundances of various
 elements.  The abundance determination is also difficult due to the
 uncertainties in the Fe-L atomic physics, because Fe-L lines couple
 with O and Ne K-lines.  However, when abundances of $\alpha$-elements
 are fixed, all the plasma codes give similar Fe abundances with a
 scattering of only 20--30\%. To relax the strong coupling among the
 elements, we included 20\% systematic errors in the Fe-L region of
 the spectra. Then, in X-ray luminous galaxies, the derived abundance
 of Fe and $\alpha$-elements both became $\sim 1$ solar within a
 factor of 2. This result relaxes the previous severe discrepancy
 between the ISM and stellar metallicities.  The ISM metallicity in
 X-ray fainter galaxies is still uncertain, but we can at least
 constrain that contribution from type-Ia SN to the ISM abundance is
 lower than in X-ray luminous systems.  These results strongly suggest
 that a large fraction of SN Ia products have escaped into
 intergalactic space.}

\kword{Galaxies:abundance --- Galaxies:ISM --- X-rays:galaxies}

\maketitle
\thispagestyle{headings}

\section{Introduction}

Discovery of X-ray emitting hot Interstellar Medium (ISM) with the
{\it Einstein observatory} has drastically renewed the view of
early-type galaxies (Forman et al.\ 1985; Trinchieri et al.\ 1986;
Canizares et al.\ 1987; Fabbiano et al.\ 1988; Fabbiano 1989).  Since
the hot ISM is considered to be gravitationally confined to a galaxy,
X-ray observations provide useful knowledge about the depth and the
shape of gravitational potential.  Furthermore, chemical composition
of the ISM tells us about the past history of the galaxy evolution,
 since stellar mass-loss products and supernova ejecta are
accumulated in the ISM\@.

The ISM is metal-enriched by stellar mass
loss and Type Ia SNe that are currently observed in elliptical
galaxies. The standard supernova rates predict the metallicity of ISM
to be as high as several times the solar value (e.g.\ Loewenstein and
Mathews 1991; Ciotti et al. 1991; Renzini et al. 1993).  However, the
previous measurements of ISM with ASCA have show that the metallicity
was less than half a solar (Awaki et al. 1994; Loewenstein et
al. 1994; Mushotzky et al. 1994; Matsushita et al. 1994; Arimoto et
al. 1997; Matsumoto et al. 1997).  Giant early-type galaxies are
estimated to have stellar iron abundance of about 1 solar, when
observed strong radial gradient of Mg$_2$ index is taken into account
(Arimoto et al.\ 1997).  Thus, the X-ray measured abundances of ISM
implied that they were even lower than the stellar metallicity.

The very low metal abundances in ICM cast some doubts in the current
scenario of supernova enrichment and chemical evolution of galaxies.
Arimoto et al.\ (1997) discussed various astrophysical aspects of the
inferred low iron abundances in the ISM, concerning the chemical
evolution of galaxies and cluster of galaxies, the evolution of gas
flows in ellipticals, and the heating of the intra-cluster medium.  In
the attempt of interpreting the observed results in a consistent way,
they explored possibilities of hiding or diluting iron in the ISM\@.
However, none of the mechanisms appeared astrophysically plausible,
and they alternatively questioned the reliability of using iron-L
lines as a diagnostic tool to infer abundances from X-ray spectra.
Recently, Buote et al. (1998), Buote et al. (1999) and Buote (1999)
reported that some X-ray luminous galaxies have about 1 solar abundance,
employing a multi-temperature plasma model.

The high quality X-ray spectra of NGC~4636 (Matsushita et al.\ 1997)
enabled us to study systematic error in our abundance determination,
such as the Fe-L atomic physics and abundance ratios.  Based on this
study, we conclude that the ISM abundance of NGC~4636 is in fact $\sim
1$ solar. This value is significantly higher than the previous ASCA
results.

{\it Einstein} observations have shown that the X-ray luminosities of
early-type galaxies scattered by nearly two orders of magnitude, even
for the objects with very similar optical properties (e.g.\ Canizares
et al.\ 1987; Fabbiano et al.\ 1992).  A clue to this long-standing
problem has been obtained recently in the deep ASCA observation of the
X-ray luminous galaxy NGC 4636, in which a largely extended X-ray
emission around the galaxy has been detected.  This indicates that NGC
4636 is sitting in the bottom of a large and deep potential structure
($M \sim10^{13} M_\odot$) filled with hot tenuous plasma (Matsushita
et al.\ 1998). This is the main reason why the system has acquired its
high X-ray luminousity.  ASCA and ROSAT observations of other X-ray
luminous galaxies also indicate greatly extended halos (Matsushita
1997; Matsushita 2000).
 Therefore, the presence or absence of such an extended X-ray
halo is considered as the main origin of the large scatter in the
X-ray luminosity.  Later, Davis and White (1996) discovered a
correlation between the metal abundance and temperature in ISM\@.
Loewenstein et al.\ (1994) report that some X-ray fainter galaxies
show significantly lower metal abundances.
These features suggest that the problem of low metal abundance may be
related with the scatter of X-ray luminosity.

In this paper, we re-examine ISM metallicity in many early-type
galaxies including NGC~4636 and discuss the origin of metals based on
the relation between ISM metallicity and ISM luminosity.  In section
2, we summarize ASCA observations of relevant galaxies.  Section 3
deals with conventional data analysis and results, and in section 4,
we examine systematic uncertainties in the abundance measurements such
as abundance ratios and Fe-L atomic physics. In section 5, we improve
the spectral analysis based on the result of section 4 and study
correlation between the ISM metallicity and the ISM
luminosity. Section 6 gives discussion of the obtained results.

In this paper we adopt for the solar iron abundance the `meteoritic'
value, Fe/H $=3.24\times 10^{-5}$ by number (Anders and Grevesse
1989).

\section{Targets and Observations}

In the ASCA archival data, we selected early-type galaxies with
distances less than 70 Mpc ($H_0 = 75$ km s$^{-1}$ Mpc$^{-1}$) and
B-band luminosities $L_{\rm{B}} >10^{10} L_\odot$\@.  Objects hosting
X-ray bright Active Galactic Nuclei (AGN) and M87, the cD galaxy of
the Virgo Cluster, are excluded.  The sample constructed in this way
consists of 27 galaxies: 6 S0's and 21 ellipticals, as listed in table
1. Our sample is not complete but contains most of the near-by bright
galaxies.  Ten galaxies are located in cluster environments (Virgo and
Fornax clusters) while the others are in the field.  The sample
contains one cD galaxy, NGC~1399 in the Fornax cluster.
Distances to the galaxies are
taken from Tully (1988) if they are listed, or otherwise calculated
assuming the Hubble constant of $H_0 = 75$ km s$^{-1}$ Mpc$^{-1}$;
which generally gives values consistent with the Tully's data.

Table 2 summarizes observational log of the sample galaxies.  Most of
them were observed for $\sim 40$ ks.  The longest observation was made
for NGC 4636 (Matsushita et al.\ 1997): $\sim 40$ ks in PV phase
(1993), and $\sim 200$ ks in AO4 (1995-1996).  All the observations
were carried out with the SIS and the GIS (Ohashi et al.\ 1996;
Makishima et al.\ 1996) together.   The SIS observing mode is
summarized in table 2.  

\section{Conventional Spectral Analysis}

\subsection{Spectral data}
The data are all screened with the standard
criteria; namely, data with cosmic-ray cutoff rigidities less than 6
GeV c$^{-1}$ or with elevation angles less than $5^\circ$ were
excluded.  We also discarded the SIS data when the elevation angle
from the sun-lit earth was less than $30^\circ$.
The response matrices consist of RMF and ARF files.  We employ
gisv4\_0.rmf for the GIS RMF file.  Since the SIS response varies with
time, the SIS RMF for each observation has to be individually prepared
using SISRMGv1.0\@. The ARF files have been produced for each
observation for the SIS and GIS, assuming that each target galaxy is a
point source.  The spectral analysis uses the XSPEC\_v9.0 package.

We accumulated on-source spectrum for each galaxy within 4 times
$r_e$, where $r_e$ is the effective radius (from the \rm{RC3} Catalog)
centered on each galaxy.  When $4 r_{e}$ is less than $3'$, the
accumulation radius is set to $3'$ which covers 80\% of the incident
X-ray photons from a point source.  The background spectrum (Cosmic
X-ray Background and Non X-ray Background) was calculated for each
galaxy by integrating the blank-sky data in the same detector region.
A 5\% systemic error in the CXB brightness fluctuation (Ishisaki 1996)
is included in the background error in a quadratic form.  

The galaxies in the Virgo and the Fornax clusters are surrounded by
strong cluster emission, which needs to be subtracted to obtain the
pure galaxy component.  Similarly, the data for NGC~499 are affected
by the extended emission in the NGC~507 group. Therefore, for NGC~499,
NGC~1404, NGC~4374, NGC~4365, NGC~4406 and NGC~4552, we derived
background spectra from the source pointing data.  The background
spectra in these cases were accumulated in off-source regions with the
same area and a similar distance from the optical axis as the
on-source integration region to match the statistical weight with the
on-source data.  For NGC~1404, the off-source region is also at the
same distance from NGC~1399 to cancel the ICM contribution peaked at
NGC~1399.

We jointly fit the SIS and GIS spectra with the standard two component
model (see Awaki et al.\ 1994; Matsushita et al.\ 1994), consisting of
a thin thermal emission and a thermal bremsstrahlung.  The former
component describes the emission from the hot ISM, and the model
parameters are temperature $kT$ and metallicity. The latter component,
with its temperature fixed at 10 keV, represents the contribution from
low-mass X-ray binaries (Matsushita et al.\ 1994).  The two components
are allowed to have free normalization, but are subjected to a common
absorption by a free column density $N_H$.

\subsection{Single-temperature fits with solar abundance ratios}

The first step of the analysis is fitting the ISM spectra (or soft
component) with the Raymond \& Smith (1977; hereafter R-S model) model
assuming solar ratios of metal abundances.  For weak sources with the
soft component flux less than $6.0\times 10^{-13} \rm{erg~s^{-1}}$~ in
0.5-10.0 keV, the hydrogen column density is fixed to the Galactic
line-of-sight value because spectral fits cannot constrain temperature
and absorption simultaneously.

Results of the spectral fits are summarized in table 3; Fig.\ 1 shows
the SIS spectra of two representive galaxies with different X-ray
luminosities.  As shown in the previous ASCA papers (e.g.\ Matsushita
et al.\ 1997), the spectra can be well described by the two-component
model.  However, the fits are statistically unacceptable for X-ray
luminous galaxies (Table 3). The large $\chi^2$ values mostly come
from the discrepancy between the data and the model around the Fe-L
complex ($0.7 \sim 1.4$ keV), which is clearly seen in the SIS data
(Figure 1).  We will address this problem later.

As shown in figure 2, X-ray luminosities of the ISM emission in
0.5--10.0 keV (i.e.\ the soft component), $L_{\rm{Xs}}$, within the
accumulated radii ($r<4 r_{e}$), differ very much from object to
object, even if their optical properties are similar.  In contrast,
the hard-component luminosities, $L_{\rm{Xh}}$, show rather good
correlation with $L_{\rm{B}}$ (figure 2).  This supports our basic
interpretation (Matsushita et al.\ 1994) that the hard component is
emitted from binary X-ray sources in the host galaxy, particularly
low-mass X-ray binaries (LMXBs), since the number of binary X-ray
sources (mostly involving neutron stars) should be roughly
proportional to its total stellar content.  We note however that
several galaxies such as NGC 507, NGC 5084, IC~4296 and IC~1459
exhibit at least a factor of 2 higher $L_{\rm{Xh}}/L_{\rm{B}}$ ratios
than other galaxies.  These excess hard X-rays are likely to come from
their low-luminosity active nuclei (e.g. Sadler et al. 1989), 
and if we exclude these 
galaxies, $L_{\rm{Xh}}$ correlates even better with $L_{\rm{B}}$.  The
average $L_{\rm{Xh}}/L_{\rm{B}}$ ratio is $10^{-3.81}$, excluding
objects with $L_{\rm{Xh}}/L_{\rm{B}} > 10^{-3.34}$.  This value agrees
very well with the X-ray luminosities of early-type spirals
($L_{\rm{X}}=10^{-3.82} L_{\rm{B}}$) obtained with {\it Einstein}
(Fabbiano et al.\ 1988; Canizares et al.\ 1987), in which X-ray
emission is considered to be dominated by the LMXB emission.
Therefore, the large scatter in the X-ray luminosities of early-type
galaxies discovered by {\it Einstein} is mainly attributed to the
scatter in $L_{\rm{Xs}}$.

Temperatures of the soft component ($kT_s$) turn out to be 0.3--1.2
keV, and roughly correlate with stellar velocity dispersions, $\sigma$ (Figure
3).  This feature implies that the ISM is primarily heated and
confined by the gravity, and, therefore, $kT_s$ reflects the depth of
the gravitational potential of each galaxy. 

Figure 4 shows relation of $L_{\rm{Xs}}$ (in $r<4r_e$) to the soft
component temperature $kT_s$. A clear positive correlation with a
correlation coefficient of 0.6, is seen. 
 Since $kT_s$ is a good indicator of the
potential depth, this correlation shows that the ISM luminosity within
the optical radii is also an indicator of the gravitational potential
configuration.  This temperature-luminosity correlation for early-type
galaxies is much steeper than, and distinct from, that of cluster of
galaxies which is plotted in the same figure.  
This suggests that there is some fundamental difference in the mechanism of
hot-gas confinement between galaxies and clusters.

The hydrogen column density $N_{\rm H}$ derived from the spectral fit
is mostly consistent with the Galactic values. Exceptions are
NGC~1399, NGC~4472, and NGC~7619, in which the measured column
densities are about $(1-2)\times 10^{21}\rm{cm^{-2}}$.

The single-temperature R-S model assuming solar abundance ratios gives
low metal abundances for the ISM, which are 0.1--0.8 solar. A strong
positive correlation is seen between the ISM abundance and the ISM
temperature, or between the abundance and the ISM luminosity, as shown
in figure 5.  These correlations are also noticed in the ROSAT data
(Davis et al.\ 1996 Fig.2). The average stellar iron abundance in
giant elliptical galaxies is about 1.0 solar (Arimoto et al.\ 1997).
Therefore, in the most X-ray luminous galaxies, the ISM abundance is
almost consistent with the stellar metallicity. However, X-ray fainter
galaxies indicate the lower ISM abundance than the stellar
metallicity, which is very difficult to interpret.
Also, there seems
no contribution from metal-rich ejacta of type Ia SNe 
(e.g. Loewenstein et al. 1991; Ciotti et al. 1991; Renzini et al. 1993) 
in the ISM of all the observed galaxies.

To find some clues for this metallicity problem, we looked at ASCA
data of NGC~5018 which indicated extremely low Mg$_2$ indices.  
However, this galaxy
turned out to be very faint in X-rays (see table 2), 
consistent with the correlation 
between $L_X/L_B$ and Mg2 index (Eskridge et al. 1995).
Thus, the data could not constrain its ISM abundance. 

The luminosity and abundance values obtained here agree with the
published ASCA results (Awaki et al.\ 1994; Matsushita et al.\ 1994;
Mushotzky et al. 1994; Loewenstein et al.\ 1994; Matsumoto et al.\
1997; Arimoto et al. 1997; Buote et al. 1998; Buote et al. 1999; Buote 1999).  This paper uses the 'meteoritic' value
for the definition of the solar Fe abundance: the number ratio of Fe/H
to be $ 3.24 \times 10^{-5}$.  The previous papers all use the 
'photospheric'
value, $4.68 \times 10^{-5}$.  The 'meteoritic' value is directly
obtained from meteorites, while the 'photospheric' one involves
spectral analysis of solar photosphere (Anders \& Grevesse, 1989).
Thus, the former is considered more reliable than the latter.

We should emphasize that the ASCA data are the first to enables us to
separate the soft and the hard spectral components unambiguously. The
previous missions did not have necessary spectral resolution or hard
X-ray sensitivity. The ASCA results obtained here are generally
consistent with the previous results by {it Ginga} (Awaki et al. 1991;
Ikebe et al.\ 1992), BBXRT (Serlemitsos et al.\ 1993), and {\it
ROSAT}\ (Trinchieri et al.\ 1994; Forman et al.\ 1993; Davis and White
1996; Fabbiano et al.\ 1994).

As a brief summary, we confirmed that the X-ray spectra of early-type
galaxies show very low metal abundances. X-ray luminous galaxies
indicate large $\chi^2$ values in the spectral fits, which is also a
problem.  In the following subsections, we attempt to address these
problems based on conventional spectral analysis.

\subsection{Multi-temperature model}

The spectral fits with single temperature models were performed in
section 3.2. However, {\it ROSAT}\  observations of X-ray luminous
galaxies show that the projected temperature profiles have significant
gradients by $\sim 0.2$ keV (e.g.\ Trinchieri et al.\ 1994; Matsushita
2000), although gradients for X-ray fainter galaxies are smaller
(Matsushita 2000).  Even when there is no large-scale inhomogeneity,
the ISM may have a temperature fluctuation in small angular scales.
The central regions of clusters of galaxies in fact often show
multi-temperature components, such as that in the Centaurus cluster
(Fukazawa et al.\ 1994; Ikebe et al.\ 1999).

Buote and Fabian (1998), Buote et al. (1999), and Buote (1999) report
that spectral fits with two-temperature or multi-temperature model for
early-type galaxies commonly give larger metal abundances than the
single-temperature fits.  We tried similar models on the SIS and GIS
spectra, assuming multiple R-S components and the 10 keV
bremsstrahlung.  Before fitting the actual data, we carried out
simulations.  Multi-temperature spectra with similar statistical
quality as the data were constructed and fitted with various
models. Single-temperature models were found to give approximately an
average temperature and metal abundance when the temperature range of
the multi component was within $\sim 0.2$ keV\@.  Based on this
result, the fitting models were made up as a sum of seven R-S
components with fixed temperatures of 0.2, 0.4, 0.6, 0.8, 1.0, 1.2 and
1.4 keV, and a 10 keV bremsstrahlung for the hard component.  The
seven R-S components were constrained to have a common abundance, but
were allowed to have separate, free normalizations.  The overall model
was subjected to absorption with a common photoelectric absorber.  The
column density was left free for X-ray brighter galaxies and fixed to
the Galactic value for fainter ones, in the same way as in the
previous subsection.  Further addition of different temperature
components did not change the result.  Thus, these 7 temperatures are
practically enough to reproduce the metal abundance for any given
multi-temperature model, and this model is more universal than the two
or three temperature model in describing the multi-temperature nature
of the ISM\@.

The fitting results are summarized in Table 3 and shown in figure 6
and 7, compared with the previous single temperature fits.  The 2
models give very similar $\chi^2$ values (figure 6). This simply
indicates that the large $\chi^2$ values for X-ray luminous galaxies
are not caused by an incorrect modeling of multi-temperature emission.
As reported by Buote et al.\ (1998), the 7-temperature fit increases
metal abundance for NGC 1399 and NGC 4472 by 50\%.  However, the
increment is only 10--20\% in other X-ray luminous galaxies, and the
average metal abundance for all the luminous galaxies increases only
by $\sim 30\%$ (figure 7).
 The uncertainty is larger for X-ray fainter galaxies
because of the increased degree of freedom in the temperature space.
However, temperature gradients are smaller in the fainter galaxies as
shown in Matsushita (2000).

We conclude that the low abundance problem in the previous section
cannot be solved by the multi-temperature model, at least in 
the several  X-ray luminous galaxies.  
\subsection{Deviation from ionization equilibrium}

The major assumption we have adopted is that the ISM is in a complete
ionization equilibrium.  Generally, the degree of ionization of a
plasma is parameterized by the quantity $nt$ (Masai 1984), which is a
product of the electron density $n$ and the elapsed time $t$ measured
from the initial time of ionization when the gas was neutral.  The
plasma reaches an ionization equilibrium when $nt$ exceeds
$\sim10^{12}\rm{\ cm^{-3}\ s}$.  Since the electron density in the ISM
of early-type galaxies is typically $n = 10^{-1}\sim 10^{-3}\rm{\
cm^{-3}}$ (Forman et al.\ 1985; Trinchieri et al.\ 1986; Canizares et
al.\ 1987), the ionization equilibrium is attained in only $3\times
10^{5\sim 7}$ yr. The ISM should, therefore, be in a complete
ionization equilibrium. 
 However, there is a possibility that the
metals are locked in dust grains, which may be able to explain the
observed low metallicity in the ISM\@. These dust grains are sputtered
by hot electrons in the ISM and gradually ionized.  The time scale for
the grain evaporation is similar to the above ionization equilibrium
time scale (Itoh 1989; Masai 1984), so the spectral features should
indicate deviation from ionization equilibrium if dust sputterings are
still going on.  

Among the plasma emission codes, only Masai model deals with plasmas
in non-ionization-equilibrium (NIE) condition.  With this NIE model,
we fitted the SIS and GIS spectra of all the sample galaxies,
similarly as in section 3.2. The derived temperature and abundance
will be separately examined later in section 4.2. Figure 8 summarizes
the derived $nt$ parameter as a function of best-fit
temperatures. Most of the objects consistent to have
 $nt \sim 10^{12}\rm{\ cm^{-3}\ s}$,
 implying that the ISM is in an ionization equilibrium.

\section{Problems in the Spectral Analysis}

As shown in section 3, no solution was given to the low-abundance
problem by the multi-temperature models or by the consideration of
non-ionization equilibrium.  The remaining points we need to examine
are deviation of abundance ratios from the solar values and
uncertainty in the Fe-L atomic physics.

\subsection{Uncertainty in the abundance ratios}

Abundance ratios provide key information about the origin of the
ISM\@. However, deriving elemental contributions from an X-ray
spectrum is a complicated task for a temperature $kT \sim 1$ keV\@.
Figure 9 shows the best-fit R-S model for the NGC~4636 data decomposed
into the constituent elements.  A major part of the emission comes
from Fe-L complex.  Considerable contribution comes from K-emission
lines of lighter elements, such as O, Ne, Mg, Si, and S\@.
Furthermore, Ni-L lines overlap with the Fe-L complex.  Another
problem is caused by free-bound emission, which gives a step-like
addition on the bremsstrahlung continuum at the energies of K and
L-edges of each element.  
These spectral features due to free-bound
continuum and emission lines in a single spectrum have to be fitted in
a consistent manner in obtaining the metal abundance.

To avoid the complexity in the spectral fit as much as possible, we
have combined the elements heavier than H and He into two groups. One
group consists of $\alpha$-elements, O, Ne, Mg, Si and S\@. A single
common abundance is assigned to these elements, which is hereafter
denoted by $A_{\rm{\alpha}}$.  The other group consists of Fe and Ni,
and their common abundance is denoted by $A_{\rm{Fe}}$. This grouping
was successfully applied to the data of NGC 4636 (Matsushita et al.\
1997).  Obviously, the division of elements reflects metal synthesis
processes; the former group elements are produced in SNe II and the
latter ones in SNe Ia, respectively.  Although Si and S are partly
synthesized in SNe Ia, at least the ASCA data of NGC 4636 indicate
that Mg to Si abundance ratio agrees well with the solar ratio
(Matsushita et al.\ 1997).  Therefore, it seems reasonable in the
present analysis to assume that all the $\alpha$-elements (O, Ne, Mg,
Si and S) have the same abundance.

The GIS and SIS spectra are jointly fitted with the same
 model as before,
with $A_{\rm{\alpha}}$ and $A_{\rm{Fe}}$ as  free parameters. All
the galaxies are analyzed in the same way, and the results are
summarized in table 4.  The $\chi^2$ values show no improvement in
this fit (figure 10).
Figure 11 shows
confidence contours of $A_{\rm{\alpha}}$ against $A_{\rm{Fe}}$ for 8
representative galaxies, in an increasing order of the ISM luminosity.
The contours all show narrow elongated shapes, indicating that the two
parameters couple strongly. This is mainly because K-lines and
free-bound emission from $\alpha$-elements push up the continuum level
for the Fe-L lines. 
As a result, the
uncertainty has increased by several times compared with the case of
the solar ratio fit.

For X-ray luminous galaxies the error regions shown in figure 11 are
very much underestimated, because their spectral fits are formally
unacceptable with reduced $\chi^2$ larger than 1.5.  To examine
whether much higher abundance is allowed by the data, we forced
$A_{\rm{\alpha}}$ to 1.0 solar and left $A_{\rm{Fe}}$ free to vary.
The results are summarized in table 4.
Figure 12 compares the two fits, 
In both cases, the data-to-model ratio deviates  to
a similar extent.  The increment of the reduced $\chi^2$ from the
previous variable abundance fit is only $\sim 10\%$ even for X-ray
luminous galaxies (figure 10). However, the previous fits were not
acceptable, either. The additional deviation of the model mainly
occurs in the O-K line region.  This O-K line problem is in fact much
smaller than the deviations already present in the variable abundance
fit around 0.7--0.8 keV, namely in the region of Fe-L complex.
Therefore, the spectral fit cannot clearly reject the case of $A_{\rm
\alpha} =1.0$ solar even for X-ray luminous galaxies.

When $A_{\rm \alpha}$ is fixed at 1.0 solar, $A_{\rm{Fe}}$ correlates
well with the ISM luminosity as shown in figure 13.  In this case,
$A_{\rm{Fe}}$ becomes about 1.0 solar in X-ray luminous ($L_X>10^{41}
\rm{~erg\ s^{-1}}$) galaxies, and 0.5 solar in X-ray fainter
ones. Thus the scatter in the iron abundance among the bright and
faint galaxies is now only within a factor of 2, while it was an
order-of-magnitude difference in the solar-abundance-ratio fit (see
figure 5).  We note that there is no difference between S0 and
elliptical galaxies.  This point will be discussed in section 6.2.3.

To summarize the fitting results, the strong coupling between
$A_{\rm{\alpha}}$ and $A_{\rm{Fe}}$ has made the abundance estimation
difficult; on one hand, this coupling hampers unique determination of
the overall metallicity. On the other hand, if we shift the
$A_{\rm{\alpha}}/A_{\rm{Fe}}$ ratio slightly off from the unity, the
allowed metallicity becomes significantly higher than the results of
solar ratio analysis.

\subsection{Problems in the iron-L atomic physics}

The spectral fit is made complicated due to the uncertainty in the
Fe-L atomic physics.  The abundances reported in the previous
subsections have been obtained with the R-S thin thermal plasma
emission code.  However, the model cannot fit the Fe-L structure of
X-ray luminous galaxies, even when we assume the multi-temperature
model or allow the abundance ratio to vary.  The similar results were
reported for the cluster emission by Fabian et al.\ (1994).  As
Arimoto et al.\ (1997) and Matsushita et al.\ (1997) have
investigated, other plasma codes, MEKA (Mewe et al.\ 1985; Mewe et
al.\ 1986; Kaastra 1992), MEKAL (Liedahl et al.\ 1995) and Masai
(Masai 1984) model, differ from the R-S model mainly in the treatment
of the Fe-L atomic physics.  When the plasma temperature is lower than
$\sim 2$ keV, these codes give considerably different spectral
features (Masai 1997).  We need to examine the goodness of the fit and
the inferred abundance for these different spectral codes.

By changing the spectral models, we fitted the SIS and GIS data of all
galaxies exactly in the same way as described in section 4.1.  The
double-component model (thermal and hard bremsstrahlung) was assumed
with $A_{\rm{\alpha}}$ and $A_{\rm{Fe}}$ as free parameters.  We
assumed an ionization equilibrium in the fit.

As summarized in Table 5, the results depend heavily on the adopted
plasma emission code.  As shown in figure 14, the reduced $\chi^2$
values are fairly model dependent for bright galaxies.  Figure 12
shows data-to-model ratio for the R-S and MEKAL models in the case of
SIS spectra for two X-ray luminous galaxies, NGC~4636 and NGC~4472.
Either of the models cannot reproduce the Fe-L complex adequately, and
the model-to-data relation is significantly different between the two
plasma emission codes.

The physical parameters derived from the spectral fits are also
strongly model dependent.  Figure 15a compares the ISM temperature and
hydrogen column density for different plasma codes. The difference in
the temperature is relatively systematic: the MEKA model gives the
lowest temperature among the four codes, about 20\% lower than the
case of R-S model, while MEKAL and Masai models predict temperatures
within 10\% of the R-S values.  As discussed in Masai (1997), the
different ionization and recombination rates of Fe mainly causes these
discrepancies in the spectral parameters.  The hydrogen column density
also depends on the plasma emission codes (figure 15b).  MEKA and MEKAL
models give consistent or slightly higher values than the R-S model,
while Masai model tends to indicate lower absorption.  For galaxies
with large column densities, the difference sometimes amounts to
$10^{21}\rm{~cm^{-2}}$.

The metal abundance shows a strong model dependence (figure 16a), with
large uncertainty in X-ray fainter galaxies.  The MEKAL and the R-S
models give similar values of $A_{\rm{\alpha}}$ and $A_{\rm{Fe}}$, and
the MEKA and Masai models indicate systematically higher abundances by
a factor of $ 2 \sim 3$ on the average.  In particular, the Masai
model gives an abundance up to $\sim 2$ solar for the most X-ray
luminous galaxies.

However, figure 17 shows that the correlation between the two
abundance groups, $A_{\rm{\alpha}}$ and $A_{\rm{Fe}}$, forms a very
similar elongated shape for all models as already seen in the case of
R-S fit (in section 4.1). In X-ray fainter galaxies, the 4 error
regions overlap and the implied abundances are consistent with each
other among the plasma codes.  In X-ray luminous galaxies, the error
regions are aligned in a narrow band which runs almost in parallel
to the elongation direction of each region.

As a result, once $A_{\rm{\alpha}}$ is given, the difference in
$A_{\rm{Fe}}$ among the models must be fairly small.  To confirm this
point in a quantitative manner, we fitted the spectra with
$A_{\rm{\alpha}}$ fixed at 1.0 solar and looked at the difference in
$A_{\rm{Fe}}$. The results are listed in Table 5 and shown in figure
16b. Clearly, $A_{\rm{Fe}}$ for different models fluctuates by only
20--30\%.

The strong coupling between Fe-L lines and the spectral features due
to other light elements is the main factor which complicates the
abundance determination. The level of coupling is significantly
different among the plasma emission models, which reflects the
uncertainty in the Fe-L atomic physics. When $A_{\rm{\alpha}}$ is
fixed, the difference with the plasma code reduces to only 20--30\% as
shown in Figure 16b.  This means that if we are able to determine the
level of the free-bound continuum, the abundance using the Fe-L lines
will be effectively constrained.

\section{Improved Spectral Analysis}

\subsection{Spectral fits above 1.6 keV}

Based on the study in section 4, we first try to determine
$A_{\rm{\alpha}}$ which is the common abundance of so-called
$\alpha$-burning elements.  Among them, lines from O and Ne are
difficult to constrain from the ASCA data
due to false coupling between the Fe-L lines and the O-K and Ne-K lines 
(see section 4.1, section 4.2).
In contrast, Si-K and S-K lines are
clearly seen in the spectra of X-ray luminous galaxies,

To utilize K-lines of Si and S,  we exclude
the Fe-L region, which is $< 1.6$
keV for the SIS and $< 1.7$ keV for the GIS data, respectively.
Because of
the effect of free-bound continuum from O, Ne, and Mg, the
determination of $A_{\rm{\alpha}}$ is not straightforward. 
We need to assume relative abundance
ratios among the $\alpha$-burning elements. The ASCA data for NGC 4636
indicate that the abundance ratio between Mg and Si agrees with the
solar value within $\sim 30\%$ (Matsushita et al.\ 1997). We assume
here the same condition for other X-ray luminous galaxies, namely O,
Ne, and Mg are all assumed to have the same abundance. 
Thus, 
the fitting model 
consists again of the 10 keV bremsstrahlung and the R-S model,
with the latter employing the same elemental groups of
$A_{\rm{\alpha}}$ and $A_{\rm{Fe}}$ (see section 4.1).

The fits turned out to be acceptable (table 6), mainly because the
Fe-L region was excluded.  The temperature shows a typical error of
$\sim0.1$ keV, and are broadly consistent with
those in the previous  fits  involving the whole energy range (table 3,4,5). 
The confidence
contours of $A_{\rm{Fe}}$ vs.\ $A_{\rm{\alpha}}$ for these hard band
fits are shown in figure 18 as shaded regions.  The omission of the
Fe-L region naturally leads to a much larger error of $A_{\rm{Fe}}$
(figure 18).  Nevertheless, $A_{\rm{\alpha}}$ has been relatively well
constrained with no strong dependence on $A_{\rm{Fe}}$. This is
because free-bound continuum of Fe-L is weaker than those of O and Ne.
In NGC 4636 and NGC 4472, $A_{\rm{\alpha}}$ is
approximately one solar when $A_{\rm{Fe}}$ is around one solar.  If
one tries to raise $A_{\rm{\alpha}}$ up to 2 solar, then $A_{\rm{Fe}}$
would have to be as high as 5 solar as implied by the shaded region in
figure 18.  In NGC 507 and NGC 1399, $A_{\rm{\alpha}}$ is smaller than
the level of NGC 4636 for the same value of $A_{\rm{Fe}}$.  Other
X-ray luminous galaxies also show a similar correlation with larger
uncertainties.

\subsection{Additional systematic errors to the Fe-L region}

The successful determination of $A_{\rm{\alpha}}$ in section 5.1 has
left $A_{\rm{Fe}}$ poorly constrained, because of the complete
ignorance of the Fe-L data which must be very informative.  We need
somehow to utilize the information carried by the low-energy spectra.
The method we take here is recovering the low-energy ($< 1.6$ keV)
data with reduced statistical weight assigned on them.  The low
statistical weight should prevent the spectral fit from imposing a
strong false constraint on $A_{\rm{\alpha}}$, which was formerly
resulted when the Fe-L band spectrum was fitted too faithfully.
Furthermore, the increased error helps to make the fit formally
acceptable and enables us to evaluate the parameter errors in a
straightforward manner.  

The best-fit models show difference from the data in the Fe-L region
with a typical amplitude of $\sim 20\%$ for X-ray luminous galaxies
(figure 12). The deviation patterns are significantly different among
the models.  When $A_{\rm{\alpha}}$ is fixed, the variation of
$A_{\rm{Fe}}$ with the plasma code also becomes $20 \sim 30\%$. This
level of variation is also expected from the theoretical consideration
by Masai (1997).  Based on these features, we will introduce an
additional systematic error of 20\% in the 0.4--1.6 keV region.  This
level of error still keeps the ability of spectral data to constrain
the model in the Fe-L region with reasonable strength, as show below.

With this procedure, all the spectral fits have become acceptable for
all emission models (table 7).  The additional 20\% error was
confirmed to be appropriate, since an error level of 10\% gives too
large $\chi^2$ (reduced $\chi^2>1.5$) and 30\% gives too low $\chi^2$
(reduced $\chi^2 <0.7$) values, respectively.  The derived temperature
and hydrogen column density are consistent with the previous results.
The confidence contours for $A_{\rm{\alpha}}$ vs.\ $A_{\rm{Fe}}$,
shown in Figure 18, are still elongated, but different plasma emission
codes now indicate overlaps with each other.  This feature enables us
to estimate confidence regions for $A_{\rm{\alpha}}$ and
$A_{\rm{Fe}}$.  NGC 4636 indicates both abundances to be $\sim 1$
solar within a factor of 2.  Other X-ray luminous galaxies show
$A_{\rm{\alpha}}$ and $A_{\rm{Fe}}$ to be $0.5 \sim 1.0$ solar.  The
ratio of $A_{\rm{\alpha}}$ to $A_{\rm{Fe}}$ is within $\pm 50\%$ of
the solar value at the 90\% confidence for all galaxies.  These
abundance values differ significantly from those obtained with the
exact solar-ratio assumption described in section 3.2. The
$A_{\rm{\alpha}}$ values agree well with the hard-band results in the
previous subsection. This consistency supports that our treatment of
the data in the Fe-L region is appropriate.

\subsection{Correlation between ISM metallicity and luminosity}

Correlation between the metallicity and the X-ray luminosity of the
ISM needs to be examined again. X-ray fainter galaxies show very weak
lines from $\alpha$ elements (see figure 1), and their soft X-ray
continua are masked by rather strong hard components. These conditions
hamper accurate determination of $A_{\rm{\alpha}}$ in these galaxies,
also making $A_{\rm{Fe}}$ values highly uncertain.

In order to cope with this difficulty in the X-ray fainter galaxies,
we have combined the data of all target galaxies into 6 groups
according to their ISM temperature and X-ray luminosity as shown in
Table 8. The spectra in each group are fitted simultaneously.  The ISM
temperature and 2 normalization parameters for the soft and hard
components are free parameters assigned for each galaxy, while all the
galaxies are assumed to have common $A_{\rm{\alpha}}$ and
$A_{\rm{Fe}}$ values.  The 20\% systematic error is included in the
Fe-L region for X-ray luminous galaxies ($L_X>10^{40}\rm{~erg\
s^{-1}}$), but X-ray fainter galaxies ($L_X<10^{40}\rm{~egs\ s^{-1}}$)
go through the fitting directly because statistical errors are much
larger.

The results from this group analysis are summarized in table 8 and
shown in figure 19.  Acceptable fits are obtained for all groups with
the parameter values consistent with those derived in section 5.2.
X-ray luminous galaxies give consistent results for all the plasma
emission codes. The confidence contours for X-ray fainter galaxies
indicate large uncertainties for both $A_{\rm{\alpha}}$ and
$A_{\rm{Fe}}$, similar to the features in figure 11. For a given
$A_{\rm{\alpha}}$, however, the power of constraining $A_{\rm{Fe}}$
has become better.

All the plasma emission codes show separation of confidence contours
for the faint and the luminous groups. X-ray luminous galaxies
indicate $A_{\rm{Fe}}$ and $A_{\rm{\alpha}}$ to be $0.5 \sim 1$ solar,
and their ratio is within $\sim$ 30\% of the solar value.  The X-ray
fainter galaxies show significantly lower $A_{\rm{Fe}}$ for the same
$A_{\rm{\alpha}}$.  When the ISM temperatures are similar, the
strength of Fe-L lines directly indicates the abundance difference.
The abundance comparison between X-ray luminous and faint
low-temperature ($0.6<kT<0.8$ keV) groups should, therefore, involve
very small systematic error. We conclude that the ISM abundance
depends more on the ISM luminosity than on the ISM temperature.

Among X-ray luminous galaxies, the high temperature group indicates
low $A_{\rm{\alpha}}$ and $A_{\rm{Fe}}$ values at $\sim$ 0.5 solar.
Representative galaxies are NGC 507 and NGC 1399, which show low
$A_{\rm{\alpha}}$ values determined from Si K-line intensity for a
given value of $A_{\rm{Fe}}$ as shown in figure 18. The abundance
ratio is similar to those in the other X-ray luminous groups.

\section{Discussion}

We have analyzed ASCA data for 27 giant ($L_B>10^{10}L_\odot$)
early-type galaxies.  
X-ray luminosities of these galaxies are $10^{40\sim 42}
\rm{~ergs^{-1}}$.

\subsection{Summary of results}
The spectra of the early-type galaxies have been fitted reasonably well
with a $ 0.3-1.0$ keV thermal model (the soft component) and a 10 keV
bremsstrahlung~model (the hard component).  The hard component should mostly
consist of emission from LMXBs such as  in early-type spirals and
in galactic bulges. 
By separating the pure ISM component,  we showed that
the large scatter in the overall X-ray luminosity of these objects
was due to fluctuations in the ISM luminosity.

Unusually low abundances
($\sim 0.3$ solar) were obtained with the R-S plasma model assuming
solar-abundance ratios. 
We carefully examined whether these results are artifacts of the data 
analysis.
We showed that multi-temperature models could
not change the results of the most X-ray luminous galaxies,  
alghouth higher abundances were allowed for
X-ray fainter galaxies simply because of poor data statistics.
Assumption of non-ionization equilibrium could not solve the problem
either.

We found that the overlap among emission lines and free-bound continuum
cause strong coupling among the abundances of various elements.  As a
result, when models with variable abundance ratios are fitted to the
data, the allowed range of abundance increases dramatically while keeping
strong correlations among different elements.  Another difficulty is
the large uncertainty in the Fe-L atomic physics. Different plasma
emission codes give significantly different abundances, and none of
them provide acceptable fits to the Fe-L spectra of X-ray luminous
galaxies.  In contrast, when abundances of $\alpha$-elements
($A_{\rm{\alpha}}$) were fixed, all the plasma codes give similar
Fe-abundances ($A_{Fe}$) within 20--30\%.  Therefore, the strong
dependence of the
abundance on the emission code is mainly attributed to the
strong coupling between Fe-L and $\alpha$-element lines (particularly
O-K and Ne-K lines).

In order to solve the above coupling problem, we have assumed
$\alpha$-elements to take the solar abundance ratios based on the
results by Matsushita et al.\ (1997).  Effect of the free-bound
emission of Fe-L on the Si abundance was estimated, and the Si
abundance was concluded to be roughly 1 solar within a factor of 2 in
X-ray luminous galaxies.  Finally, we included additional systematic
errors by 20\% in the Fe-L region of the spectra and relaxed the
strong coupling between Fe and $\alpha$-elements (particularly O and
Ne).  This has allowed the Si-K lines to make a significant 
contribution to  the $\chi^2$
minimization for several X-ray luminous galaxies, and as a
consequence, the derived $A_{\rm{\alpha}}$ and $A_{\rm{Fe}}$ both became $\sim 1$ solar
within a factor of 2. In addition, discrepancy among the plasma
emission codes has been reduced, and the spectral fits have become
mostly acceptable with all plasma models.

In contrast to the luminous objects, the ISM metallicity in X-ray
faint galaxies remained highly uncertain because of very large
 statistical errors in
the Si line intensity.  However, strong correlation
was again seen between Fe and $\alpha$-element abundances, and 
the $A_{\rm{Fe}}$ vs.\ $A_{\rm{\alpha}}$  confidence
regions showed statistically significant  dependences on the X-ray luminosity.

\subsection{ISM metallicity and chemical evolution}

The hot ISM is considered to be a mixture of the stellar mass-loss products
and the supernova ejecta.  Therefore, elemental abundances in the ISM
measured with X-ray spectra give almost unique information on the
contribution from these enrichment processes.

More specifically, the ISM abundances can be expressed as,
\begin{equation}
z^i =\frac{\alpha_*y^i_{*}+\alpha_{\rm SN}y_{\rm
 SN}^i}{\alpha_*+\alpha_{\rm SN}} \sim y_{*}^i+\langle
 \frac{\alpha_{\rm SN}}{\alpha_{*}} \rangle y_{\rm SN}^i
\end{equation}
(Loewenstein and Mathews 1991; Ciotti et al.\ 1991), where $z^i$ is
the mass fraction of the $i$th element, $\alpha_*$ and $\alpha_{\rm
SN}$ are the mass loss rates of stars and SNe respectively, and
$y_{*}^i$ and $y_{\rm SN}^i$ are their yields.

\subsubsection{Is the ISM Diluted by the ICM?}

We first check the validity of 
equation (1) in which 
we assume that the whole ISM in a galaxy consists of matter supplied
 by the stellar component of the galaxy itself. 
Elliptical galaxies are often surrounded by  a hot ICM,  which
 can be accreted if an inflow is
established, and dilute the indigenous ISM (Renzini et al.\ 1993).

Eskridge et al. (1995) discovered a strong correlation between $L_X/L_B$
and potential depth, or stellar velocity dispersion.
Later,    
the existence of largely extended X-ray emission around some X-ray
luminous galaxies such as NGC 4636 (Matsushita et al.\ 1998) and NGC
1399 (Ikebe et al.\ 1996) suggests that these galaxies are located in
the bottom of a largely extended potential well filled with a tenuous
plasma.  If surrounding metal poor gas is accreting onto a galaxy, the
galaxy becomes overluminous in X-rays, as the additional $PdV$ work by
the infalling material needs to be radiated away in a quasi-stationary
cooling flow and become metal poor.  Actually, the most luminous
objects, NGC 507, which is surrounded by diffuse extended group ICM
(Trinchieri et al.\ 1997 ), and NGC 1399 tends to have slightly lower
ISM abundance than the other X-ray luminous galaxies.  The similarity
of $A_{\rm{\alpha}}$ to $A_{\rm{Fe}}$ ratio among the X-ray luminous
objects suggests that the low metallicity in the most luminous objects
may be slightly more diluted by metal poor gas than other X-ray
luminous galaxies.

X-ray fainter objects show small scatter of $A_{\rm{Fe}}$ for a given
level of $A_{\rm{\alpha}}$, as shown in figure 13.  The sample
contains objects surrounded by dense ICM, such as NGC 1404 and NGC
4374, and also relatively isolated galaxies, such as NGC 720 and IC
1459.  This suggest that the ISM in the X-ray faint objects are not
significantly diluted by ICM\@.  As summarized in figure 19, X-ray
luminous objects show higher metal abundance than X-ray fainter ones.
Therefore, dilution may well occur in the X-ray luminous objects, and
in this case the abundance difference between the luminous and faint
galaxies becomes even larger.

\subsubsection{Constraints on the SNe Ia rate}

From the equation (1),
 combining the latest observed SNe Ia rate with the Hubble constant 
of  $75~ {\rm km s^{-1} Mpc^{-1}}$
(Cappellaro et al.\ 1993; Cappellaro et al.\ 1997; Tammann et al.\ 1995; van den Bergh et al.\ 1991),
with  a nucleosynthesis
calculation by Thielemann et al.\ (1996) and a stellar mass loss rate
by Ciotti et al.\ (1991), 
the abundance of Fe enriched only by
the SNe Ia is predicted to be  2 solar.

The measured Fe abundance actually consists of contributions from SNe
II and SNe Ia; the former one is supplied via stellar mass loss, while
the latter are by SN mass ejection and by stellar mass loss, since we
do not know whether stars contain SNe Ia products or not.  If we would
explain the observed Fe abundance in terms of SNe Ia only, then it
gives an upper limit of the SNe Ia rate.  The SNe II ejecta are
estimated to have an abundance ratio $A_{\rm{\alpha}}/A_{\rm{Fe}}\sim
3$ (e.g.\ Thielemann et al.\ 1996), while SNe Ia products have
$A_{\rm{\alpha}} \ll A_{\rm{Fe}}$.  Si and S can be supplied by both
SNe II and SNe Ia, but the latter contribution should be negligible as
long as the SN Ia rate is low.  Thus, contribution from SNe Ia to
$A_{\rm{Fe}}$ (${A_{\rm{Fe}}}_{\rm SN Ia}$) is estimated as
\begin{equation}
 {A_{\rm{Fe}}}_{\rm SN Ia}=A_{\rm{Fe}}-1/3A_{\rm{\alpha}}.
\end{equation}
On the $A_{\rm{\alpha}}$ vs.\ $A_{\rm{Fe}}$ plane, we draw lines of
$A_{\rm{Fe}} = \rm{const.} + 1/3 A_{\rm{\alpha}}$ as shown in figure
19 by dotted lines. Each line indicates a constant contribution from
SNe Ia.

The large uncertainty in $A_{\rm{\alpha}}$ in figure 19 indicates that
the contribution of SN II is highly ambiguous. However, the confidence
contours for $A_{\rm{\alpha}}$ vs.\ $A_{\rm{Fe}}$ are rather narrow
along the lines of constant SN Ia contribution.  This is most clearly
seen for X-ray fainter galaxies.  This means that the contribution of
SN Ia is well constrained by the data. By extrapolating these contours
along the dotted lines in figure 19 to a small $A_{\rm{\alpha}}$
value, we can estimate the SN Ia contribution to $A_{\rm{Fe}}$.  X-ray
fainter galaxies indicate the iron abundance attributable to SN Ia,
${A_{\rm{Fe}}}_{\rm SN Ia}$ inferred from equation (1), to be less
than 0.2 solar for R-S, MEKA, and MEKAL models, and $0.2 \sim 0.5$
solar for Masai model, respectively.  In contrast, X-ray luminous
galaxies show ${A_{\rm{Fe}}}_{\rm SN Ia}$ to be about 0.5 solar for
RS, MEKA, and MEKAL and $\sim 1$ solar for Masai model, respectively.
With the slight difference among the plasma emission codes, we can
conclude that the SN Ia contribution to $A_{\rm{Fe}}$ is positively
correlated with the X-ray luminosity.

Taking the result by Masai model, which gives the highest Fe
abundance, the upper limit of the inferred SNe Ia rate is a factor of
2 and 5 smaller than the Cappellaro's value for the X-ray luminous and
fainter galaxies, respectively.  As mentioned in section 6.2.1, the
ISM in X-ray luminous objects may be simply diluted by ICM\@.
However, the low Fe abundance in X-ray fainter galaxies is a severe
problem which contradicts with the SN rate.

\subsubsection{Constraint on stellar metallicity}

Optical measurements of stellar metallicity in giant early-type
galaxies have been performed only within $\sim 1 r_e$.  In our sample
galaxies, the average stellar metallicity over an entire galaxy is
typically $0.5\sim 1 $ solar taking into account the observed
metallicity gradient beyond $\sim 1r_e$ (Arimoto et al.\ 1997;
Kobayashi et al.\ 1999).  Stellar metal ratios are reported to be SNe
II-like at least in the center of galaxies (Worthey et al.\ 1992;
Kobayashi et al.\ 1999).

X-ray observations are able to constrain stellar metallicity over an
entire galaxy, which is hardly performed with optical observations.
Based on equation (1), $A_{\rm{Fe}}$ gives an upper limit of the
stellar Fe abundance.  $A_{\rm{\alpha}}$, on the other hand, directly
reflects the stellar $\alpha$-element abundance, since SNe Ia
contribution is negligible even for Si and S\@. X-ray luminous
galaxies show both $A_{\rm{Fe}}$ and $A_{\rm{\alpha}}$ to be $\sim
1.0$ solar, which is fairly close to the stellar levels. A few
galaxies such as NGC 507 indicate somewhat lower ISM abundance.  Thus,
the previous severe discrepancy, that the ISM metallicity is even
lower than the stellar metallicity (e.g.\ Awaki et al.\ 1994; Arimoto
et al.\ 1997), has been mostly removed based on our improved analysis.

Since Mg$_2$ indices and colors are similar for all the observed
galaxies (Kodama and Matsushita 2000), it is reasonable to assume that
X-ray fainter galaxies have similar stellar metallicity to those in
X-ray lumious galaxies.  In this case, $A_{\rm{\alpha}}$, which
directly reflects the stellar $\alpha$-element abundance, should be
$\sim 1.0$ solar in these galaxies.  If ISM is diluted in X-ray
luminous objects, the $\alpha$-element abundance in the fainter
galaxies may be even higher.  Now, figure 19 implies a larger SN II
contribution to ISM in more luminous systems, which should mainly come
from stellar mass loss. This feature leads us to predict that stellar
metallicity in giant early-type galaxies should be SNeII-like even
when averaged over the whole galaxy.

We can also constrain stellar metallicity in S0 galaxies.  There are 5
S0 galaxies in the X-ray fainter group, and we do not detect any
difference in the ISM abundance between S0 and elliptical galaxies
(figure 13).  These S0 galaxies show Mg$_2$ indices close to those of
elliptical galaxies.  Therefore, abundances of $\alpha$-elements in
the stars in S0 galaxies should be similar within a few tens of \% to
those in elliptical galaxies.  If stellar iron abundance is different
between S0 and elliptical galaxies, it should be reflected as a
significant difference in the ISM abundance, contrary to the observed
feature. Therefore, stellar metallicity in S0 galaxies is considered
to be similar to that in elliptical galaxies.  This conclusion gives
an important constraint in modeling the origin and evolution of S0
galaxies.

\subsubsection{Relation to metals in the ICM}

It has been recognized that the hot ICM in clusters of galaxies
contains a large amount of iron (Hatsukade 1989; Tsuru 1993; Fukazawa
et al.\ 1998).  Clearly, this iron must have been supplied by member
galaxies, with early-type galaxies as main contributors (Arnaud et al.\ 1992).

In evaluating the metal enrichment process independently of the amount
primordial gas, it is convenient to introduce ``iron mass to light
ratio'' (IMLR), which is the ratio of the total iron mass in a cluster
ICM over the total optical luminosity of cluster galaxies.  The
observed value of IMLR appears to be remarkably constant among the
clusters at $0.01-0.02$ (Tsuru 1993; Arnaud et al.\ 1992;
Renzini 1993). This IMLR level is too high to be explained by the
stellar mass loss from member galaxies (Tsuru 1993).  The total
released mass from stars over the Hubble time is $10 \sim20\%$ of the
stellar mass in a galaxy, and its contribution to the cluster IMLR is
estimated to be $(1-2) \times 10^{-3}$ assuming the metallicity of the
mass-loss gas to be about 1 solar.  This is only $ \sim 1/10$ of
the observed IMLR in the ICM\@.  Therefore, the supernova enrichment
is definitely required to explain the observed IMLR\@.

The IMLR values are calculated for the ISM, based on our updated
abundance determination, and compared with the cluster level in figure
20.  The ISM mass within $4 r_{e}$ is estimated from a single $\beta$
model, assuming $\beta=0.5$ and a core radius = 1 kpc which are
typical values obtained by {\it Einstein} (Forman et al.\ 1985;
Trinchieri et al 1986).  Individual galaxies indicate low IMLR values
compared with clusters.  This is not surprising in a sense, because
the ISM mass is at most a few percent of the stellar mass while the
ICM is 2--5 times more massive than the stellar content in clusters.
An important feature seen in figure 20 is that the stellar mass-loss
and SNe Ia can release only $ \sim 20$\% of the total stellar mass in
the Hubble time (e.g.\ Renzini et al.\ 1993). Therefore, with the
present ISM metallicity in the outflow, the accumulated IMLR would
reach at most 10 times higher level and unable to explain the observed
cluster metals.  In this view, the ISM metallicity is too low
regardless of the estimated supernova rate.

This consideration suggests that the ISM metallicity may not directly
reflect the SN Ia rate, even though the measured values themselves are
secure quantities.

\subsection{Interpretation of the remaining problem, low SNe Ia
contribution}

The discrepancy between stellar and ISM abundances has been mostly
removed. There are various models which try to explain how a large
amount of Fe in ICM could have been synthesized even with the present
SNe Ia rate, including the Capperaro's value (e.g.\ Renzini et al.\
1993; Arimoto et al.\ 1997; Ishimaru 1997).  Considering the large Fe
contribution from SNe Ia to the ICM (Fukazawa et al.\ 1998), the SNe
Ia rate must have been higher in the past.  We need to examine if this
picture can explain why SNe Ia contribution is lower in the X-ray
fainter galaxies, as shown in figure 19.

Since X-ray luminous and X-ray fainter galaxies show similar optical
properties (Kodama and Matsushita 2000), the difference in the metal
abundance cannot be simply explained by the present SN Ia frequency.

The larger amount of ISM in X-ray luminous galaxies suggests a longer
accumulation time.  In this case, the higher metal abundance may
reflect a gradual change in the SN Ia rate in the past.  However,
as shown in figure 20, within the optical radii, the gas mass in X-ray luminous galaxies is
only a few \% of the stellar mass.
  This means that it takes only a few Gyr to build up
the ISM with the present stellar mass loss rate (e.g.\ Ciotti et al.\
1991) even in the luminous galaxies. A quick decline in the type Ia
supernova rate in such a short time scale is very unlikely.

We propose here that the low SN Ia contribution in the ISM of X-ray
fainter galaxies is due to a loss of SN Ia products to the
intergalactic space (see also Fujita 1997). This is mainly caused by
the shallow gravitational potential. The stellar mass-loss component
may also escape from these systems in the form of a mild outflow
(Ciotti et al.\ 1994).  The SN Ia products would escape together with
the mass-loss gas or may have higher escaping probability than it.  In
X-ray luminous galaxies, the extended gravitational potential
(Matsushita 1997; Matsushita et al.\ 1998) would trap considerable
fraction of the supernova product.  Therefore, the spatial extent of
SN Ia products probably reflect the overall potential structure beyond
the optical scale. In closing, we should note that the dynamical
motion of supernova bubbles in early-type galaxies may be
significantly different from that in spirals. Progenitors follow a
random velocity field in early-type systems instead of the disk
rotation, and the ambient pressure is dominated by the hot plasma
(i.e.\ ISM) rather than by cold interstellar gas.  We hope future high
resolution imaging and spectroscopy in X-rays will bring us rich
information about the motion and structure of metal-rich gas in early
galaxies.

\bigskip
We would like to thank Nobuo Arimoto and Kuniaki Masai for valuable
suggestions on this work.  This work was supported by the Japan
Society for the Promotion of Science (JSPS) through its Postdoctoral
Fellowship for Research Abroad and Research Fellowships for Young
Scientists.

\clearpage
\section*{References}
\re
Anders E., \& Grevesse~N., 1989, gca, 53,197
\re 
Arimoto,~N., Matsushita,~K., Ishimaru,~Y., Ohashi,~T., \& Renzini,~A. 1997, ApJ, 477, 128
\re
Arnaud, M., et al. 1992, A\&A, 254, 49
\re
Awaki, H., Koyama, K., Kunieda, H., Takano, S., Tawara, Y., \& Ohashi, T.
     1991, ApJ, 366, 88
\re
 Awaki,~H.  et al. 1994, PASJ, 46, L65
\re
Buote, D. A., and Fabian A.C., 1998, MNRAS, 296,977
\re
Buote, D. A.,  1999, MNRAS, 309, 685
\re
Boute, D. A., Canizares, C. R., \& A. C. Fabian,2000, MNRAS, 310, 483
\re
Canizares, C.R., Fabbiano, G., \& Trinchieri, G. 1987, ApJ, 312, 503
 \re
 Cappellaro,~E., Turatto,~M., Benetti,~S., Tsvetkov,~D.~Yu.,      Bartunov,~O.~S., \& Makarova,~I.~N. 1993, A\&A   , 273, 383
\re
Cappellaro, E.,  Turatto, M.; Tsvetkov, D. Yu.;
 Bartunov, O. S.; Pollas, C.; Evans, R.; Hamuy, M., 1997, A\&A, 322, 431
 \re
 Ciotti,~L., Pellegrini,~S., Renzini,~A., \&  D'Ercole,~A. 1991, ApJ, 376,     380
\re
Davis D.S., White R.E., 1996, ApJ, 470, 35
\re
de Vaucouleurs G., de Vaucouleurs A., Corwin Jr. H.G., Buta R.J.,
     Paturel G., \&Fouque P., 1991, 
Third Reference Catalogue of Bright Galaxies (RC3 Catalog)
\re
Eskridge,  P.B., Fabbiano, G., and Kim, D.-W., 1995, ApJ,  97, 141
\re
Edge, A.C., \& Stewart, G.C.,  1991, MNRAS, 252, 414
\re
 Faber,~S.M., Wegner,~G., Burstein,~D.,
 Davies,~R.L., Dresller,~A., Lynden-Bell,~D., Terlevich,~R.J. 1989 ApJS, 69,763
\re
Fabbiano, G., 1989, ARA\&A, 27,87
\re
Fabbiano, G., Gioia, I.M., Trinchieri, G., 1988, ApJ, 324, 749
\re
Fabbiano, G., Kim, D.-W., \&,Trinchieri, G. 1992, ApJS, 80, 531
\re
Fabbiano, G., Kim, D.-W., \&,Trinchieri, G.,
1994, ApJ, 429, 94
 \re
 Fabian,~A.C., Arnaud,~K.A., Bautz,~M.W., \& Tawara,~Y. 1994, ApJ,     436, L63
\re
 Forman,~W., Jones,~C., \& Tucker,~W.  1985, ApJ, 293, 102
\re
Forman, W., Jones, C., David, L., Franx, M., Makishima, K., \& Ohashi, T.
     1993, ApJL, 418, 55
\re
Fujita, K. et al. 1997, Ph. D. thesis, University of Kyoto
\re
Fukazawa, Y. et al. 1994, PASJ, 46, L141
\re
Fukazawa, Y. et al. 1998, PASJ, 50, 187
\re
Hatsukade, I., 1989, Ph.D. Thesis, Osaka University 
\re
Ikebe, Y., et al. 1992, ApJ, 384, L5
\re
Ikebe, Y., et al. 1996, Nature, 379, 427
\re
Ikebe Y., et al. 1999, ApJ, 525, 581
\re
Ishiamaru Y., 1997, Ph.D. thesis, University of Tokyo
\re
Ishisaki Y., 1996, Ph.D. thesis, University of Tokyo
\re
Itoh, H. 1989, PASJ, 41, 853
 \re
 Kaastra,~J.S. 1992, An X-Ray Spectral Code for Optically Thin Plasmas (Internal SRON-Leiden Report, updated version 2.0)
\re
Kobayashi C., and Arimoto N., 1999, accepted to ApJ
\re
Kodama, K., and Matsushita, K., 2000, submitted to ApJ
 \re
Liedahl,~D. A., Osterheld,~A.L., \& Goldstein,~W.H.  1995, ApJ, 438,     L115
\re
Loewenstein,~M., \& Mathews,~W.G. 1991, ApJ, 373, 445
\re
Loewenstein,~M.,
 Mushotzky,~R., Tamura,~T., Ikebe,~Y., Makishima,~K., Matsushita,~K.,
 Awaki,~H., \& Serlemitsos,~P. 1994, ApJL, 436, 75
\re
Makishima, K. et al. 1996, PASJ, 48, 171
 \re
Masai, K. 1984,Ap\&SS, 98, 367
  \re
 Masai, K. 1997, A\&A, 324, 410
 \re
 Matsumoto,~H., Koyama,~K., Awaki,~H., Tsuru,~T., Loewenstein,~M., \& Matsushita,~K. 1997, ApJ, 482, 133
 \re
Matsushita,~K. et al. 1994, ApJL, 436, 41
 \re
Matsushita,~K. 1997, Ph.D. thesis, University of Tokyo
\re
Matsushita, K., Makishima, K., Rokutanda, E.,
     Yamasaki, N., \& Ohashi, T., 1997, ApJ, 488, 125
\re
Matsushita, K., Makishima, K., Ikebe, Y., Rokutanda, E.,
     Yamasaki, N., \& Ohashi, T., 1998, ApJL, 499, 13
\re
Matsushita, K, 2000, submitted to ApJ
\re
Mewe,~R., Gronenschild,~E.H.B.M., \& van~den~Oord,~G.H.J. 1985, A\&AS, 62,197
 \re
Mewe,~R., Lemen,~J.R., \& van~den~Oord,~G.H.J. 1986, A\&AS, 65,511
 \re
Mushotzky,~R.F., Loewenstein,~M., Awaki,~H., Makishima,~K., Matsushita,~K., \& Matsumoto,~H.  1994, ApJL, 436, 79
 \re
 Ohashi,~T. et al. 1996, PASJ, 48, 157
\re
 Raymond,~J.C., \& Smith,~B.W. 1977, ApJS, 35, 419
 \re
 Renzini,~A., Ciotti,~L., D'Ercole,~A., \& Pellegrini,~S. 1993, ApJ,    419, 52
\re
Sadler, E.M., Jenkins, C. R., \& Kotanyi, C.G., 1989, MNRAS, 240, 591
\re
Serlemitsos,~P.J., Loewenstein,~M., Mushotzky,~R.F., Marshall,~F.E,    \& Petre,~R. 1993, ApJ, 413, 518
\re
Tammann, G., \& Sandage, A., 1995, ApJ, 452, 16
\re
Thielemann,~F-K., Nomoto,~K., \& Hashimoto,~M. 1996, ApJ, 460, 408
 \re
Trinchieri, G., Fabbiano, G., \& Canizares, C.R., 1986, ApJ, 310, 637
\re
Trinchieri,~G., Kim,~D.-W., Fabbiano,~G., \& Canizares,~C.R. 1994, ApJ, 428, 555
\re
Trinchieri,~G., Fabbiano,~G. and Kim,~D.-W., 
1997, A\&A, 318, 361
\re
Tsuru, T.  1993, PhD Thesis, University of Tokyo, ISAS RN 528
\re
Tully, R.B., 1988, Nearby Calaxies Catalog, 
     Cambridge: Cambridge University Press)\\
\re
van den Bergh, S., \& Tammann, G. 1991, ARA\&A, 29, 363
\re
Whitemore, B.C., McElroy, D.B., \& Tonry, J.L., 1985, ApJS, 59, 1\\
\re
Worthey, G., Faber, S.M., \& Gonz\'alez, J.J. 1992, ApJ, 398, 69

\clearpage
\scriptsize
\begin{table*}[t]
\begin{center}
Table~1.\hspace{4pt} The sample\\
\begin{tabular}[t]{rcrrrlrc}
galaxy & &B  & D & $L_B$ &$\sigma$ & $r_e$ & $L_X$  \\
& (1) &(2)& (3) &(4) &(5) & (6) & (7)  \\ 
 \hline
 NGC 499  &-3 &12.79 &68.0 &10.60 &237    &21  &42.20\\
 NGC 507  &-2 &12.19 &68.0 &10.94 &365    &77  &42.69\\
 NGC 720  &-5 &11.15 &20.3 &10.34 &247    &40  &40.93\\
 NGC 1316 &-2 & 9.67 &16.9 &10.78 &252$^*$&81  &40.83\\
 NGC 1332 &-2 &11.17 &17.7 &10.22 &306$^*$&28$^*$  &40.40\\
 NGC 1399 &-5 &10.85 &16.9 &10.31 &310    &42  &41.90\\
 NGC 1404 &-5 &11.20 &16.9 &10.17 &225    &27  &41.06\\
 NGC 1407 &-5 &10.65 &21.6 &10.60 &285    &72  &40.90\\
 NGC 3607 &-2 &10.95 &19.9 &10.41 &248    &66  &40.47\\
 NGC 3923 &-5 &10.83 &25.8 &10.68 &216    &53  &40.94\\
 NGC 4125 &-5 &10.57 &24.2 &10.73 &229    &60  &-----\\
 NGC 4365 &-5 &10.61 &16.8 &10.40 &248    &57  &40.01\\
 NGC 4374 &-5 &10.18 &16.8 &10.57 &287    &55  &40.75\\
 NGC 4382 &-2 &10.02 &16.8 &10.64 &200$^*$&55$^*$  &40.31\\
 NGC 4406 &-5 & 9.98 &16.8 &10.65 &250    &91  &41.83\\
 NGC 4472 &-5 & 9.31 &16.8 &10.92 &287    &104 &41.65\\
 NGC 4552 &-5 &10.73 &16.8 &10.35 &261    &30  &40.51\\
 NGC 4636 &-5 &10.47 &17.0 &10.46 &191    &102 &41.58\\
 NGC 4649 &-5 & 9.77 &16.8 &10.74 &341    &74  &41.20\\
 NGC 4697 &-5 &10.17 &23.3 &10.85 &165    &75  &40.56\\
 NGC 5018 &-6 &11.60 &40.9 &10.78 &223    &25  & -----\\
 NGC 5084 &-2 &11.70 &26.4 &10.36 &213$^*$&51$^*$  &40.80\\
 NGC 5846 &-5 &11.11 &28.5 &10.66 &278    &83  &41.70\\
 NGC 7619 &-5 &11.93 &49.9 &10.82 &337    &32  &41.65\\
 NGC 7626 &-5 &12.06 &49.9 &10.67 &234    &38  &41.23\\
 IC 1459  &-5 &10.93 &20.0 &10.42 &308    &39  &40.60\\
 IC 4296  &-5 &11.42 &53.4 &11.01 &323    &57  &41.63\\
\end{tabular}
\end{center}
 \begin{tabular}[t]{ll}
   Column (1)~: & Morphology (RSA) -2: S0; -5: E       \\
   Column (2)~: & Total B band magnitude (RC3)\\
   Column (3)~: & Distance (Mpc) (Tully 1988)\\
   Column (4)~: & B band luminosity ($L_\odot$) (Tully 1988) \\
   Column (5)~: & stellar velocity dispersion at galaxy center (km/s)
  (Faber et al. 1989), *: Whitemore et al. 1985 \\
   Column (6)~: & effective radius (arcsec) (Faber et al. 1989),
     *: RC 3 Catalog\\
   Column (7):& X-ray luminosity (0.2-3.5 keV) (Fabbiano 1992) corrected
  distance using column (3)\\
 \end{tabular}
\end{table*}
\scriptsize
\begin{table*}[t]
 \begin{center}
Table 2.~~  Observatinal log.
 \end{center}
\begin{tabular}[t]{l|c|rl|rllc}
   & & \multicolumn{2}{c|}{GIS}&\multicolumn{4}{c}{SIS}\\
target   & date$^a$ &time$^b$ &counts/s &time$^b$ & counts/s &  mode$^c$ &  clock$^d$    \\\hline
 NGC 499      &95 01 28 & 32.9 & 0.024 & 44.1 & 0.067 & Faint/Faint  &2CCD\\
 NGC 499/507  &94 01 24 & 37.1 & 0.092 & 32.9 & 0.21  & Faint/Faint  &2CCD\\
 NGC 720      &93 07 17& 34.8 & 0.009 & 35.9 & 0.028 & Faint/Bright &4CCD\\
 NGC 1316     &94 01 12& 33.0 & 0.023 & 31.8 & 0.058 & Faint/Faint  &2CCD\\
 NGC 1332     &95 08 05& 50.1 & 0.005 & 55.0 & 0.013 &  Faint/Faint& 1CCD\\
 NGC 1399/1404&93 07 15& 17.2 & 0.090 & 23.7 & 0.25  & Faint/Bright &4CCD\\
\end{tabular}
 \vspace*{0.3cm}

 \begin{tabular}[t]{cl}
a & Observation start date \\
b & Exposure time (ks)\\
c & SIS mode. Mode in BIT High/ Mode in Bit Medium\\
d & CCD clocking mode\\
e & CCD clocking mode in BIT High/BIT Med\\
 \end{tabular}
\end{table*}

\tiny
\begin{table*}[t]
\begin{center}
 Table 3. \hspace{4pt} Summary of spectral fits with the two component
 model, R-S model with solar abundance ratio and a 10 keV Bremstrallung ,
 for early-type galaxies.
Errors show 90\% confidence limits for a single parameter.
\end{center}
\begin{center}
 \begin{tabular}[t]{r| ll l rrr | llr }
        & \multicolumn{6}{c}{single kT (\S3.2)} &\multicolumn{3}{c}{multi kT (\S.3.3)}\\
 galaxy & kT  & Abundance & $N_H$&${F_X}^*$&${F_X}_{\rm{h}}^*$&$\chi^2/\nu$  & Abundance & $N_N$ &$\chi^2/\nu$\\
   & (keV) & (solar) & ($10^{22}\rm{cm^{-2}}$)  & \multicolumn{2}{c}{($10^{-13}\rm{erg s^{-1}{cm^{-2}}}$)} & & (solar) & ($10^{22}\rm{cm^{-2}}$) & \\\hline
NGC~499 & 0.71 $^{+0.03 }_{-0.02 }$& 1.22 $^{+3.5 }_{-0.62 }$& 0.20 $^{+0.06 }_{-0.07 }$& 16.0 & 1.5 & 52.3/61        &  1.30 $ (>0.71)$& 0.15 $^{+0.06 }_{-0.04 }$&62.6/56 \\                         
NGC~507 & 1.04 $^{+0.02 }_{-0.02 }$& 0.75 $^{+0.15 }_{-0.17 }$& 0.15 $^{+0.01 }_{-0.01 }$& 58.5 & 11.3 & 162.4/113   &  0.75 $^{+0.15 }_{-0.14 }$& 0.15 $^{+0.01 }_{-0.01 }$& 162.3/108  \\           
NGC~720 & 0.58 $^{+0.04 }_{-0.07 }$& 0.14 $^{+0.07 }_{-0.05 }$& 0.01 (fix)& 8.6 & 3.9 & 35.6/31  &  0.26 $^{+0.60 }_{-0.14 }$& 0.01 (fix)&  31.4/26 \\                            
NGC~1316 & 0.74 $^{+0.03 }_{-0.02 }$& 0.24 $^{+0.08 }_{-0.09 }$& 0.00 $^{+0.03 }_{-0.00 }$& 13.5 & 4.9 & 78.9/67  & 0.29 $^{+0.28 }_{-0.09 }$& 0.00 $^{+0.07 }_{-0.00 }$& 67.5/62 \\               
NGC~1332 & 0.49 $^{+0.13 }_{-0.05 }$& 0.16 $^{+0.24 }_{-0.09 }$& 0.02 (fix)& 5.0 & 2.7 & 31.8/21  & 0.66 $ (>0.22)$& 0.02 (fix)&  30.6/16 \\
\end{tabular} 
\end{center} 
 *: X-ray flux (0.5-10.0 keV). h means hard component.
\end{table*}
\begin{table*}[h]
\begin{center}
Table 4.~~ Summary of spectral fits with variable abundance R-S model.
\end{center}
 \begin{tabular}[t]{r| llllr | lllr }
        & \multicolumn{5}{c|}{varying $A_{\rm{\alpha}}$~ and $A_{\rm{Fe}}$~} & \multicolumn{4}{c}{$A_{\rm{\alpha}}$=1.0 solar (fixed)}\\
   galaxy & kT  & $A_{\rm{\alpha}}$ &$A_{\rm{Fe}}$  & $N_H$& $\chi^2/\nu$  & kT  & $A_{\rm{Fe}}$& $N_H$& $\chi^2/\nu$ \\
   & (keV) & (solar) & (solar) &($10^{22}\rm{cm^{-2}}$)  && (keV) & (solar) & ($10^{22}\rm{cm^{-2}}$)  & \\\hline
      NGC~499& $0.72^{+0.03}_{-0.03}$ &  $1.09 (>0.46)$  &$1.05 (>0.56)$&  $0.17^{+0.10}_{-0.05}$ &  52.5/60
             & $0.72^{+0.03}_{-0.03}$ &  $1.00^{+0.29}_{-0.15}$ &  $0.18^{+0.09}_{-0.05}$ &  52.4/61\\
      NGC~507& $1.04^{+0.03}_{-0.02}$ &  $0.70^{+0.31}_{-0.16}$ & $0.71^{+0.24}_{-0.14}$&  $0.15^{+0.03}_{-0.04}$ &162.6/112
             & $1.05^{+0.02}_{-0.02}$ &  $0.92^{+0.09}_{-0.08}$ &  $0.14^{+0.03}_{-0.03}$ &  165.0/113\\ 
      NGC~720& $0.53^{+0.10}_{-0.05}$ &  $0.27^{+0.23}_{-0.22}$ &$0.16^{+0.09}_{-0.06}$&   0.01 (fix)            &   25.7/31
             & $0.60^{+0.05}_{-0.10}$ &  $0.51^{+0.10}_{-0.08}$ &   0.01 (fix)            &   30.9/32\\ 
     NGC~1316& $0.74^{+0.04}_{-0.05}$ &  $0.28^{+0.24}_{-0.17}$ & $0.24^{+0.10}_{-0.08}$& $0.00^{+0.04}_{-0.00}$ &   78.7/66
             & $0.70^{+0.04}_{-0.04}$ &  $0.44^{+0.10}_{-0.07}$ &  $0.00^{+0.02}_{-0.00}$ &  88.0/67\\ 
     NGC~1332& $0.48^{+0.09}_{-0.08}$ &  $1.76 (>0.10)$ & $0.88 (>0.08)$& 0.02 (fix)             &   29.9/20
             & $0.48^{+0.07}_{-0.07}$ &  $0.52^{+0.10}_{-0.17}$ &  0.02 (fix)             &   29.6/21\\
\end{tabular}
\end{table*}
\tiny
\begin{table*}[p]
\begin{center}
 Table 5.~~ Summary of spectral fits with MEKA, MEKAL and Masai codes.
\end{center}
\hspace*{-2cm}
 \begin{tabular}[t]{r|l llll  r|llllr}
        &     & \multicolumn{5}{c|}{varying $A_{\rm{\alpha}}$~ and $A_{\rm{Fe}}$} & \multicolumn{4}{c}{$A_{\rm{\alpha}}$=1.0 solar (fixed)}\\
 galaxy &model& kT                   & $A_{\rm{\alpha}}$                  &$A_{\rm{Fe}}$                    & $N_H$         &$\chi^2/\nu$ &
kT                   &$A_{\rm{Fe}}$                    & $N_H$         &$\chi^2/\nu$ \\
        &     &(keV)                 & (solar)              & (solar)                & ($10^{22}\rm{cm^{-2}}$)&
  \\\hline
 NGC~499&MEKA &0.59$^{+0.02}_{-0.03}$&2.3($>1.4$)&2.4($>1.5$)&0.18$^{+0.05}_{-0.05}$&68.5/60 &
 0.60$^{+0.02}_{-0.02}$&1.14$^{+0.19}_{-0.15}$&0.14$^{+0.05}_{-0.05}$&74.1/61\\
 NGC~499&MEKAL&$0.64^{+0.02}_{-0.02}$&1.8($>0.60)$&$1.9 (>0.76)$&$0.23^{+0.07}_{-0.05}$&58.7/61&
 $0.64^{+0.02}_{-0.02}$& $1.15^{+0.24}_{-0.17}$&$0.22^{+0.06}_{-0.05}$&59.2/61\\
 NGC~499&Masai&$0.73^{+0.02}_{-0.03}$&2.4 ($>1.6$)&2.1 ($>1.6$)&0.11$^{+0.06}_{-0.04}$&55.8/60 &
$0.73^{+0.02}_{-0.02}$& $1.47^{+0.25}_{-0.19}$&$0.10^{+0.05}_{-0.05}$ &  61.8/61\\
  \hline
 NGC~507&MEKA &0.84$^{+0.01}_{-0.02}$&0.56$^{+0.23}_{-0.18}$&0.52$^{+0.18}_{-0.11}$&0.16$^{+0.03}_{-0.02}$&236.3/112 &
 0.84$^{+0.01}_{-0.02}$&0.82$^{+0.07}_{-0.08}$&0.18$^{+0.03}_{-0.04}$&245.2/113 \\
\hline
  \end{tabular} 
 \end{table*}

\begin{table*}
\begin{center}
Table 5. (continued)\end{center}
\hspace*{-2cm}
 \begin{tabular}[t]{r|l llll  r|llllr}
        &     & \multicolumn{5}{c|}{$A_{\rm{\alpha}}$ , $A_{\rm{Fe}}$ free parameter} & \multicolumn{4}{c}{$A_{\rm{\alpha}}$=1.0 solar (fixed)}\\
 galaxy &model& kT                   & $A_{\rm{\alpha}}$                  &$A_{\rm{Fe}}$                    & $N_H$         &$\chi^2/\nu$ &
kT                             &$A_{\rm{Fe}}$                    & $N_H$         &$\chi^2/\nu$ \\
        &     &(keV)                 & (solar)              & (solar)                & ($10^{22}\rm{cm^{-2}}$)&
  \\\hline  
NGC~4382&MEKA &$0.32^{+0.04}_{-0.05}$&19.3 ($>0.02$) &26.9 ($>0.15$)  & 0.03 (fix)     &38.9/29
 &$0.32^{+0.05}_{-0.04}$             &$1.46 ({>0.68})$ &0.03 (fix)            &39.0/30\\
NGC~4382&MEKAL& $0.29^{+0.09}_{-0.06}$&0.33 $(>0.0)$&1.9 $(>0.30)$ &0.03 (fix)  &37.2/29
& $0.29^{+0.09}_{-0.07}$ &  $6.14 ({>0.62})$ &  0.03 (fix) &  37.3/30\\
NGC~4382&Masai& $0.35^{+0.05}_{-0.05}$ & 2.5 $(>0.08)$& 3.4 $(>0.15)$&  0.03 (fix) &  36.5/29
& $0.35^{+0.06}_{-0.05}$  & $1.44^{+3.05}_{-0.90}$ &  0.03 (fix) &  36.5/30\\
  \hline
NGC~4406&MEKA &$0.65^{+0.01}_{-0.01}$&1.02$^{+0.88}_{-0.37}$&1.12$^{+0.85}_{-0.35}$&$0.09^{+0.01}_{-0.01}$&276.4/143
 &$0.65^{+0.01}_{-0.01}$            &$1.13^{+0.06}_{-0.06}$&$0.09^{+0.02}_{-0.02}$&276.5/144\\
\hline
 \end{tabular} 
\end{table*}
\scriptsize
\begin{table*}[t]
\begin{center}
Table 6.
Temperature (keV) and $\chi^2$
 derived from fitting the spectra above 1.6 keV (SIS)
 and 1.7 keV (GIS) with vR-S model.
 \begin{tabular}[t]{c|cr}
 & temperature (keV) & $\chi^2$/dof. 
  \\
  \hline
 NGC~~507 &1.05$\pm0.12$&49.2/42\\
 NGC~1399 &1.06$\pm0.10$&47.2/28\\
 NGC~4406 &0.80$\pm0.11$&57.6/52\\
 NGC~4472 &1.05$\pm$0.07 &116.9/81\\
 NGC~4636 &0.73$\pm$0.04 &210.6/202\\
 NGC~5846 &0.80$\pm$0.12 &42.4/37\\
 \end{tabular}
\end{center}
\end{table*}
\begin{table*}
\begin{center}
 Table 7. The fitting results for the SIS spectra by adding 
 20 \% systematic errors in   0.4-1.6 keV.
 The Raymond-Smith model has been used. 
\end{center}
\begin{center}
\begin{tabular}{rcllllr}
  galaxy &model & $kT$    & $A_{\rm{\alpha}}$ & $A_{\rm{Fe}}$ & $N_H$ & $\chi^2/\nu$\\
 & &(keV)& (solar)  & (solar) & ($10^{22}\rm{cm^{-2}}$) & \\
\hline
NGC~~507&R-S  &1.02$^{+0.04}_{-0.08}$&0.48$^{+0.30}_{-0.16}$&0.43$^{+0.27}_{-0.15}$&0.16$^{+0.08}_{-0.04}$&77/101\\
NGC~~507&MEKA &0.87$^{+0.06}_{-0.04}$&0.44$^{+0.22}_{-0.14}$&0.44$^{+0.16}_{-0.14}$&0.21$^{+0.04}_{-0.10}$&94/101\\
NGC~~507&MEKAL&1.02$^{+0.08}_{-0.06}$&0.52$^{+0.28}_{-0.16}$&0.39$^{+0.18}_{-0.12}$&0.18$^{+0.04}_{-0.06}$&72/101\\
NGC~~507&Masai&0.97$^{+0.06}_{-0.05}$&0.57$^{+0.34}_{-0.16}$&0.63$^{+0.30}_{-0.19}$&0.18$^{+0.04}_{-0.06}$&87/101\\
 \hline 
NGC~1399&R-S  &1.09$^{+0.04}_{-0.03}$&0.51$^{+0.16}_{-0.16}$&0.72$^{+0.24}_{-0.18}$&0.10$^{+0.04}_{-0.04}$&96/83\\ 
NGC~1399&MEKA &0.94$^{+0.05}_{-0.06}$&0.37$^{+0.12}_{-0.12}$&0.56$^{+0.15}_{-0.15}$&0.14$^{+0.04}_{-0.04}$&100/83\\
\hline
\end{tabular}
\end{center}

\end{table*}
\begin{table*}[h]
\begin{center}
 Table 8.~~  The results from combined fit to spectra of galaxies
with similar ISM luminosity and temperature.

\end{center}
\begin{center}
\begin{tabular}[t]{c|lllr}
        & Model & $A_{\rm{\alpha}}$~ & $A_{\rm{Fe}}$~ & $\chi^2$/bin \\\hline
Group1  & R-S   & 0.62$^{+0.17}_{-0.19}$ & 0.69$^{+0.17}_{-0.19}$ & 453.7/397\\
        & MEKAL & 0.66$^{+0.15}_{-0.16}$ &0.61$^{+0.10}_{-0.10}$ & 483.4/397\\
        & MEKA  & 0.50$^{+0.22}_{-0.13}$ &0.53$^{+0.18}_{-0.11}$ & 606.9/397\\
        & Masai & 0.72$^{+0.21}_{-0.13}$ &0.88$^{+0.12}_{-0.10}$ & 498.7/397\\ \hline
 Group2 & R-S   & 0.69$^{+0.23}_{-0.24}$ & 0.60$^{+0.22}_{-0.20}$ & 370.2/379\\
        & MEKAL & 0.77$^{+0.21}_{-0.29}$ & 0.64$^{+0.19}_{-0.22}$ & 360.8/379\\
        & MEKA  & 1.01$^{+0.53}_{-0.20}$ & 0.80$^{+0.30}_{-0.15}$  & 411.1/379\\
\hline
\end{tabular}
\tiny
\begin{tabular}[t]{crll}
        & $kT$ & $L_X$ & member \\
 Group1 & $>$ 0.9 keV        &$>10^{41}\rm{erg s^{-1}}$ & NGC 507, NGC 1399, \& NGC 4472\\
 Group2 & 0.8 $\sim$ 0.9 keV &  $>10^{41}\rm{erg s^{-1}}$ & NGC 4406, NGC 5846, NGC 7619, \& IC 4296\\
 Group3 & 0.8 $\sim$ 0.9 keV &  $5\times 10^{40}\sim 10^{41}\rm{erg s^{-1}}$ & NGC 1407, NGC 4649 \& NGC 7626\\
 Group4 & 0.6 $\sim$ 0.8 keV &  $>10^{41}\rm{erg s^{-1}}$ & NGC 4636, \& NGC 499 \\
 Group5 & 0.6 $\sim$ 0.8 keV &  $<5\times 10^{40}\rm{erg s^{-1}}$ &
   NGC 1332, NGC 1404, NGC 4552, NGC 3923, NGC 4374, IC 1459, NGC 5018, \& NGC 1316\\
 Group6 & 0.4 $\sim$ 0.6 keV & $<5\times 10^{40}\rm{erg s^{-1}}$& NGC 720, NGC 4125, NGC 4365 \& NGC 3607 \\
\end{tabular}
\end{center}
\end{table*}

\newpage
\normalsize
\centerline{
\psfig{file=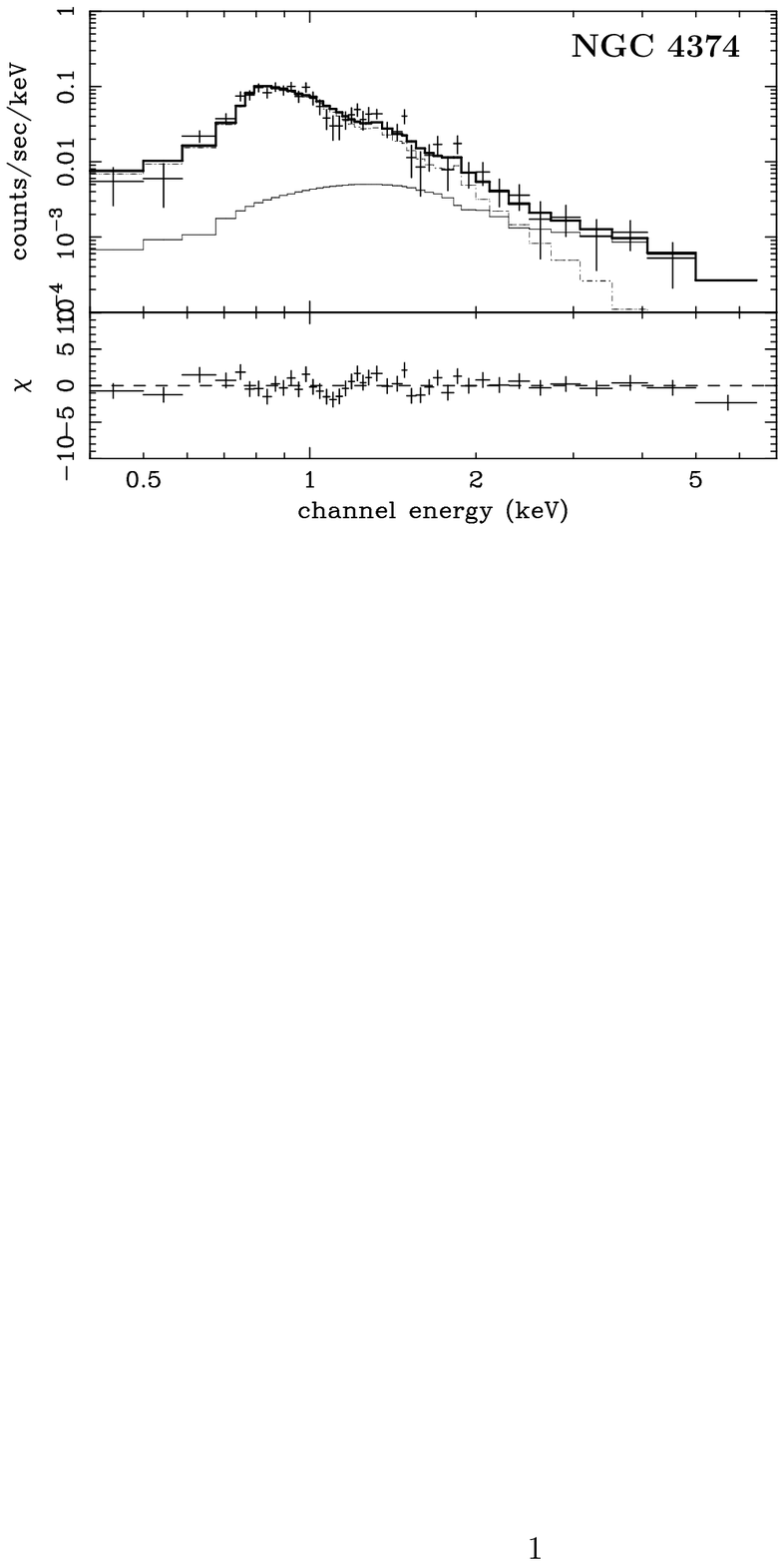,width=8cm,clip=}
\psfig{file=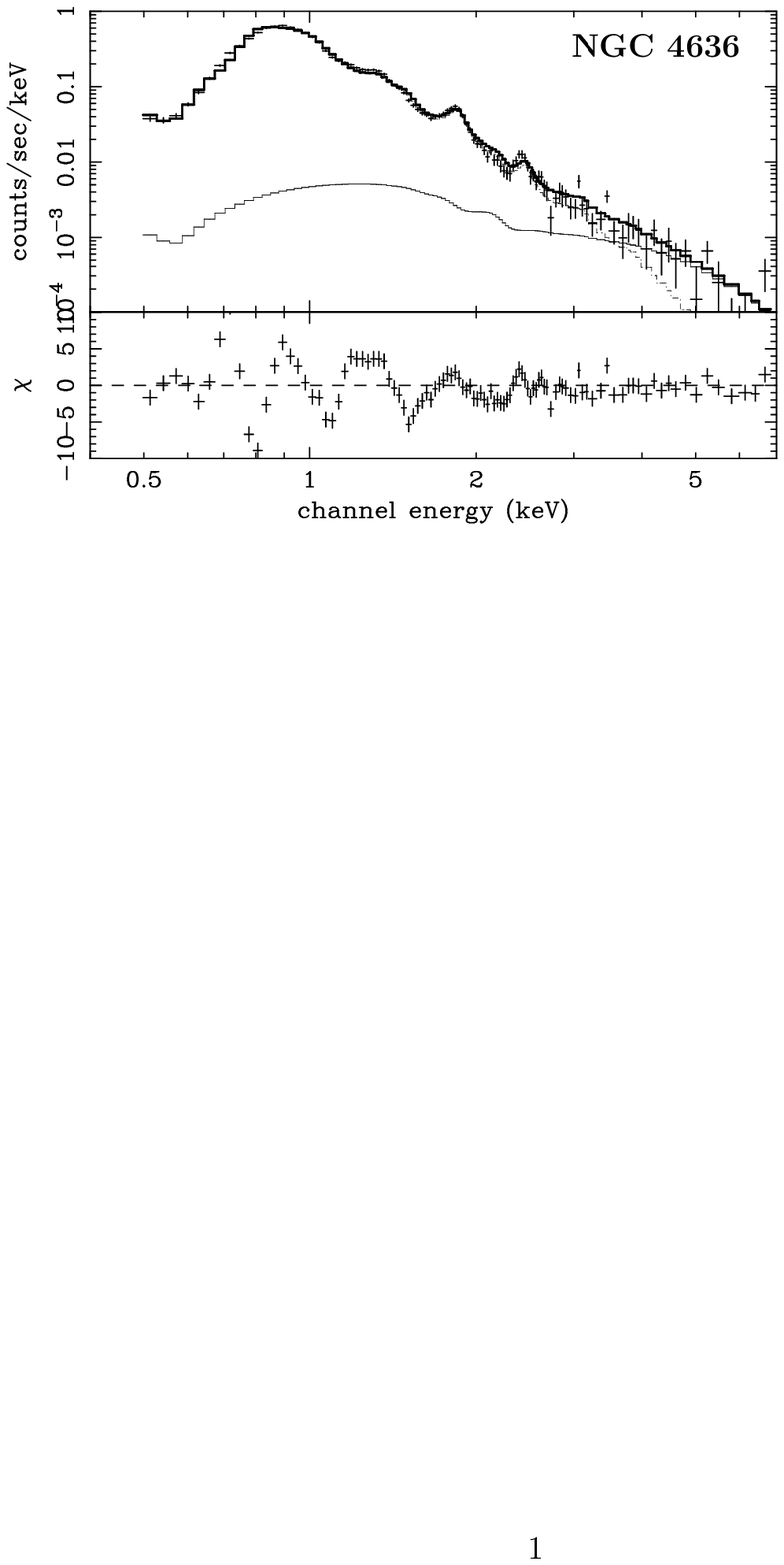,width=8cm,clip=}
}
\begin{fv}{1}{1cm}
{The SIS spectra (crosses) of  (a) NGC~4374
  and (b) NGC~4636  fitted with the double-component 
   model, consisting of a soft R-S component (dot-dash line) 
   and a hard bremsstrahlung component (thin solid line).
The bottom panels show residuals of the fit.}
\end{fv}

\psfig{file=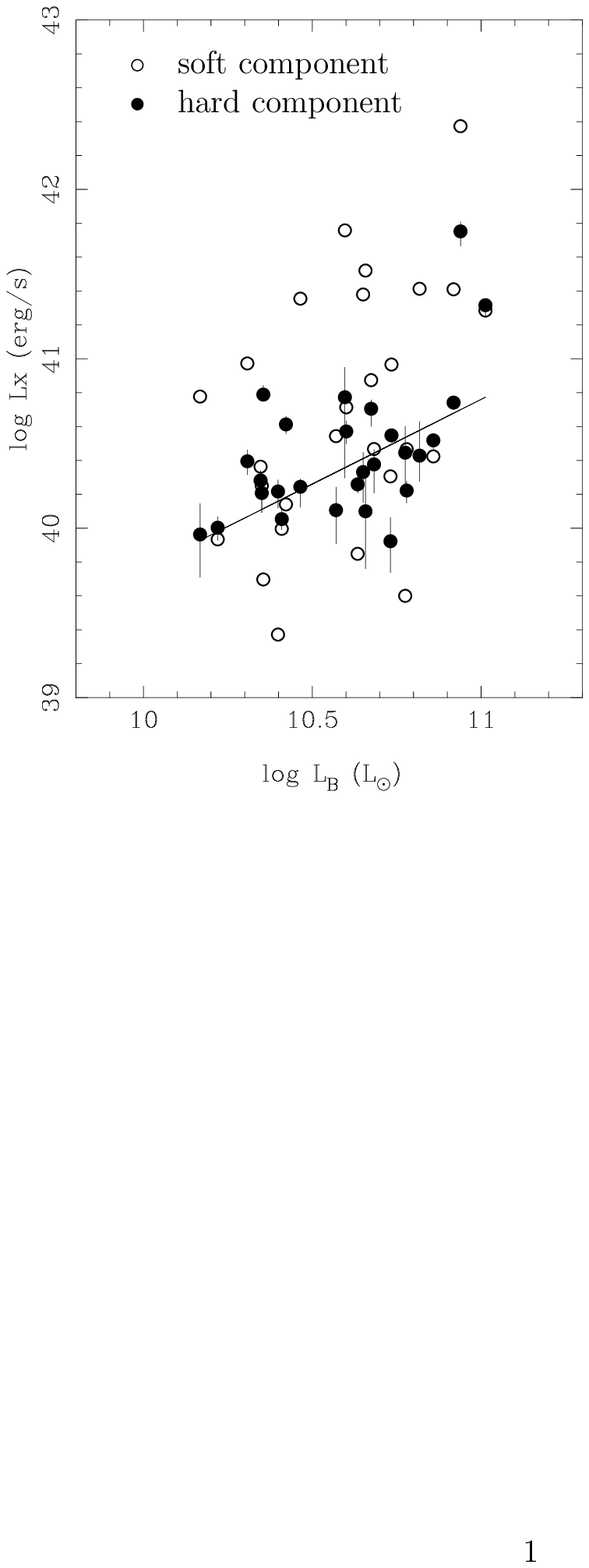}
\begin{fv}{2}{1cm}
{X-ray (0.5--10.0 keV) vs. B-band luminosities of early-type galaxies.
      Open and closed circles represent
      the soft and hard components, respectively.
      Solid line represents the contribution of discrete sources 
      estimated by Canizares et al. (1987) 
    from X-ray luminosities of bulges of early spiral galaxies. }
\end{fv}

\psfig{file=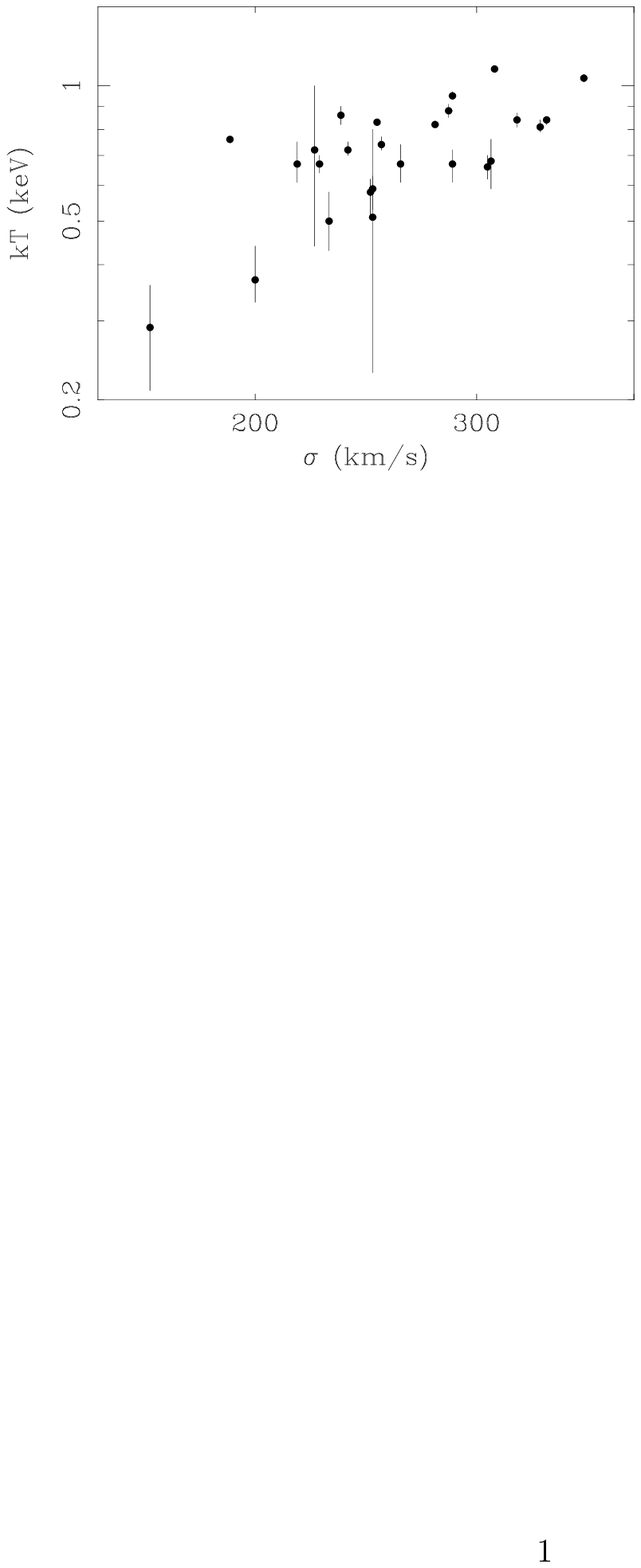}
\begin{fv}{3}{1cm}
{ The ISM temperature vs. the central stellar velocity dispersion ($\sigma$)
  of the galaxies.
}
\end{fv}

\psfig{file=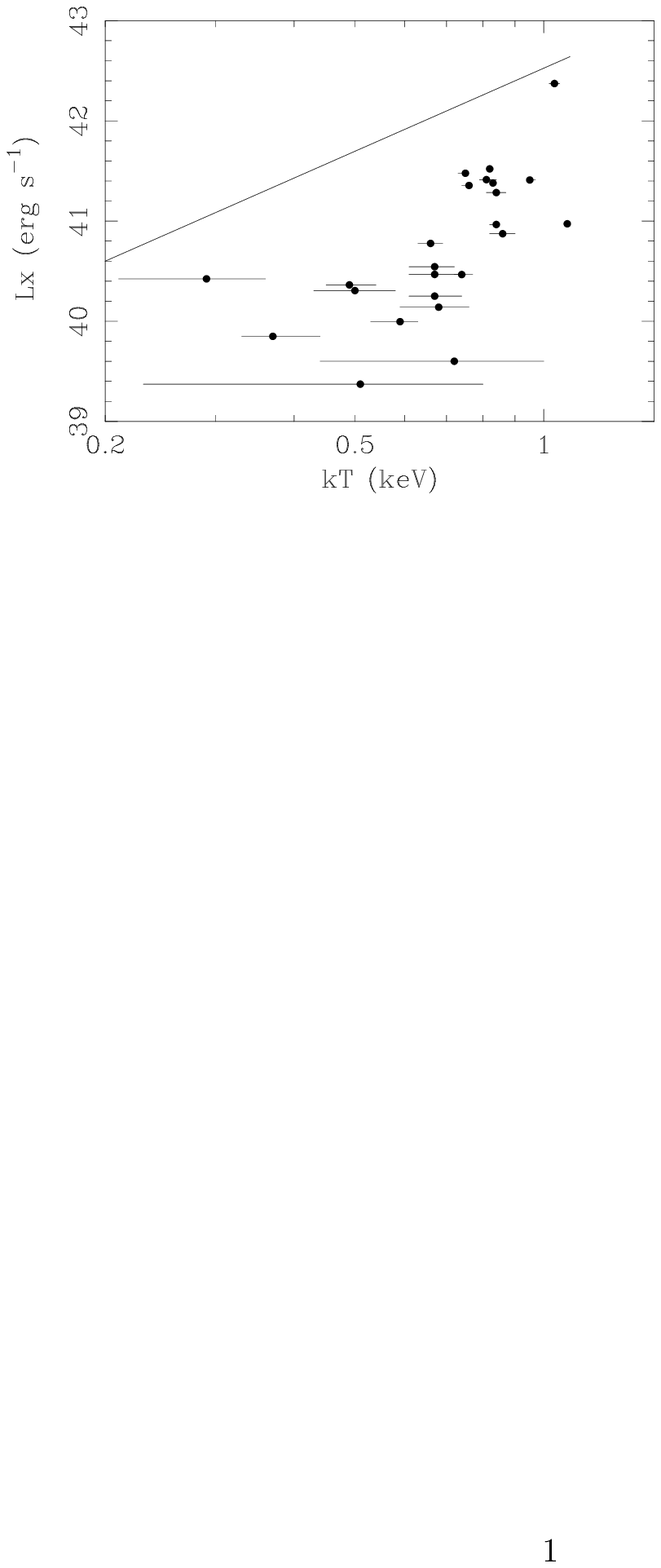}
\begin{fv}{4}{1cm}
{The X-ray luminosities vs. temperature of the ISM (filled circles). 
       The solid line represents an extrapolation of  
       the average $kT$  vs. $L_X$ (bolometric) relation 
       for clusters of galaxies (Edge et al. 1991).}
\end{fv}

\centerline{
\psfig{file=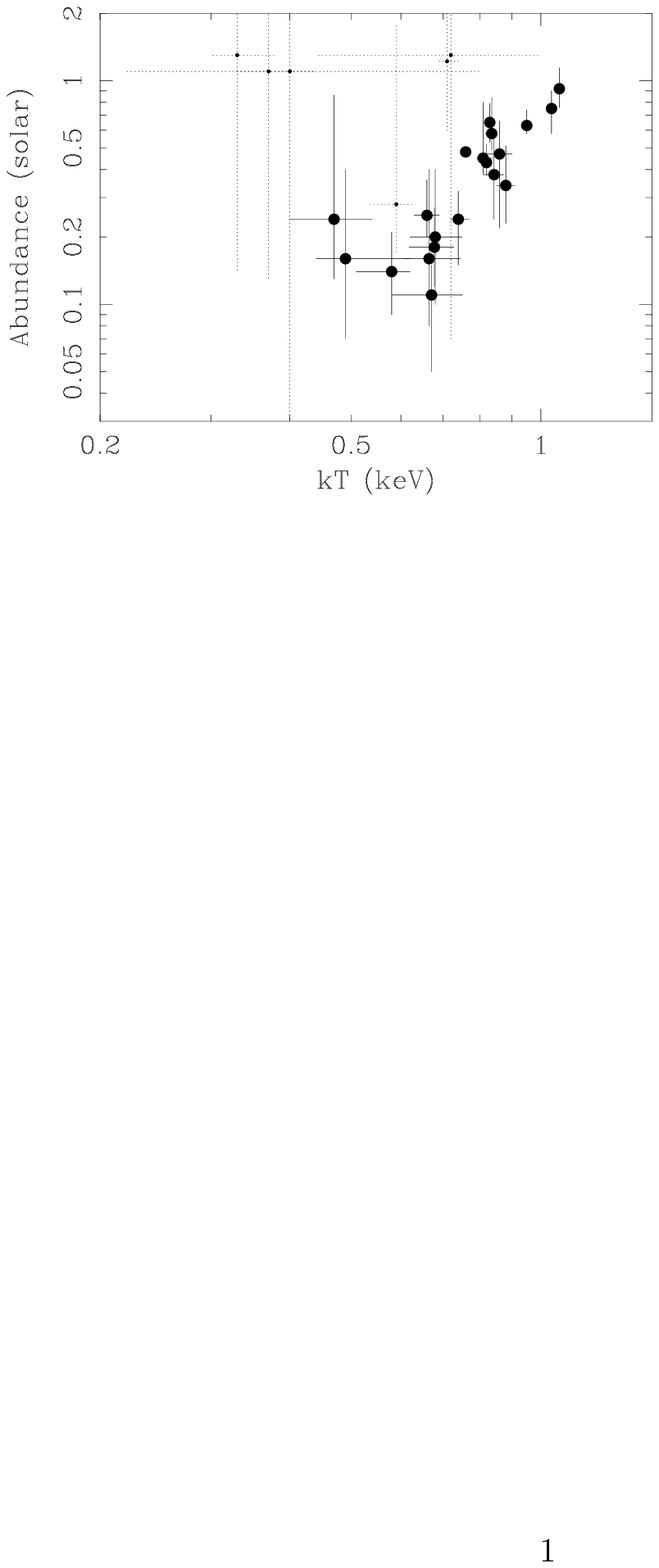,clip=}
\psfig{file=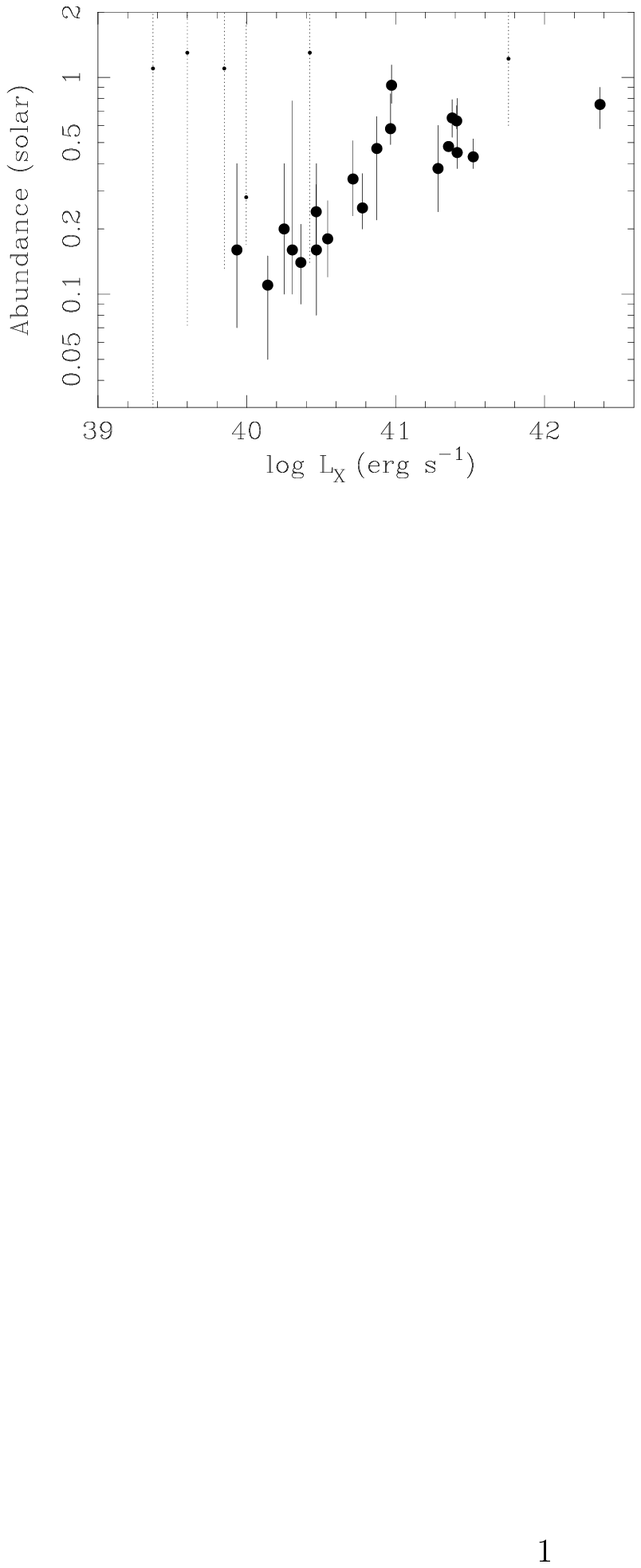,clip=}
}
\begin{fv}{5}{1cm}
{(a) The abundance vs. temperature of the ISM.
      We show the data with large uncertainty (relative error exceeding 50\%)
in dotted lines in order to highlight the high-quality data points.
  (b) The abundance vs. ISM luminosity (0.5--10.0 keV; $r<4 r_e$).}
\end{fv}

\psfig{file=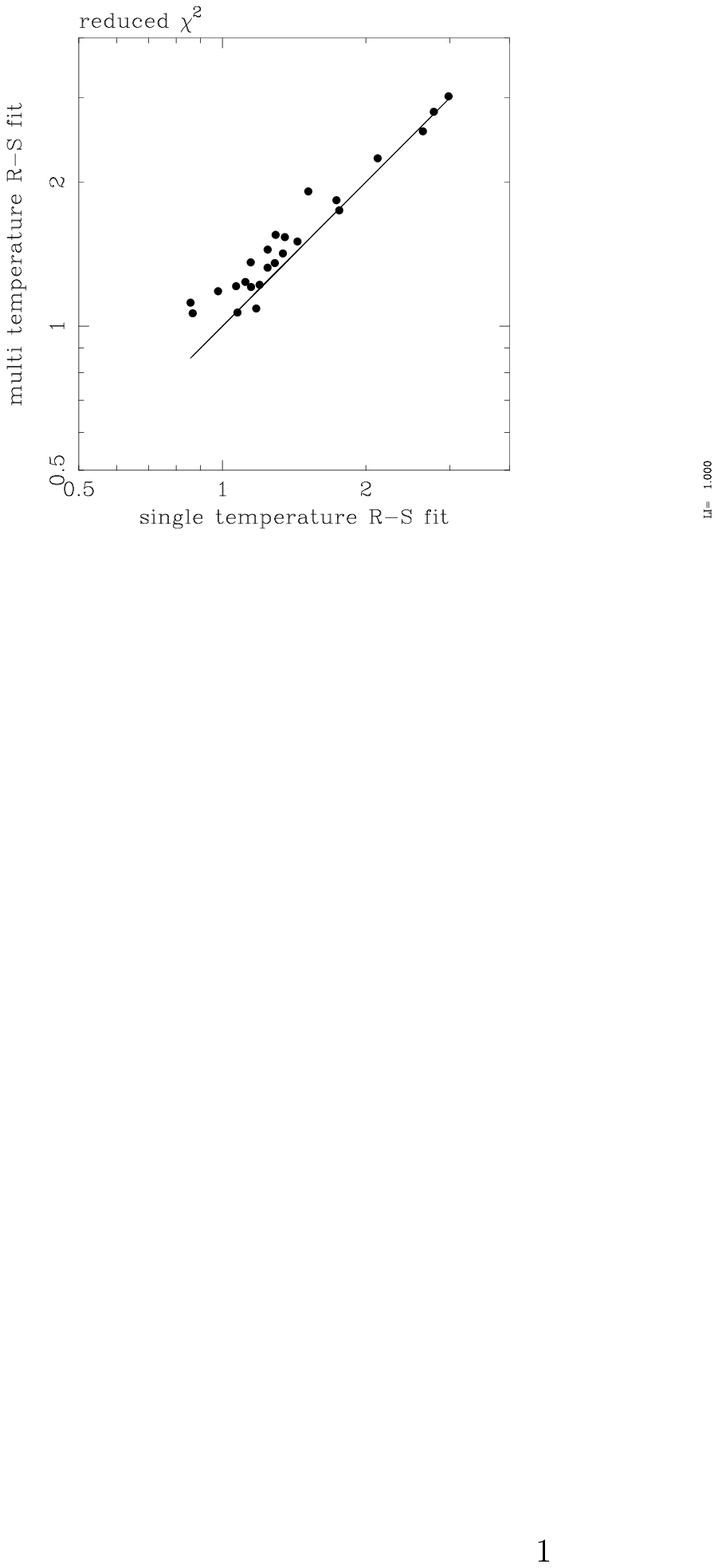,clip=}
\begin{fv}{6}{1cm}
{  Comparison of the reduced $\chi^2$
    between the 1 temperature R-S + bremsstrahlung fit vs.
    multi-temperature R-S + bremsstrahlung
     fit.   Solid line indicates the equal value between the two fits.}
\end{fv}

\psfig{file=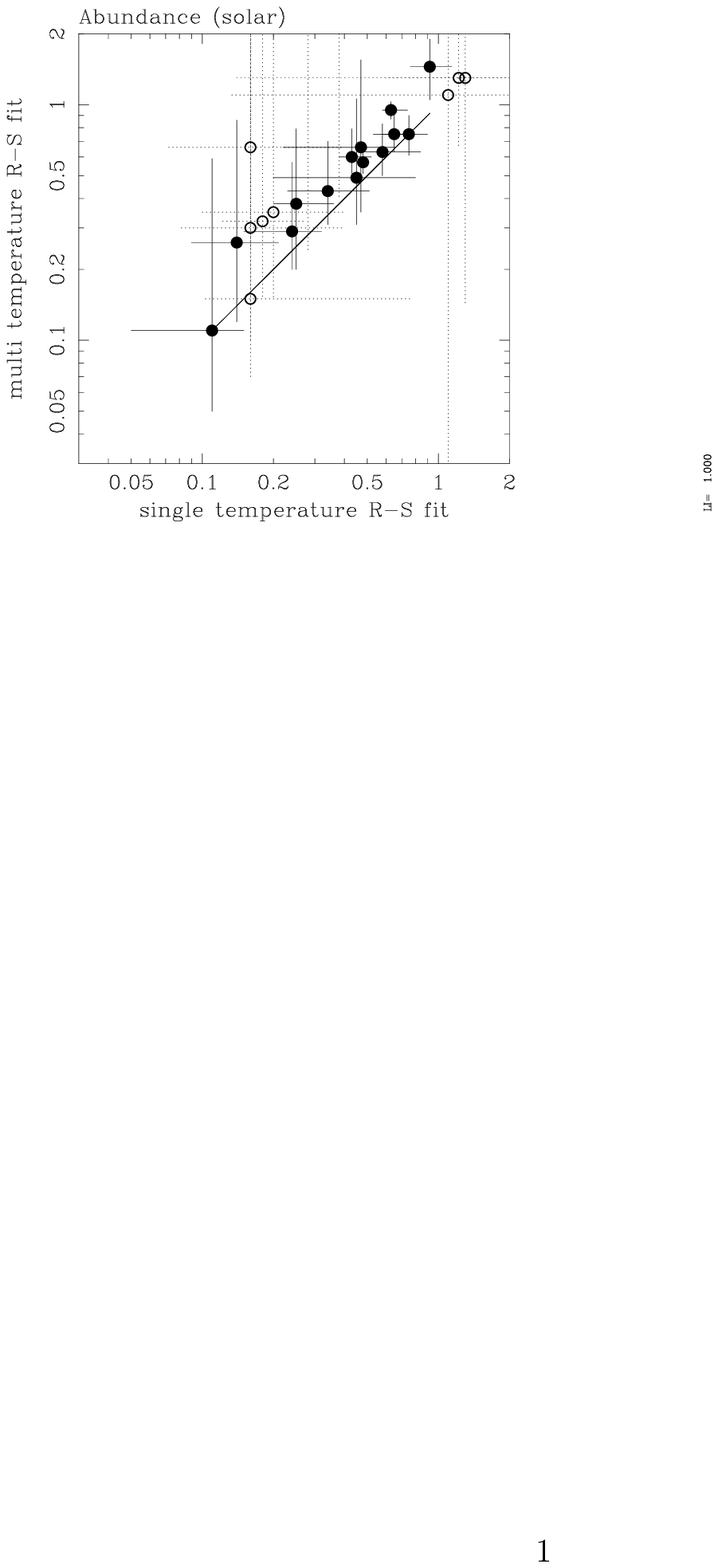,clip=}
\begin{fv}{7}{1cm}
{  Comparison of the derived abundances
    between the 1 temperature R-S + bremsstrahlung fit vs.
    multi-temperature R-S + bremsstrahlung
     fit.   Solid line indicates the equal value between the two fits.
 In order to emphasize the high quality data, the points with large
 error bars are represented in dotted lines.
}
\end{fv}

\psfig{file=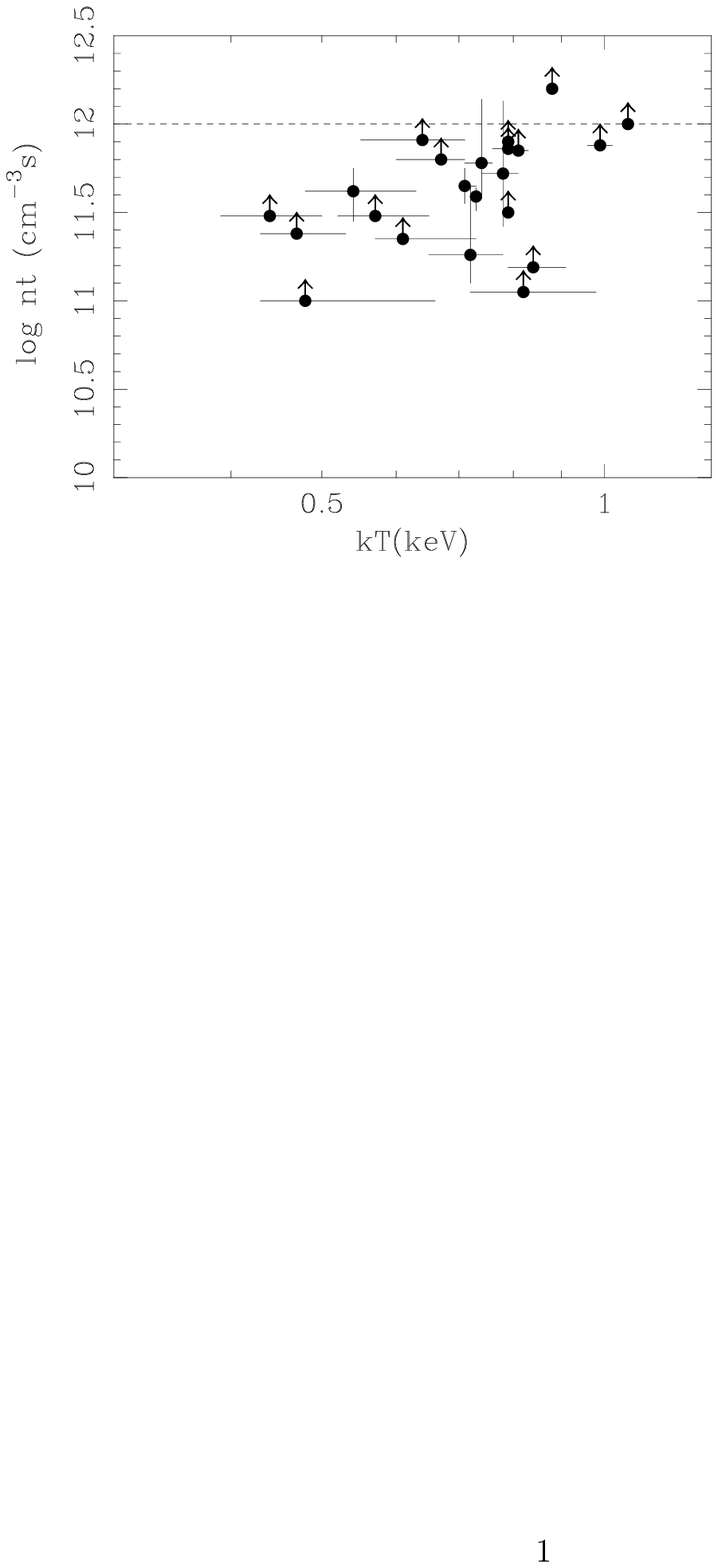,clip=}
\begin{fv}{8}{1cm}
{  nt vs. the ISM temperature derived from the NEI Masai model. }
\end{fv}

\psfig{file=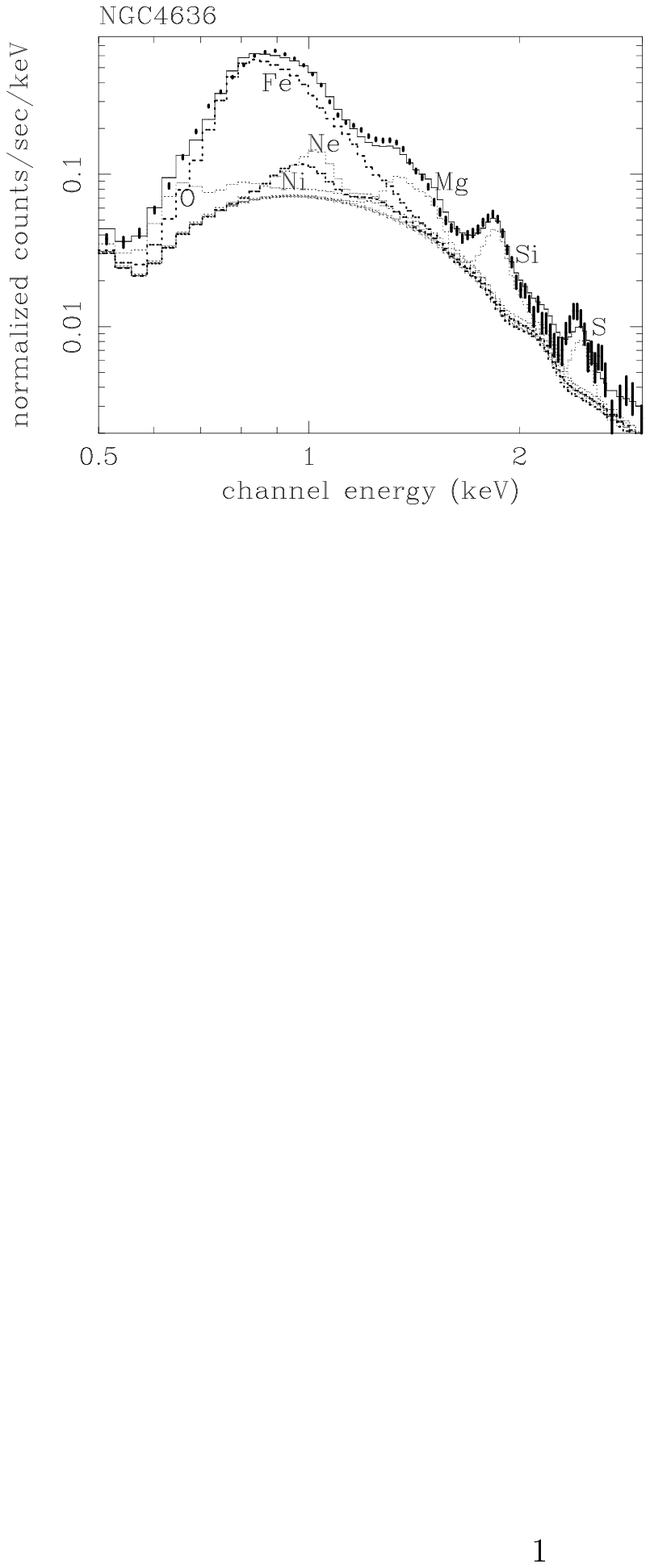,clip=}
\begin{fv}{9}{1cm}
{The SIS spectrum of NGC~4636 and the best-fit model prediction,
   decomposed into contributions from individual elements.}
\end{fv}

\psfig{file=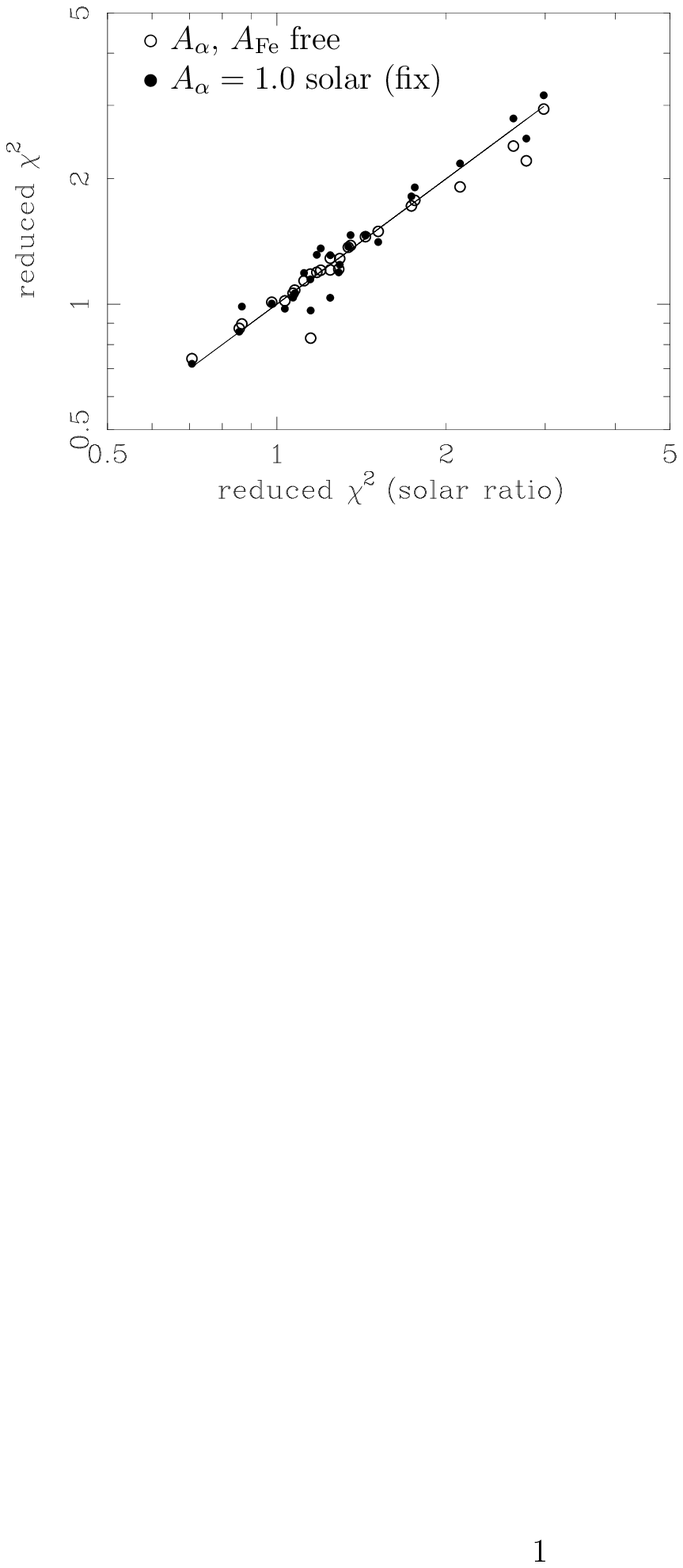,clip=}
\begin{fv}{10}{1cm}
{ Reduced $\chi^2$ derived with $A_{\alpha}$ and $A_{\rm Fe}$ allowed to be separate
(open circles) 
or asuming $A_{\rm{\alpha}}$  =1.0 solar (closed circles), plotted in the ordinate,  against
 those derived assuming solar abundance ratio in the abscissa.}
\end{fv}

\psfig{file=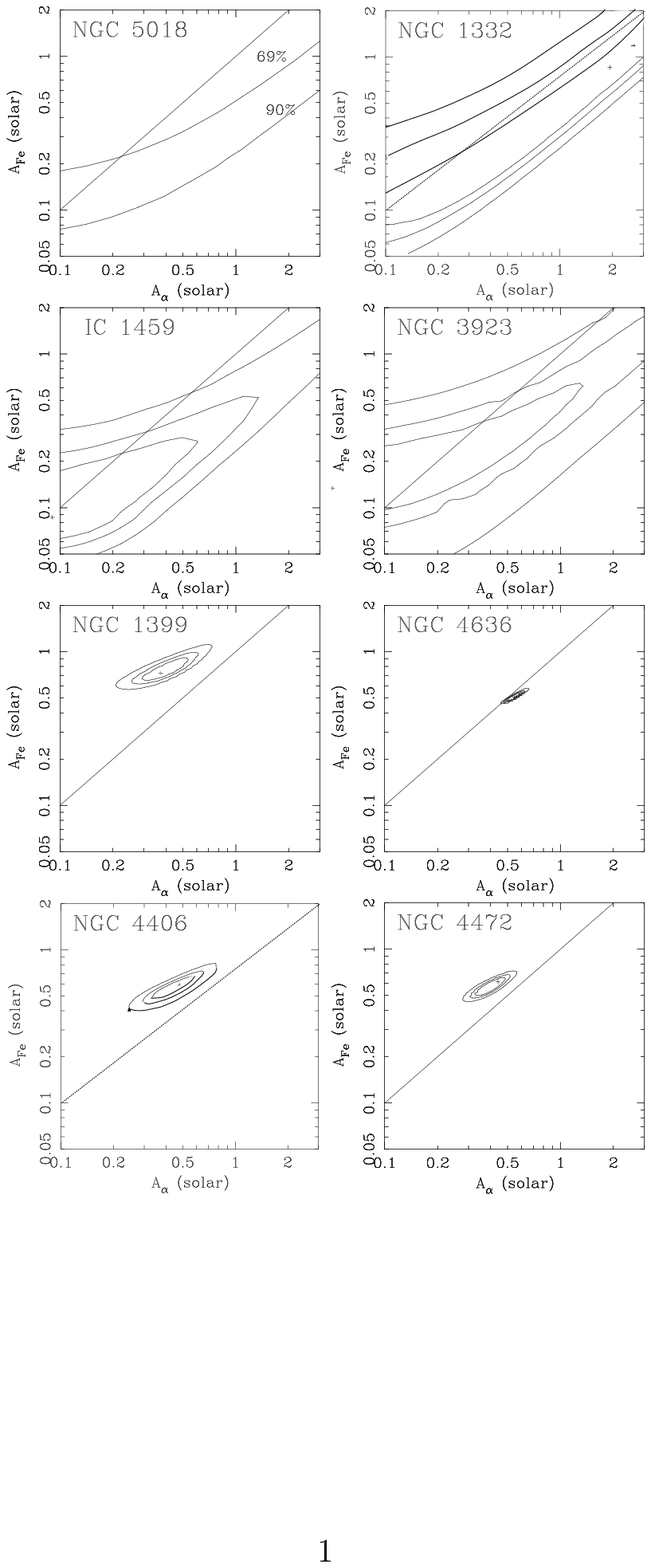,clip=}
\begin{fv}{11}{1cm}
{  Confidence contours (69\%, 90\%, and 99\%)  for $A_{\rm Fe}$ vs. $A_{\alpha}$ 
   for the ISM of representive early-type galaxies, obtained with the R-S model.
   Solid lines indicate the condition of solar abundance ratio.
   The area of confidence regions is
   heavily under-estimated for X-ray luminous galaxies
   because their fits are not formally acceptable.}
\end{fv}

\centerline{
\psfig{file=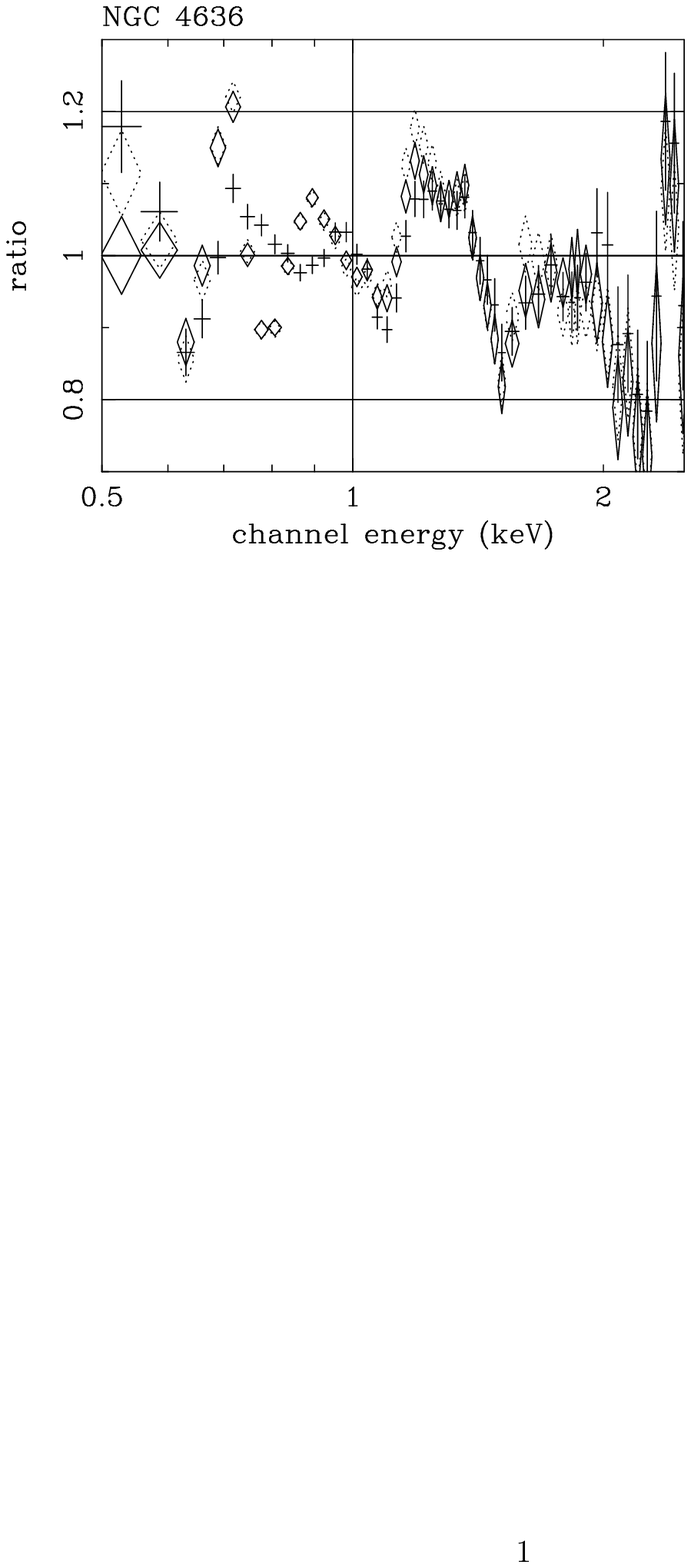,width=8cm,clip=}
\psfig{file=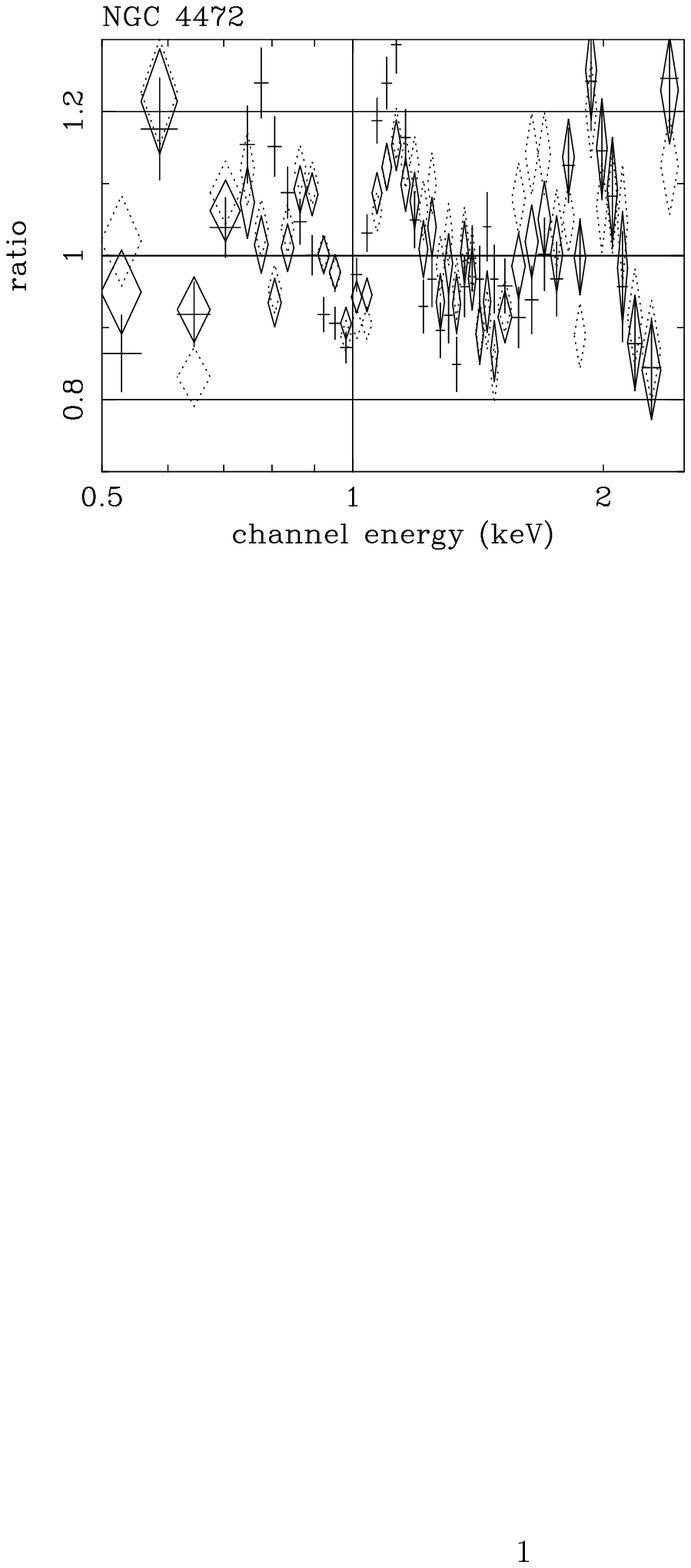,width=8cm,clip=}
}
\begin{fv}{12}{1cm}
{ The data-to-model ratio of the SIS spectra of  NGC~4636 (observed in AO5)
and NGC~4472.
Abundances of  heavy elements are grouped into $\alpha$-elements and iron group.
The solid diamonds  represent the fit allowing both $A_{\rm{\alpha}}$  and $A_{\rm{Fe}}$ to be free,
and dotted diamonds represent the model fixing  $A_{\rm{\alpha}}$ to be 1.0 solar.
Those for the best fit  MEKAL model (both $A_{\rm{\alpha}}$ and $A_{\rm{Fe}}$ are allowed to be free) are also shown (crosses).}
\end{fv}

\psfig{file=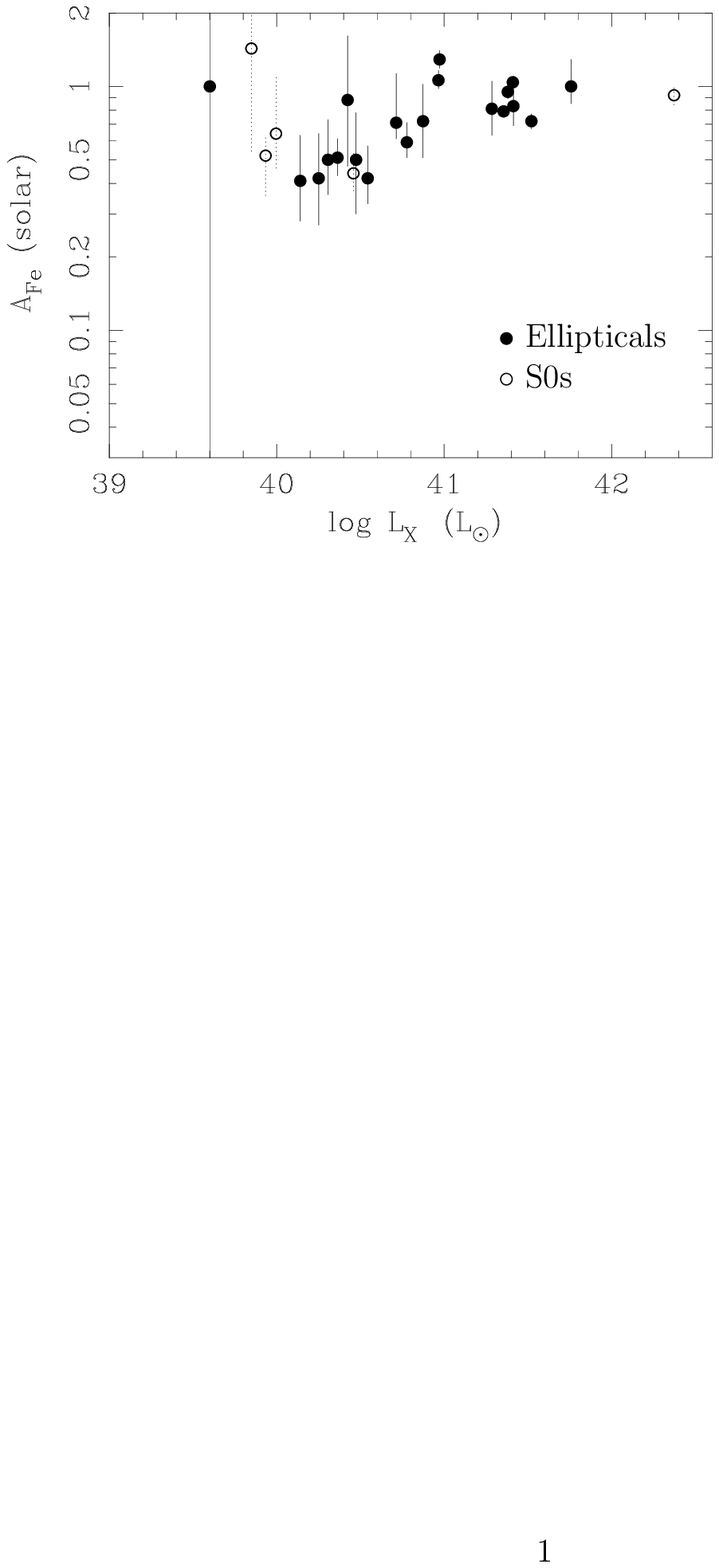,clip=}
\begin{fv}{13}{1cm}
{  $A_{\rm{Fe}}$ vs. ISM luminosity (0.5-10.0 keV) within $r<4 r_e$ when
$A_{\rm{\alpha}}$  is fixed to 1.0 solar. 
Open circles indicate S0 galaxies, and filled circles are elliptical galaxies.}
\end{fv}

\psfig{file=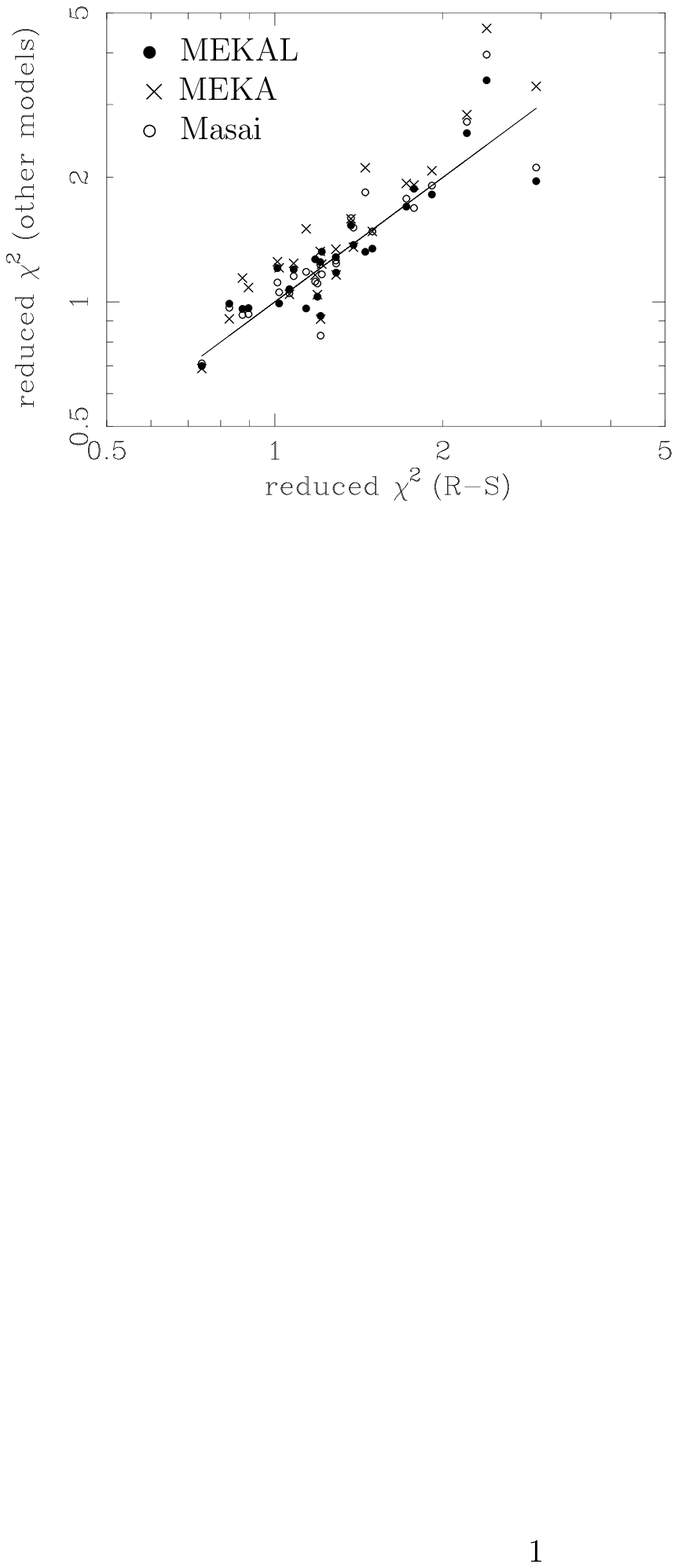,clip=}
\begin{fv}{14}{1cm}
{ The reduced $\chi^2$ of the spectral fits, obtained by MEKAL (closed circle),
  MEKA (crosses), and Masai (open circles) codes, plotted against those obtained by the
  R-S model. }
\end{fv}

\psfig{file=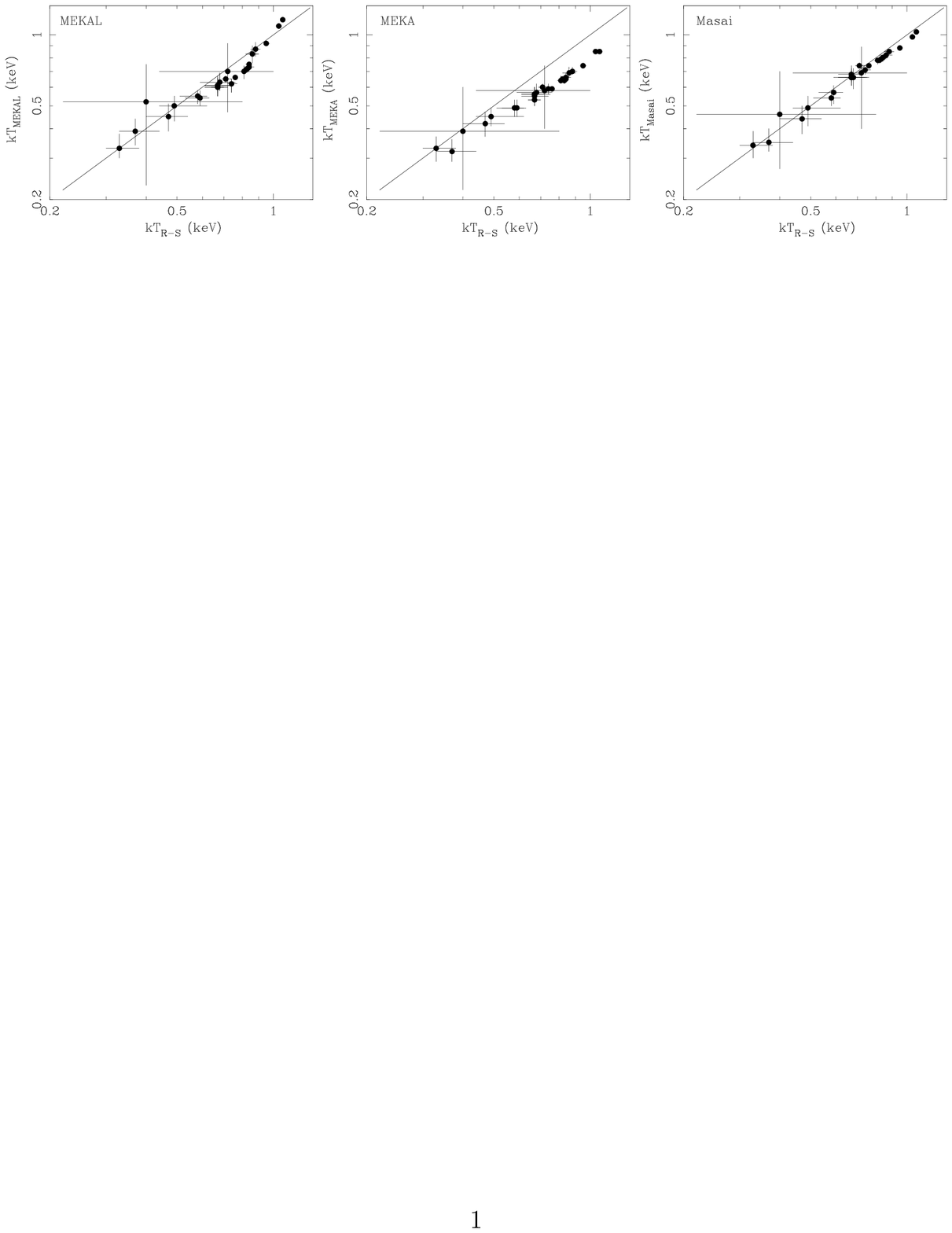,clip=}
\psfig{file=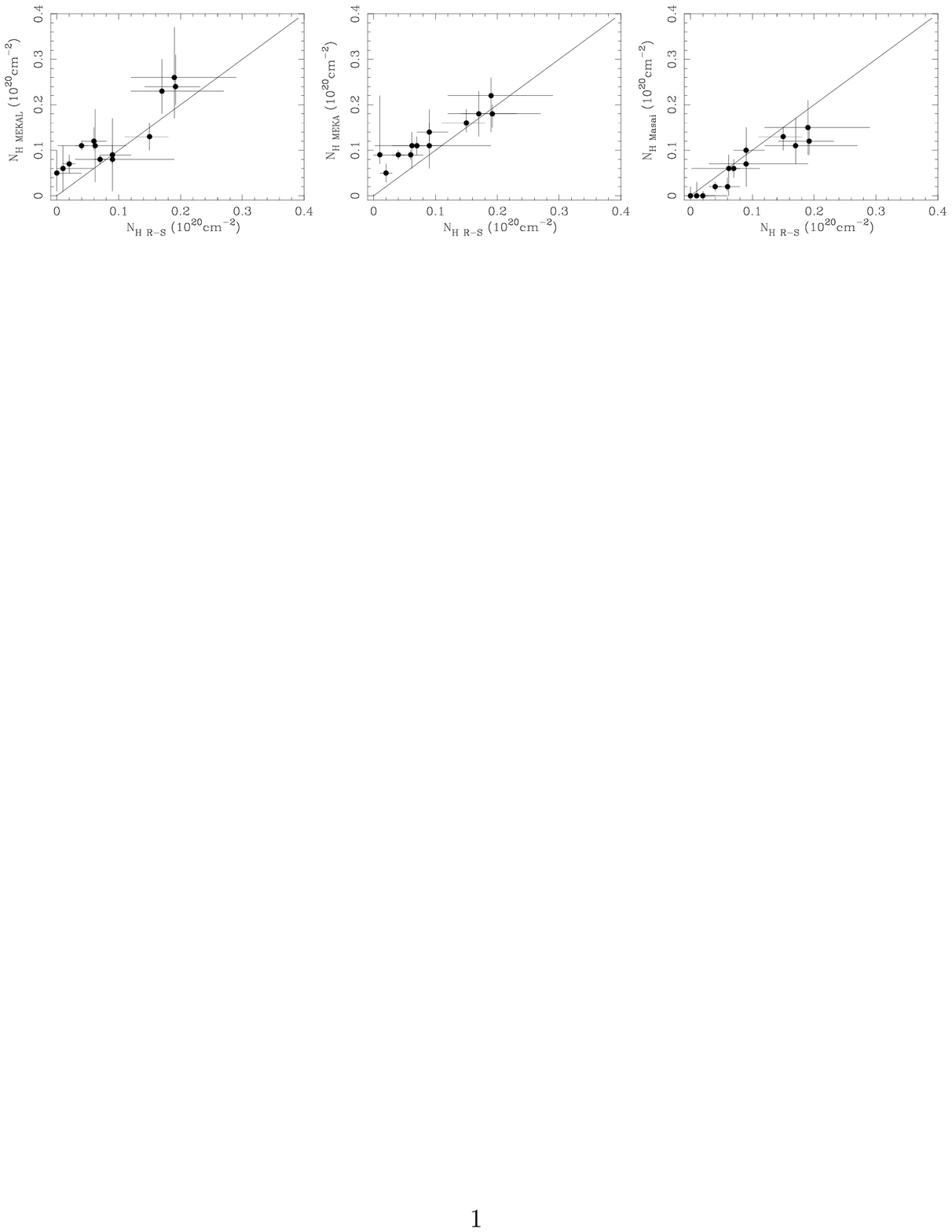,clip=}
\begin{fv}{15}{1cm}
{ The results from the MEKAL, MEKA, and Masai model fits,
  shown against those obtained with the R-S model for
  (a) temperature and (b) hydrogen column density.}
\end{fv}

\psfig{file=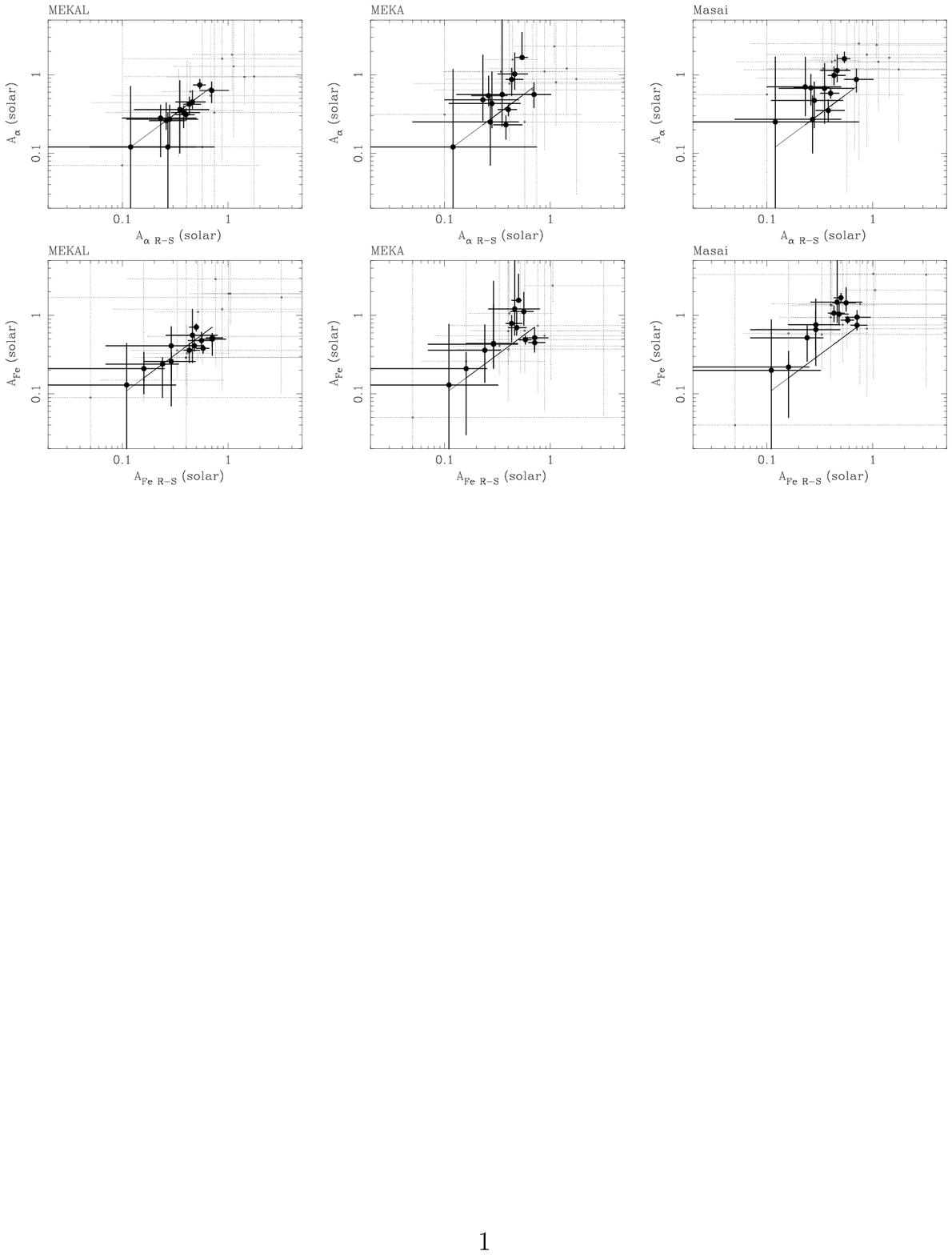,clip=}
\psfig{file=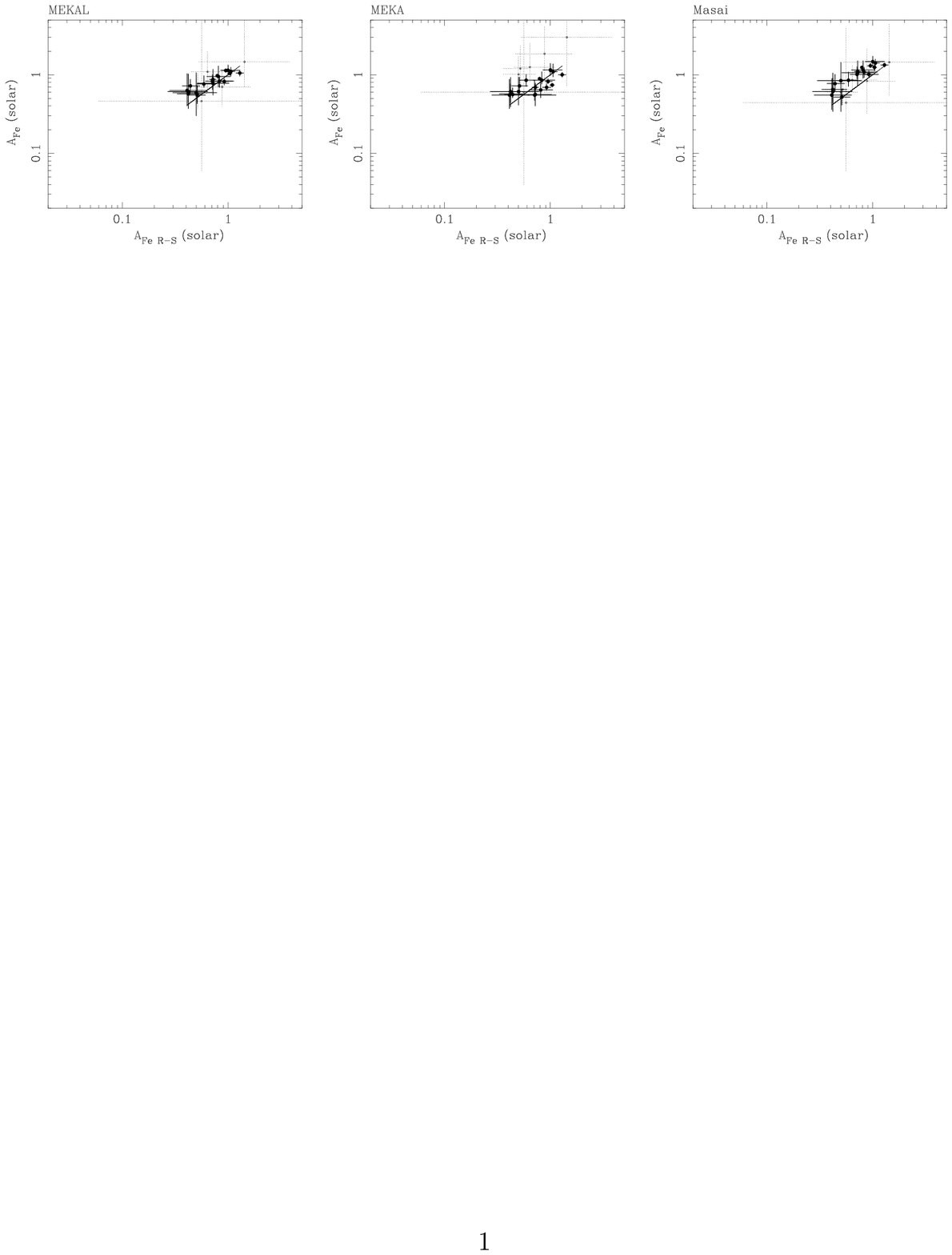,clip=}
\begin{fv}{16}{1cm}
{  The abundances derived by MEKAL, MEKA, and Masai models,
   shown against those obtained with the R-S model.
   In order to emphasize the high quality data, the points with large
 error bars are represented in dotted lines.
   (a) Both $A_{\rm{\alpha}}$  and  $A_{\rm{Fe}}$  are allowed to be free.
   (b) $A_{\rm{\alpha}}$  is fixed to be 1 solar,  and  $A_{\rm{Fe}}$  is plotted.}
\end{fv}

\centerline{
\psfig{file=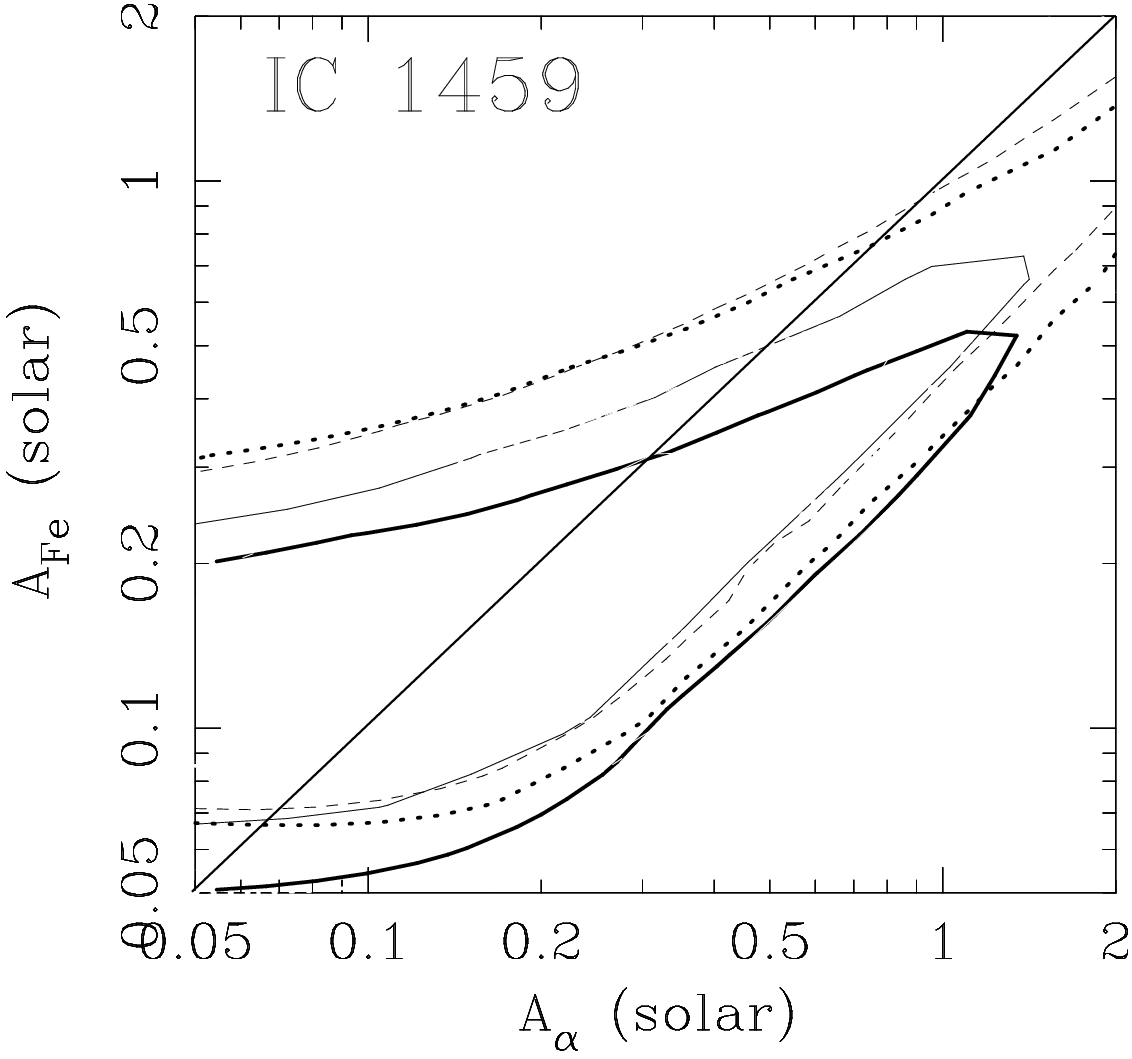,width=6cm}
\psfig{file=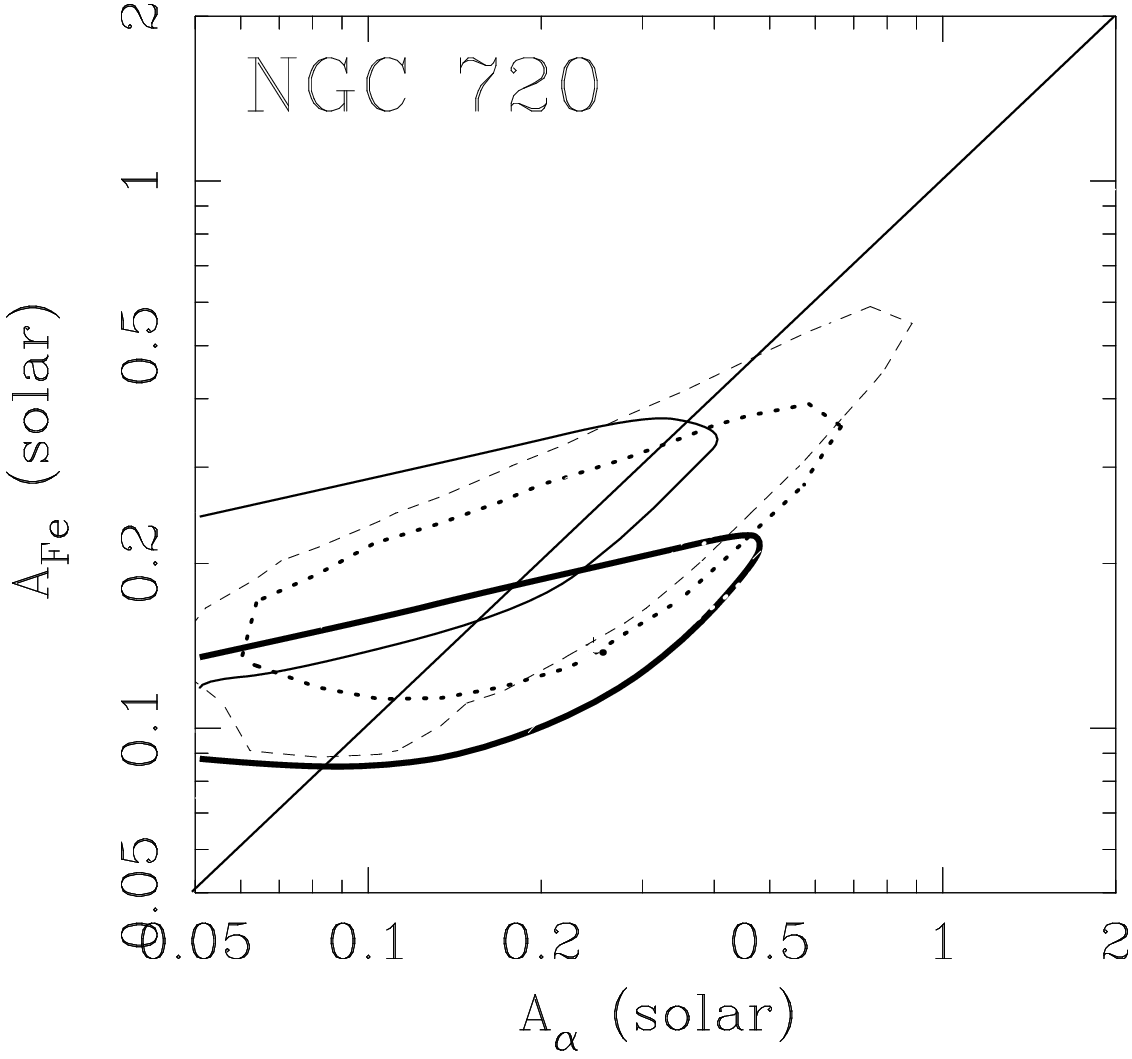,width=6cm}}
\centerline{
\psfig{file=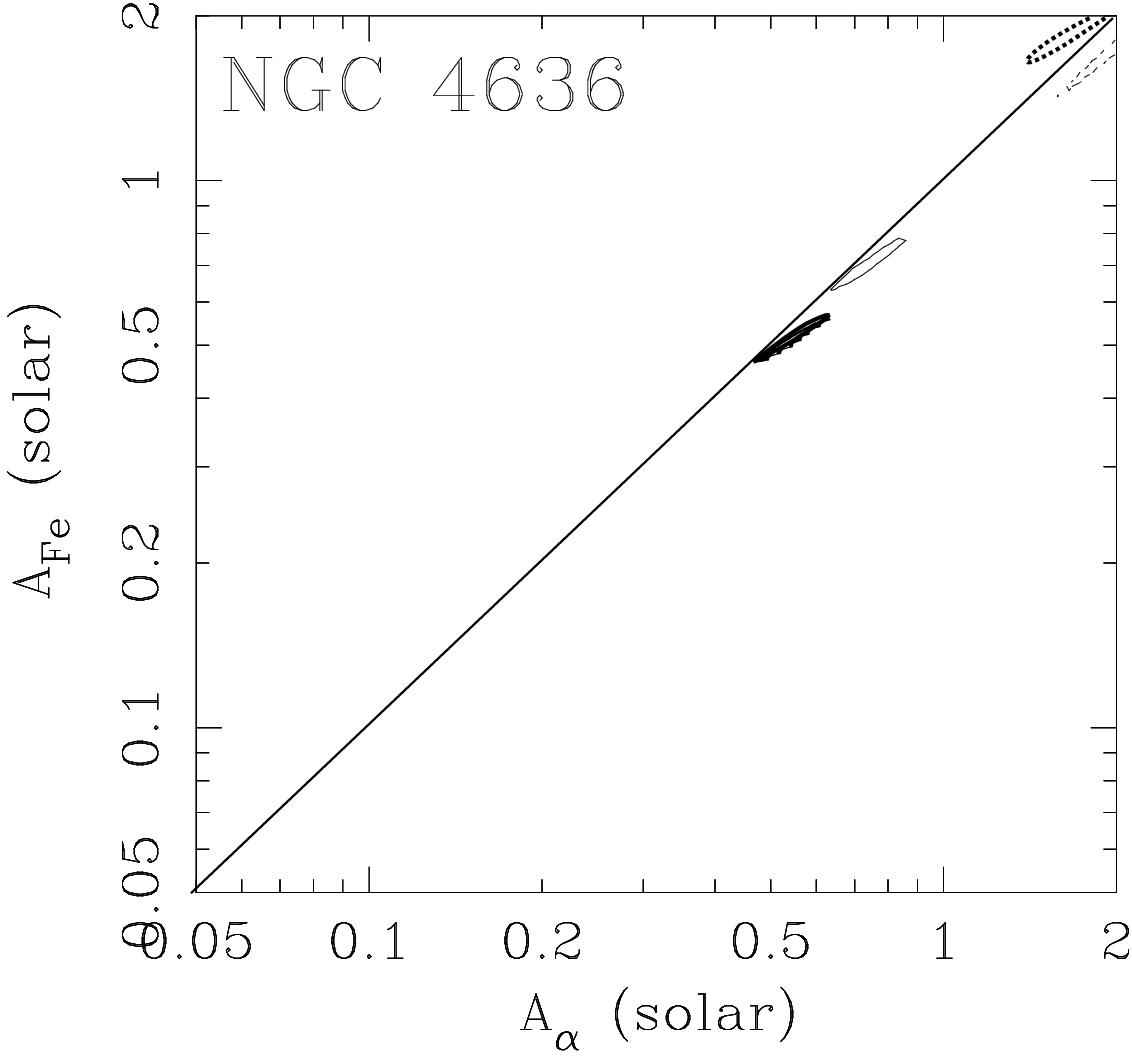,width=6cm}
\psfig{file=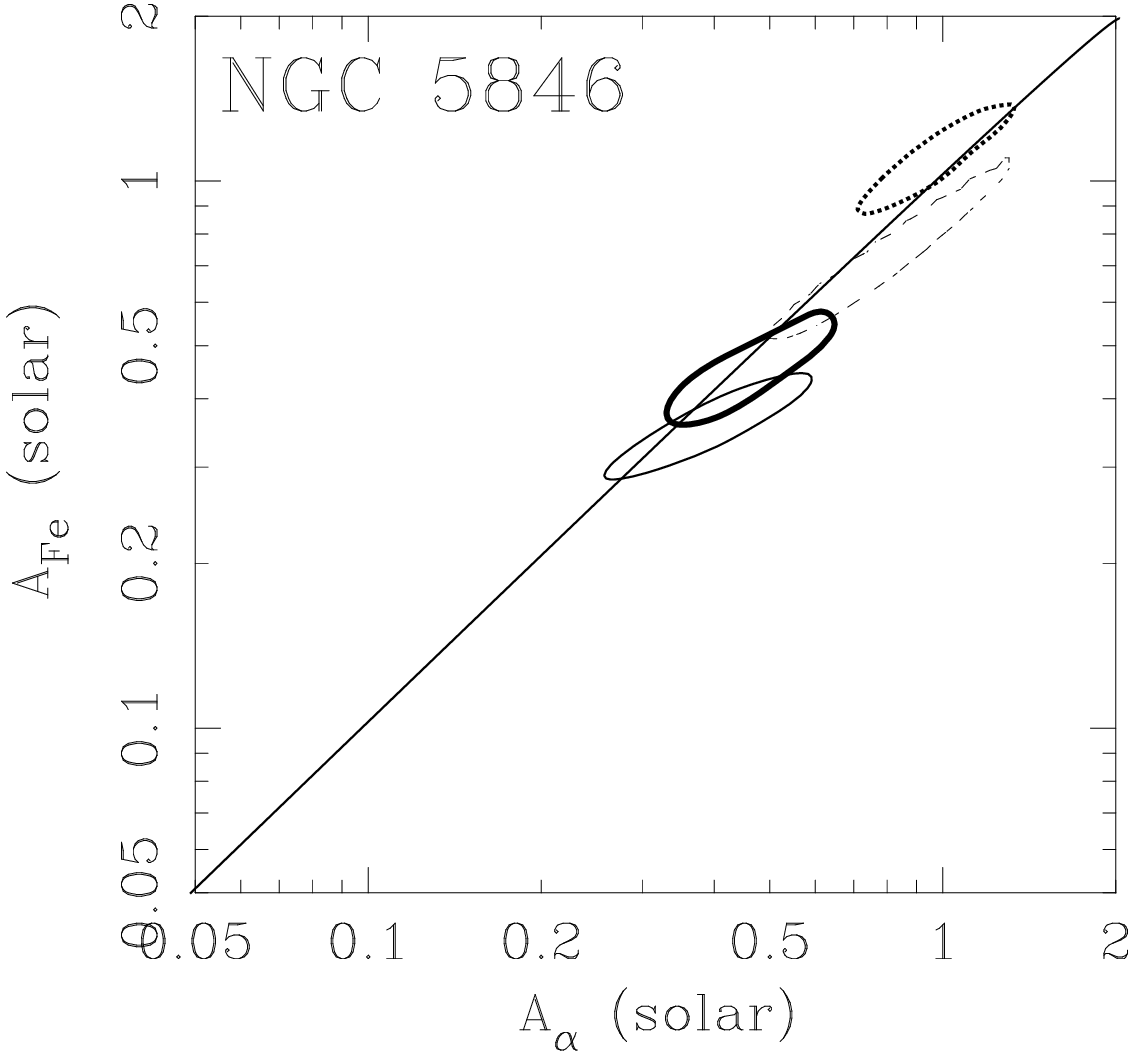,width=6cm}}
\begin{fv}{17}{1cm}
{ 90 \% confidence contours  of the correlation
    between $A_{\rm{Fe}}$ vs.  $A_{\rm{\alpha}}$ for four  galaxies, 
    obtained by R-S (thick-solid), MEKA (dashed), MEKAL (thin-solid), 
    and Masai (dotted) models.}
\end{fv}

\newpage
\centerline{
\psfig{file=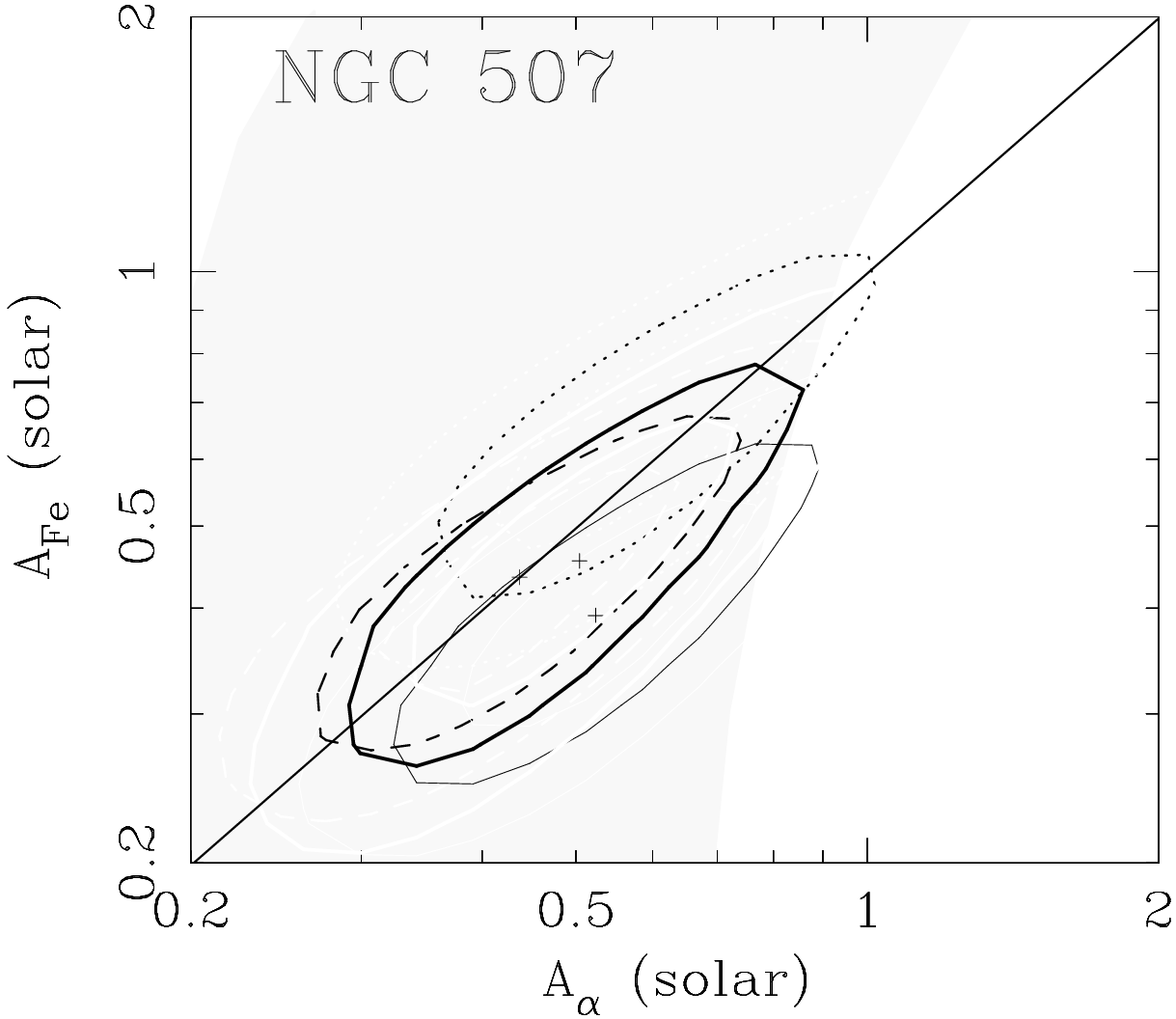,width=6cm,angle=-90}
\psfig{file=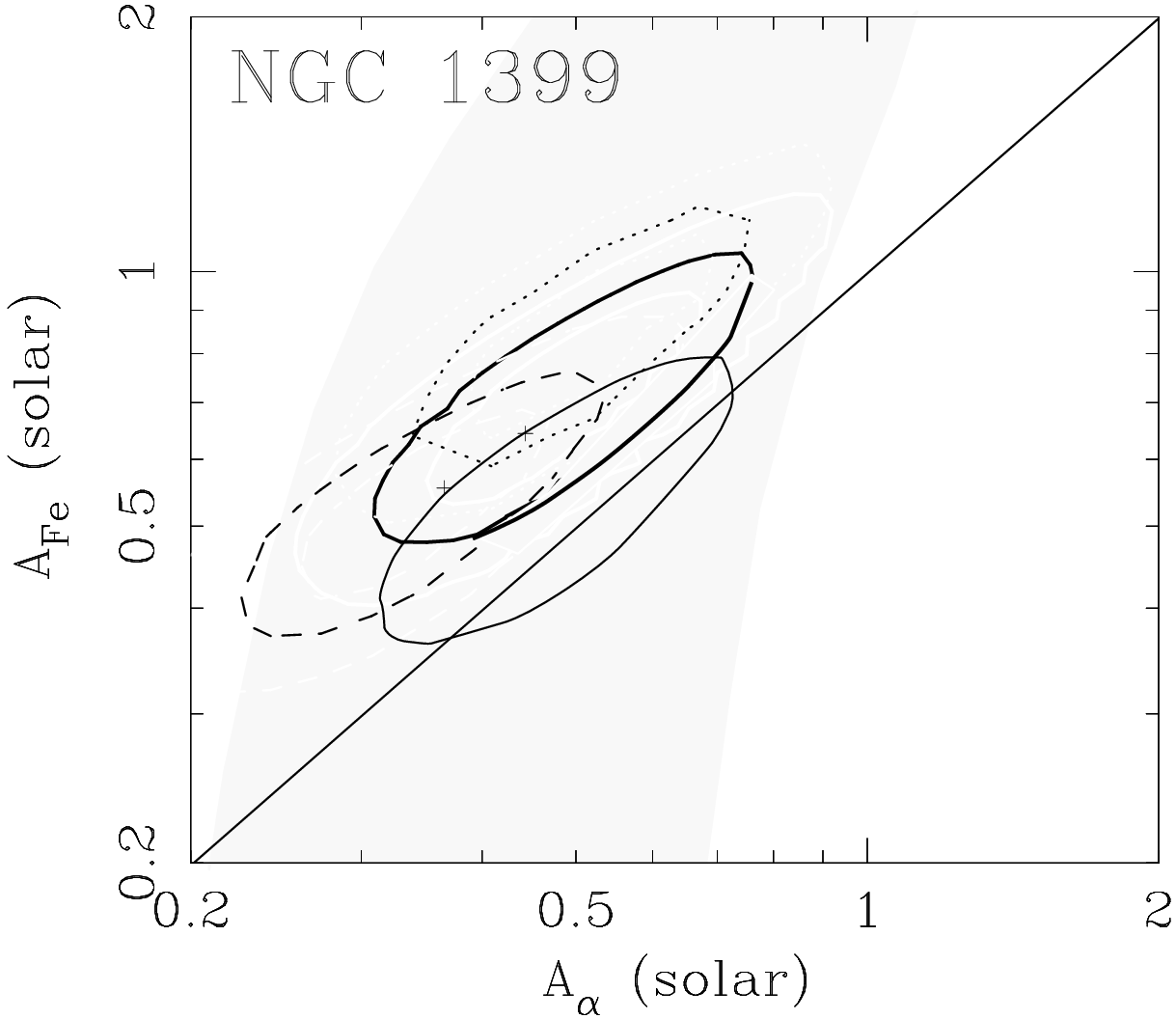,width=6cm,angle=-90}}
\centerline{
\psfig{file=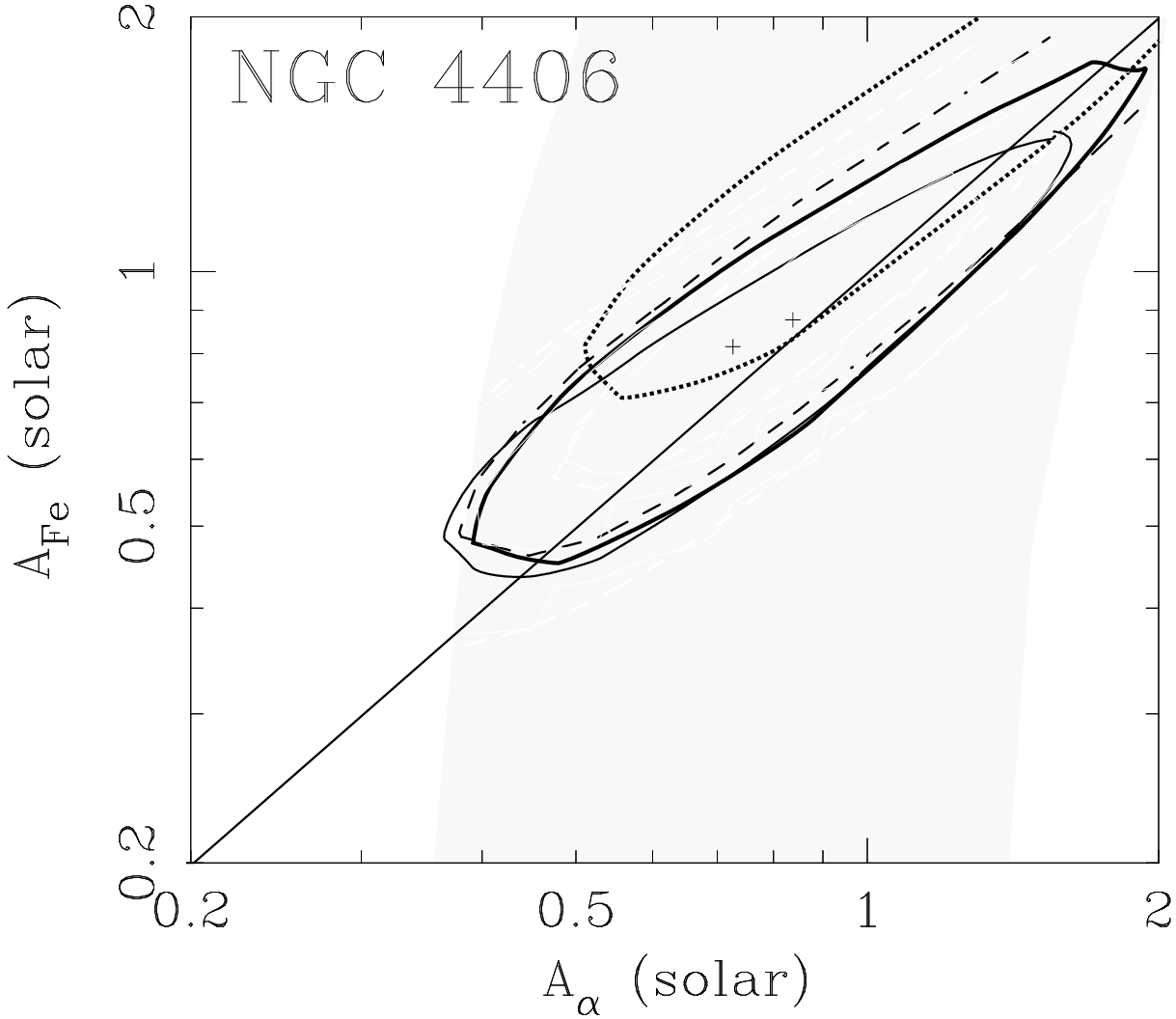,width=6cm,angle=-90}
\psfig{file=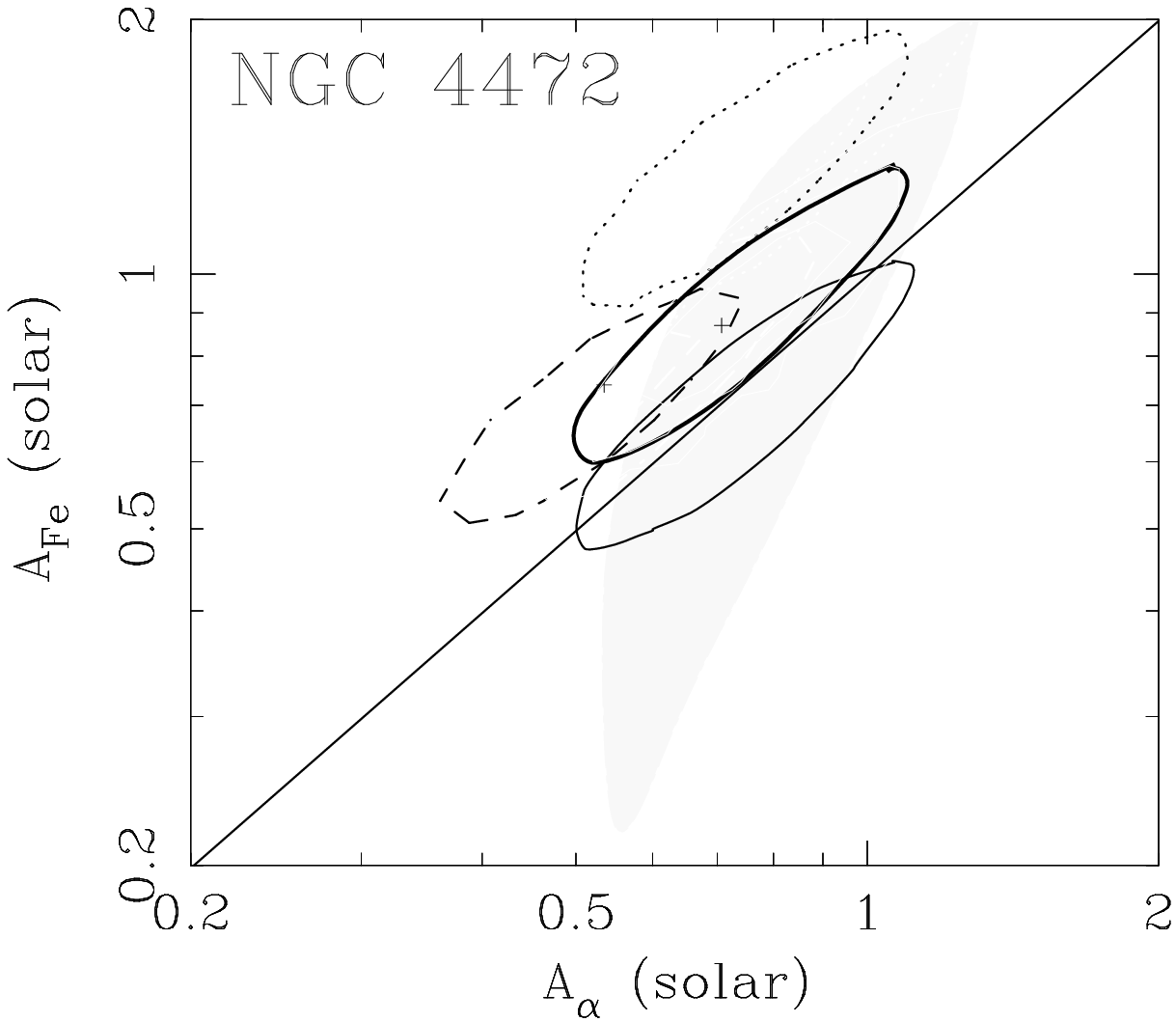,width=6cm,angle=-90}}
\centerline{
\psfig{file=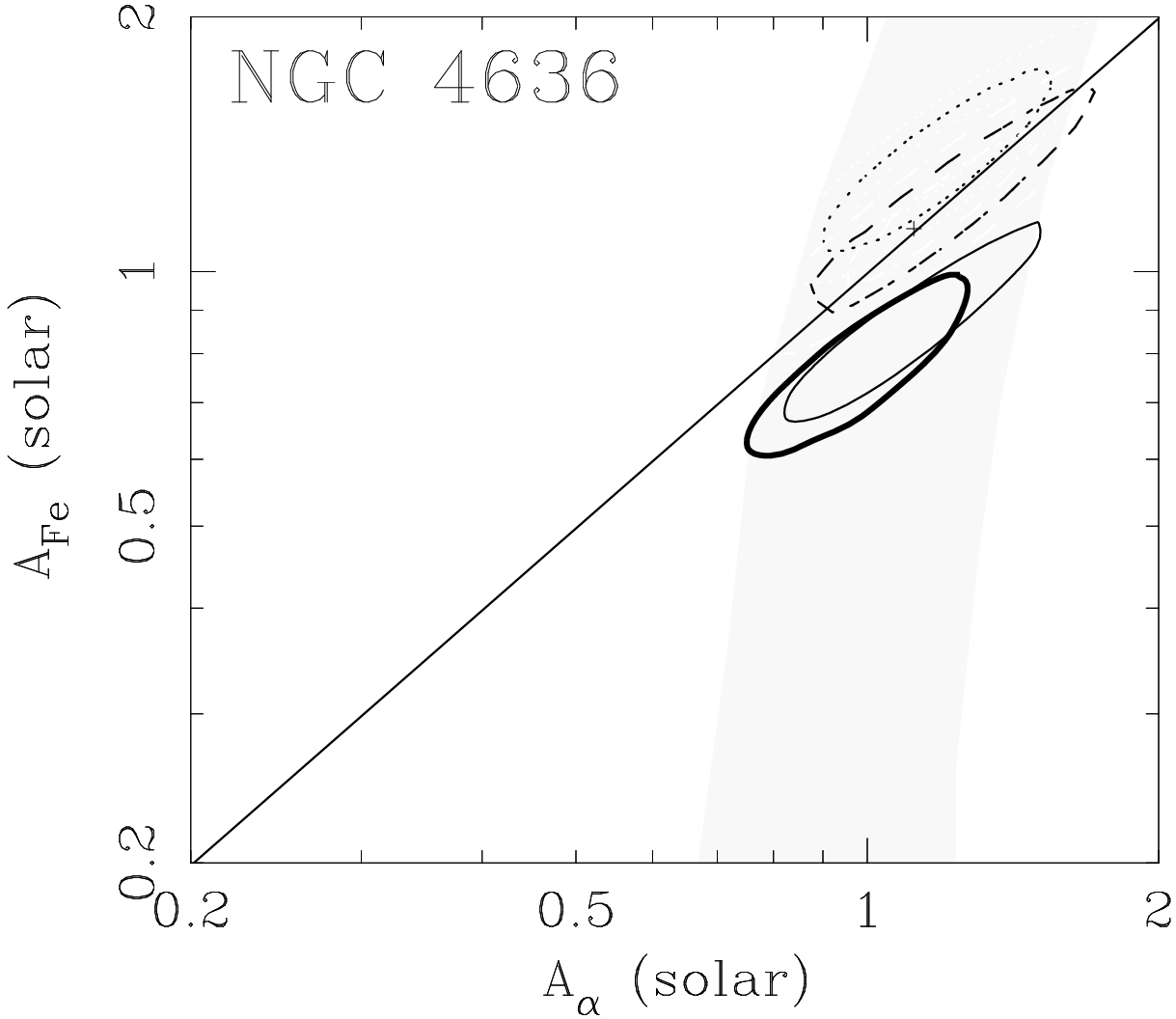,width=6cm,angle=-90}
\psfig{file=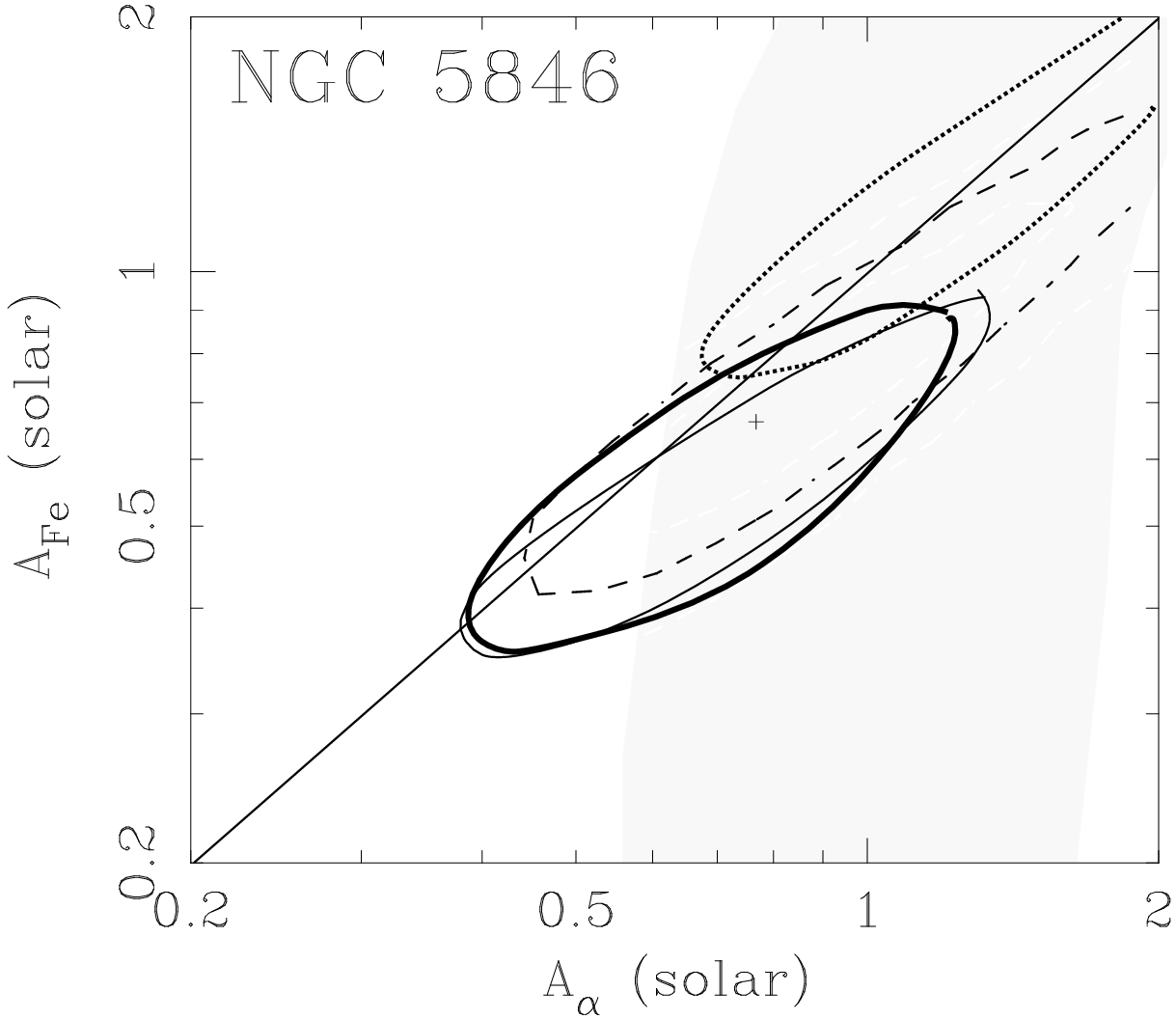,width=6cm,angle=-90}}
\begin{fv}{18}{1cm}
{ Confidence regions (90\%) of  $A_{\rm{Fe}}$ vs. $A_{\rm{\alpha}}$  obtained by fitting the
    spectra only above 1.6 keV (SIS) and 1.7 keV (GIS), using the R-S model
    (shaded region). Contours show the results obtained by
    R-S (solid), MEKAL (thin), MEKA (dash) and Masai (dot) models, when 
    20 \% systematic errors are included in the energy range of  0.4--1.6 keV.}
\end{fv}

\centerline{
\psfig{file=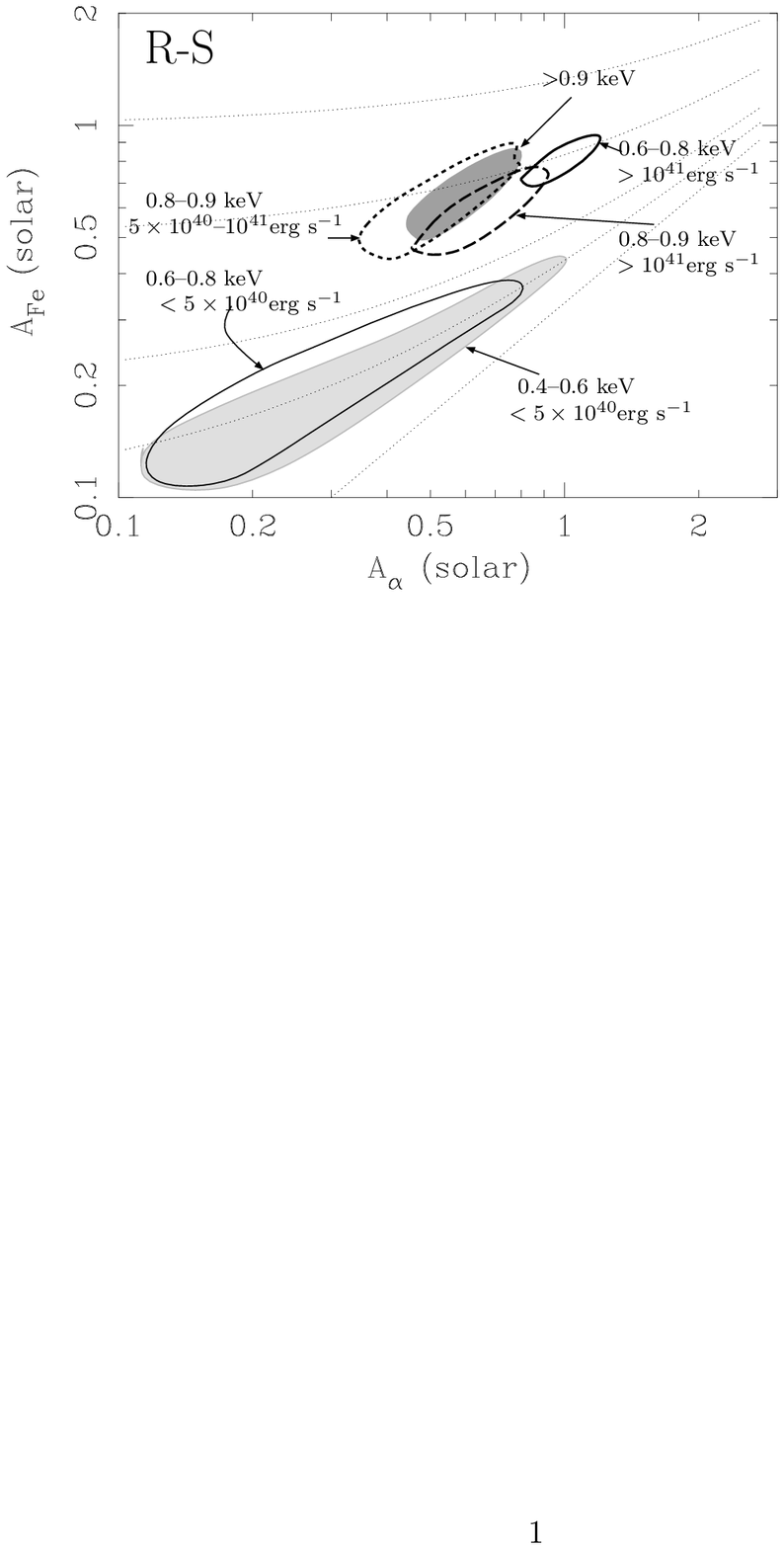,width=8cm,clip=}
\psfig{file=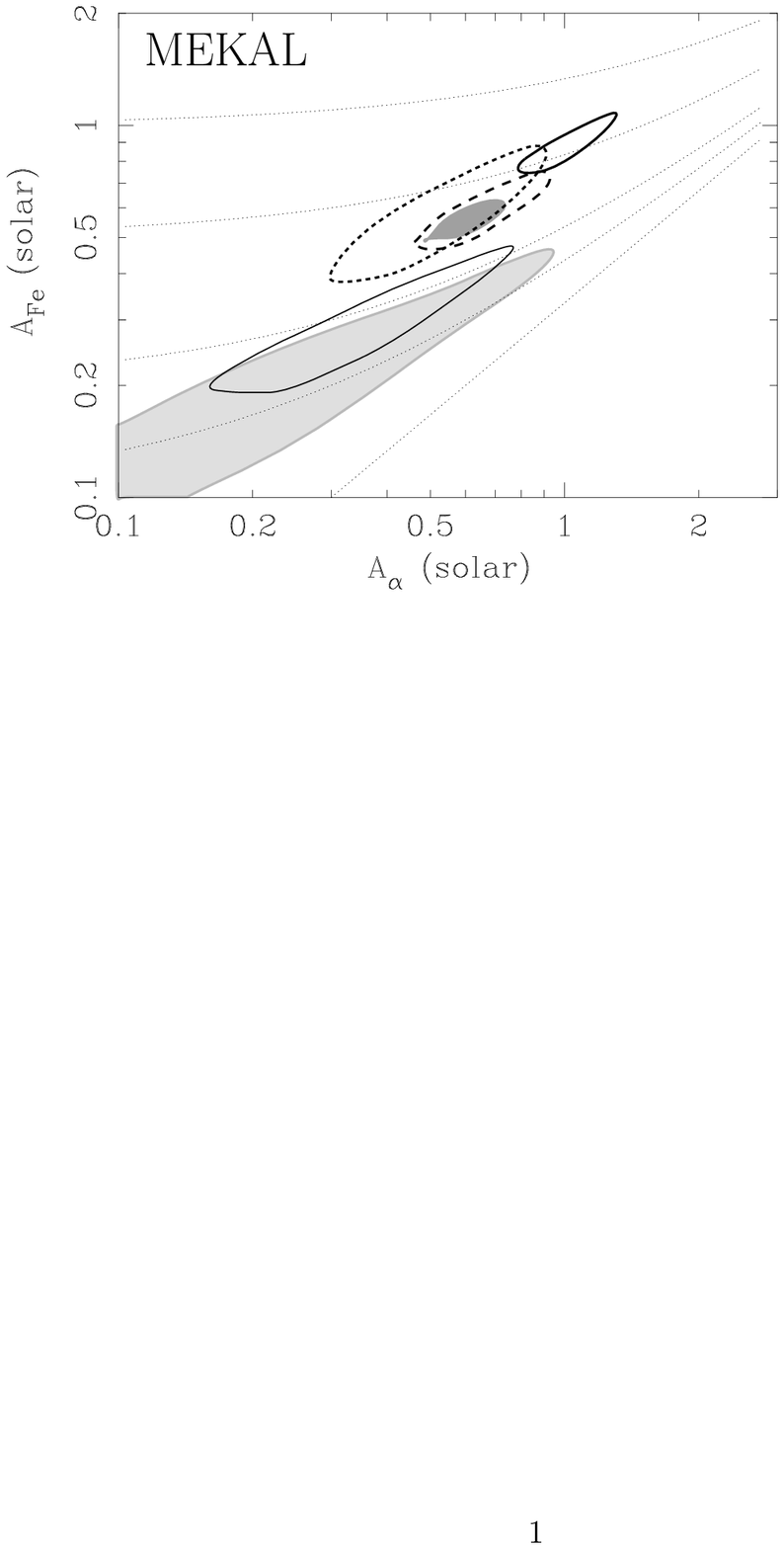,width=8cm,clip=}}
\centerline{
\psfig{file=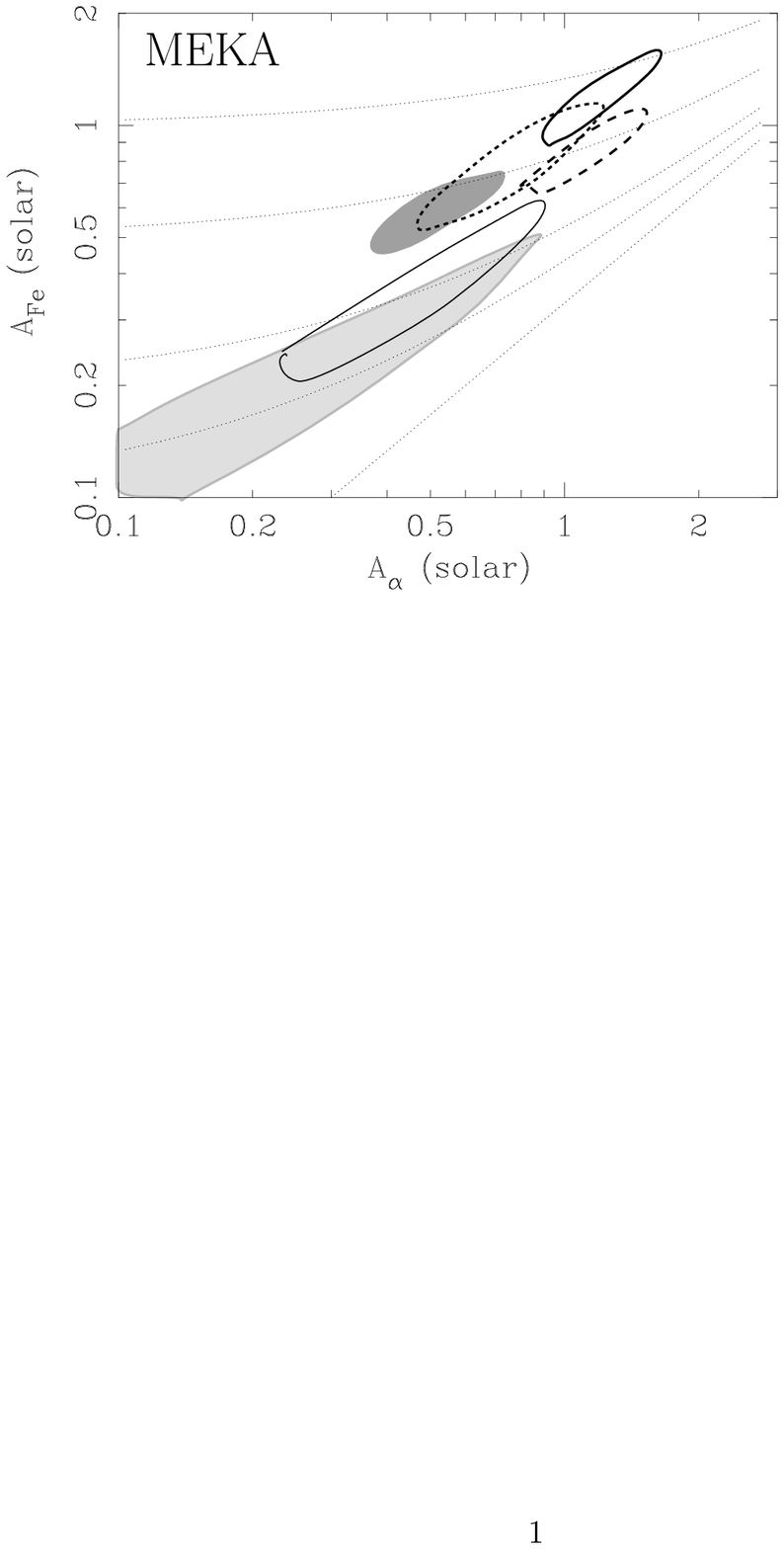,width=8cm,clip=}
\psfig{file=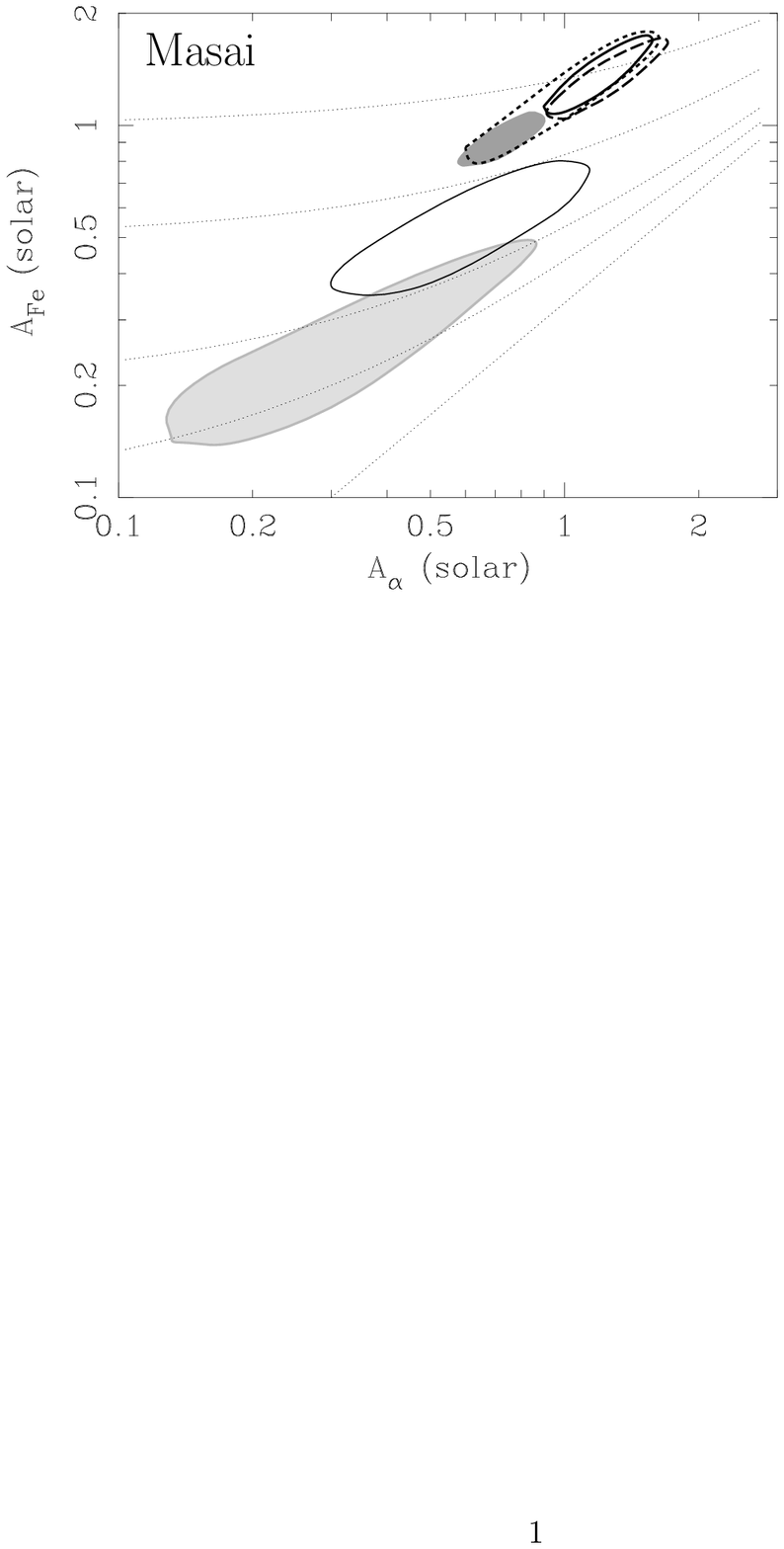,width=8cm,clip=}}
\begin{fv}{19}{1cm}
{Confidence contours of   $A_{\rm{\alpha}}$ vs $A_{\rm{Fe}}$ of galaxies.
Galaxies are allocated into 6 groups:
$kT>0.9$ keV (hatched region),
$0.8<kT<0.9$ keV and $L_X>10^{41}\rm{erg~ s^{-1}}$ (dased lines),
$0.8< kT<0.9$ keV and $5\times10^{40}<L_X<10^{41}\rm{erg~ s^{-1}}$ (dotted lines),
$0.6< kT<0.8$ keV and $L_X>10^{41}\rm{erg~ s^{-1}}$ (bold lines),
$0.6< kT<0.8$ keV and $L_X<5\times 10^{40}\rm{erg~ s^{-1}}$ (thin lines),
and $0.4< kT<0.6$ keV and $L_X<5\times 10^{40}\rm{erg~ s^{-1}}$
(dotted region).
Each dotted curve traces the abundance change when contribution
from SN II is varied with a fixed SN Ia imput. Different curves, on
the other hand, correspond to the change in the SN Ia contribution.}
\end{fv}

\psfig{file=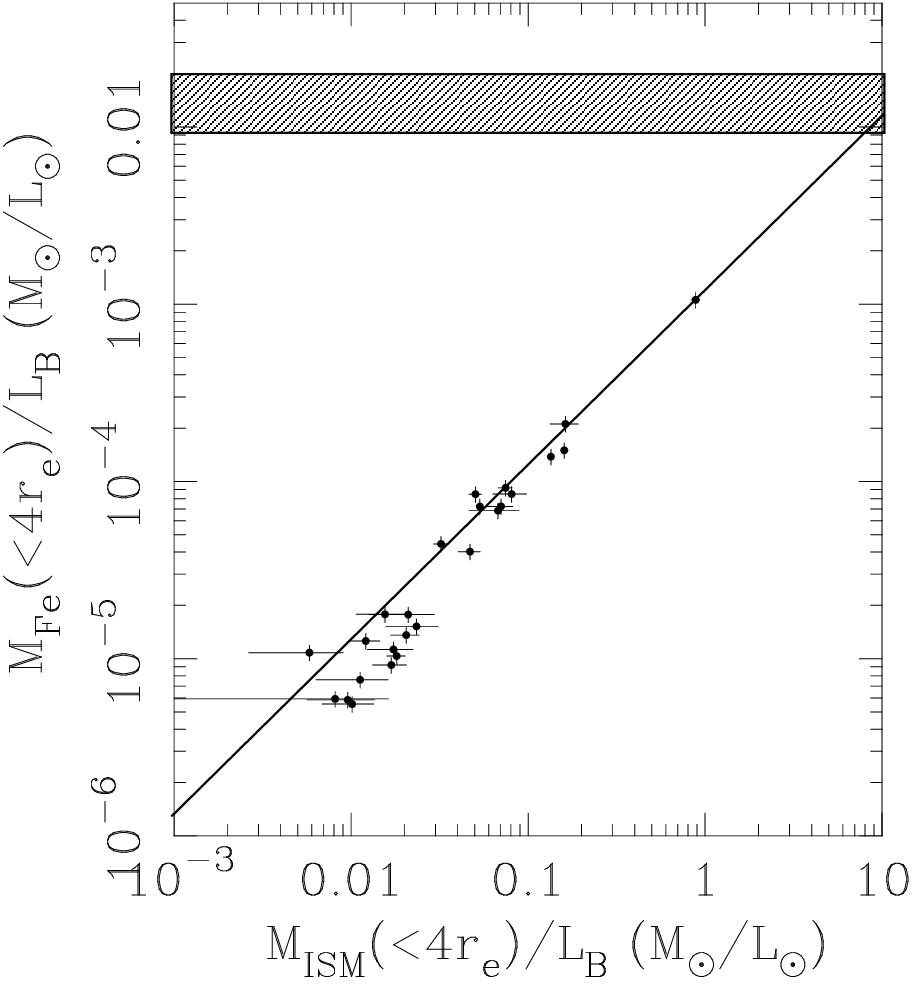,clip=}
\begin{fv}{20}{1cm}
{The ratio $M_{\rm Fe}(R<4r_e)/L_B$ (IMLR) are plotted against 
$M_{\rm{ISM}}(R<4r_e)/L_B$. 
The hatched region represents the IMLR for clusters.
The solid line represents the relation of
IMLR $\propto M_{\rm ISM}(R<4r_e)/L_B$ whch fits the data for X-ray luminous objects. }
\end{fv}

\end{document}